\documentclass{emulateapj}
\usepackage[dvips]{epsfig}
\usepackage[dvips]{color}
\usepackage{amsmath}    % need for subequations

\begin{document}
\slugcomment{To appear in Vol. 5 of ``Planets, Stars and Stellar Systems'', by Springer, in 2012}
\shorttitle{\sc METAL--POOR STARS}

\shortauthors{FREBEL \& NORRIS}

\newcommand{\my}   {metal-poor}
\newcommand{\lya}  {Lyman-$\alpha$}
\newcommand{\cd}   {CD$-38^{\circ}\,245$}
\newcommand{\bd}   {BD$-18^{\circ}\,5550$}
\newcommand{\hen}  {HE~0107--5240}
\newcommand{\hea}  {HE~1327--2326}
\newcommand{\hej}  {HE~0557--4840}
\newcommand{\sdss} {SDSS~J102915+172927}
\newcommand{\kms}  {\rm km~s$^{-1}$}
\newcommand{\feh}  {[Fe/H]} 
\newcommand{\teff} {$T_{\rm eff}$} 
\newcommand{\logg} {log~$g$} 
\newcommand{\loggf} {log~$gf$} 

\title{METAL-POOR STARS AND THE CHEMICAL ENRICHMENT OF THE UNIVERSE}

\author {ANNA FREBEL}
\affil {Harvard-Smithsonian Center for Astrophysics, 60 Garden St., Cambridge, MA
02138, USA; email: afrebel@cfa.harvard.edu}

\author {JOHN E. NORRIS}

\affil {Research School of Astronomy \& Astrophysics, The
Australian National University, \\Mount Stromlo Observatory, Cotter
Road, Weston, ACT 2611, Australia; email: jen@mso.anu.edu.au}

\begin{abstract}   

Metal-poor stars hold the key to our understanding of the origin of
the elements and the chemical evolution of the Universe. This chapter
describes the process of discovery of these rare stars, the manner in
which their surface abundances (produced in supernovae and other
evolved stars) are determined from the analysis of their spectra, and
the interpretation of their abundance patterns to elucidate questions
of origin and evolution.

More generally, studies of these stars contribute to other fundamental
areas that include nuclear astrophysics, conditions at the earliest
times, the nature of the first stars, and the formation and evolution
of galaxies -- including our own Milky Way.  This is
  illustrated with results from studies of lithium formed during the
Big Bang; of stars dated to within $\sim$1 Gyr of that event; of the
most metal-poor stars, with abundance signatures very different from
all other stars; and of the build-up of the elements over the first
several Gyr.  The combination of abundance and kinematic signatures
constrains how the Milky Way formed, while recent discoveries of
extremely metal-poor stars in the Milky Way's dwarf galaxy satellites
constrain the hierarchical build-up of its stellar halo from small
dark-matter dominated systems.

Two areas needing priority consideration are discussed.
The first is improvement of abundance analysis techniques. While
one-dimensional, Local Thermodynamic Equilibrium (1D/LTE) model
atmospheres provide a mature and precise formalism, proponents of more
physically realistic 3D/non-LTE techniques argue that 1D/LTE results
are not accurate, with systematic errors often of order $\sim$0.5\,dex
or even more in some cases.  Self-consistent 3D/non-LTE analysis as a
standard tool is essential for meaningful comparison between the
abundances of metal-poor stars and models of chemical enrichment.

The second need is for larger samples of metal-poor stars, in
particular those with [Fe/H] $<$ --4 and those at large distances
($20-50$\,kpc), including the Galaxy's ultra-faint dwarf satellites.
With future astronomical surveys and facilities these endeavors will
become possible. This will provide new insights into small-scale
details of nucleosynthesis as well as large-scale issues such as
galactic formation.

\end{abstract}

\keywords {Galaxy: formation $-$ Galaxy: halo $-$ stars: abundances
$-$ early Universe $-$ nuclear reactions, nucleosynthesis, abundances}

\section{INTRODUCTION} 

A few minutes after the beginning of the Universe, the only chemical
elements that existed were hydrogen ($\sim$0.75 by mass fraction),
helium ($\sim$0.25), and a miniscule amount of lithium
($\sim$2${\times}10^{-9}$).  Today, some 13.7\,Gyr later, the mass
fraction of the elements Li -- U in the Milky Way galaxy stands at
$\sim$0.02, essentially all of it created by stellar nucleosynthesis.
Metal-poor stars provide the foundation for our understanding of the
intricate details of the manner in which this enrichment occurred.

The astronomer Carl Sagan summarized cosmic chemical evolution in just
one sentence, ``We are made from star stuff''.  Studying stars that
are extremely underabundant in their heavy elements (collectively
referred to as ``metals'') takes us right to the heart of this
statement. These objects allow us to study the origin of the elements
that were subsequently recycled in stellar generations over billions
of years until ending up in the human body.

The rationale for analyzing metal-poor stars is that they are
long-lived, low-mass objects, the majority of which are main-sequence
and giant stars that have preserved in their atmospheres the chemical
signatures of the gas from which they formed.  Given that the overall
Universe was largely devoid of metals at the earliest times, it is
generally assumed (and borne out by analysis) that low metallicity
indicates old age.  For these objects to be still observable, their
masses are of order 0.6 -- 0.8\,M$_{\odot}$.  By measuring their
surface composition today, one can ``look back'' in time and learn about
the nature of the early Universe.  Another vital assumption is that
the stellar surface composition has not been significantly altered by
any internal ``mixing'' processes or by external influences such as
accretion of interstellar material that would change the original
surface abundance.

Analysis of old, metal-poor stars to study the early Universe is often
referred to as ``stellar archaeology'' and ``near-field cosmology''.
This fossil record of local Galactic metal-poor stars provides unique
insight into the enrichment of the Universe, complementing direct
studies of high-redshift galaxies.

\subsection{The Role of Metal-Poor Stars}

The abundances of the elements in stars more metal-poor than the Sun
have the potential to inform our understanding of conditions from the
beginning of time -- the Big Bang -- through the formation of the
first stars and galaxies, and up to the relatively recent time when
the Sun formed.  An incomplete list of the rationale for studying
metal-poor stars includes the following.

\begin{itemize}

\item

The most metal-poor stars ([Fe/H] $\la-4.0$), with primitive
abundances of the heavy elements (atomic number Z $>$ 3), are most
likely the oldest stars so far observed.

\item

The lithium abundances of extremely metal-poor near
main-sequence-turnoff stars have the potential to directly constrain
conditions of the Big Bang.

\item

The most metal-poor objects were formed at epochs corresponding to
redshifts z $>$ 6, and probe conditions when the first heavy element
producing objects formed.  The study of objects with [Fe/H] $<$
--3.5 permits insight into conditions at the earliest times that is
not readily afforded by the study of objects at high redshift.

\item

They constrain our understanding of the nature of the first stars, the
initial mass function, the explosion of super- and hyper-novae,
and the manner in which their ejecta were incorporated into subsequent
early generations of stars.

\item

Comparison of detailed observed abundance patterns with the results of
stellar evolution calculations and models of galactic chemical
enrichment strongly constrains the physics of the formation and
evolution of stars and their host galaxies.

\item

In some stars with [Fe/H] $\sim-3.0$, the overabundances of the
heavy-neutron-capture elements are so large that measurement of Th and U is
possible and leads to independent estimates of their ages and hence of the
Galaxy.

\item 

Stars with [Fe/H] $\la$ --0.5 inform our understanding of the evolution
of the Milky Way system.  Relationships between abundance, kinematic,
and age distributions -- the defining characteristics of stellar
populations -- permit choices between the various paradigms of how the
system formed and has evolved.

\end{itemize} 

\subsection{Background Matters} \label{Sec:terms}

\subsubsection {Essential Reading}

The study of metal-poor stars for insight into the chemical evolution of
the Universe has resulted in a rich literature, embracing diverse
areas.  The reader will find the following topics and reviews of
considerable interest.

For the context of the early chemical enrichment of the Universe, and
how one might use metal-poor stars to explore back in time to the Big
Bang see \citet{bromm&larson04}, \citet{frebel10}, and
\citet{pagel97}.  To understand how one determines the chemical
abundances of stars, the important abundance patterns, and how
reliable the results are, refer to \citet{wheeleretal89},
\citet{snedenetal08}, and \citet{asplund05}.  Other relevant questions
and reviews include the following.  How does one discover metal-poor
stars: \citet{beers&christlieb05}.  What is the role of abundance in
the stellar population paradigm: \citet{sandage86},
\citet{gilmoreetal89}, and \citet{freemanbh02}.  How do the abundances
constrain galactic chemical enrichment: \citet{mcwilliam97}.  What
progress has been made in understanding the supernovae and hypernovae
that produce the chemical elements: \citet{timmesetal95},
\citet{arnett96}, and \citet{kobayashietal06}, and references therein.
These reviews are of course not one-dimensional, and in many cases
they describe matters in several of the topics highlighted above.
They will repay close reading by the interested student.

\subsubsection {Abundance Definitions} \label{Sec:definitions}

Most basically, $\epsilon$(A), the abundance of element A is presented
logrithmically, relative to that of hydrogen (H), in terms of N$_{\rm
  A}$ and N$_{\rm H}$, the numbers of atoms of A and H.

\begin{equation*}
      \log_{10}{\epsilon} {\rm{(A)}} = \log_{10}(N_{\rm A}/N_{\rm H}) + 12
\end{equation*}

\noindent (For lithium, the abundance is mostly expressed as A(Li) =
log$\epsilon$(Li); and for hydrogen, by definition,
log$_{10}{\epsilon}$(H) = 12.)  For stellar abundances in the
literature, results are generally presented relative to their values
in the Sun, using the so-called ``bracket notation'',

\begin{equation*}
{\rm{[A/H]}} = \log_{10}(N_{\rm A}/N_{\rm H})_\star - \log_{10}(N_{\rm A}/N_{\rm H})_\odot 
\end{equation*}
\vspace{-0.5cm}
\begin{equation*}
 = \log_{10}\epsilon(A)_\star - \log_{10}{\epsilon}(A)_\odot, 
\end{equation*}

\noindent and for two elements A and B, one then has

\begin{equation*}
{\rm {[A/B]}} = \log_{10}(N_{\rm A}/N_{\rm B})_\star - \log_{10}(N_{\rm A}/N_{\rm B})_\odot 
\end{equation*}

\noindent 
In the case of the Fe metallicity, [Fe/H] = $\log_{10}(N_{\rm
  Fe}/N_{\rm H})_\star - \log_{10}(N_{\rm Fe}/N_{\rm H})_\odot$. For
  example, [Fe/H] = --4.0 corresponds to an iron abundance 1/10000
  that of the Sun.

For completeness, it should be noted that with the bracket
notation one needs to know the abundance not only of the star being
analyzed, but also of the Sun, the chemical composition of which has
recently been revised substantially for some elements
\citep{asplund09}.

%%%%%%%%%%%%%%%%%%%%%%%%%%%%%%%%%%%%%%%%%%%%%%%%%%%%%%%%%%%%%%%%%%%%%%%%%%%%%%%%%%
% Table 1

\begin{deluxetable*}{lcc}
\tabletypesize{\small}
\tablecolumns{3}
\tablewidth{0pt}
\tablecaption{\label{Tab:definitions} Metal-Poor Star Related Definitions}
\tablehead{
\colhead{Description} & 
\colhead{Definition}    & 
\colhead{Abbreviation\tablenotemark{a}} 
  }
\startdata
Population\,III stars & postulated first stars, formed from zero-metallicity gas& Pop\,III\\
Population\,II stars & old (halo) stars formed from low-metallicity gas & Pop\,II\\
Population\,I stars & young (disk) metal-rich stars & Pop\,I\\\\ \hline\\

  Solar               & $\mbox{[Fe/H]}=0.0$  &      \\
  Metal-poor          & $\mbox{[Fe/H]}<-1.0$  &   MP \\
  Very metal-poor     & $\mbox{[Fe/H]}<-2.0$  &   VMP \\
  Extremely metal-poor& $\mbox{[Fe/H]}<-3.0$  &   EMP \\
  Ultra metal-poor    & $\mbox{[Fe/H]}<-4.0$  &   UMP \\
  Hyper metal-poor    & $\mbox{[Fe/H]}<-5.0$  &   HMP \\\\\hline\\

Carbon-rich stars     &  $\mbox{[C/Fe]} > +0.7$ for $\log (L/L_{\odot} ) \le 2.3$  &  CEMP\\
                      &  $\mbox{[C/Fe]} \ge (+3.0 - \log(L/L_{\odot} ))$ \mbox{for}
  $\log(L/L_{\odot}) > 2.3$&  CEMP \\
n-capture-rich stars  &  $0.3 \le \mbox{[Eu/Fe]} \le +1.0$ and $\mbox{[Ba/Eu]} < 0$ & r-I  \\
n-capture-rich stars  &  $\mbox{[Eu/Fe]} > +1.0$ and $\mbox{[Ba/Eu]} < 0$           & r-II \\
n-capture-rich stars  &  $\mbox{[Ba/Fe]} > +1.0$ and $\mbox{[Ba/Eu]} > +0.5$        & s    \\
n-capture-rich stars  &  $0.0 < \mbox{[Ba/Eu]} < +0.5$                              & r/s \\ 
n-capture-normal stars  &  $\mbox{[Ba/Fe]} < 0$                    & no \\ 
\enddata
%\vspace{-0.8cm}
\tablecomments{Carbon-rich stars appear with r- and s-process
enhancements also. The CEMP definitions are from
\citet{aoki_cemp_2007} and differ somewhat  from \cite{beers&christlieb05}.}
\tablenotetext{a}{Commonly used in the literature.}
\end{deluxetable*}

%%%%%%%%%%%%%%%%%%%%%%%%%%%%%%%%%%%%%%%%%%%%%%%%%%%%%%%%%%%%%%%%%%%%%%%%%%%%%%%%%%

\subsubsection {Nomenclature}

\citet{baade44}, in his seminal paper on the subject, defined two
groups of stars, Type\,I and Type\,II, which today are referred to as
Population\,I and Population\,II.  The first referred to young stars,
including open clusters, which reside in the disk of the Galaxy, while
the second includes its globular clusters, and essentially all of its
known metal-poor stars.  In what follows, Population\,II will be
referred to as the ``halo'', which defines the spatial distribution of
the population. It has been speculated that a so-called
Population\,III exists, which comprises the elusive first stars.  With
the advent of detailed cosmological simulations of primordial star
formation, the term ``Population\,III'' is now widely used only for
stars that first formed from zero-metallicity gas that consisted only
of hydrogen, helium and traces of lithium.  The most metal-poor stars
currently known are thus extreme members of Population\,II.

Following \citet{beers&christlieb05} (with some modifications and
additions) the nomenclature listed in
Table~\ref{Tab:definitions} will be adopted for different
types of metal-poor stars in terms of population, metallicity and
chemical signatures. As can be seen, the main metallicity indicator is
the iron abundance, [Fe/H].  Iron has the advantage that among the
elements it has the richest absorption line spectrum in the optical
region, facilitating determination of Fe abundance independent of the
wavelength range covered by the spectrum.  With few exceptions, [Fe/H]
traces the overall metallicity of the stars fairly well.

\subsection  {Plan of Attack} \label{Sec:plan_attack}

For convenience, and the purposes of this chapter, the term
``metal-poor'' will be taken to mean stars in the Milky Way system
having [Fe/H] $<$ --1.0.  This embraces all of the ``metal-poor''
categories of \citet{beers&christlieb05} shown here in Table 1.  It
will confine our attention principally to field stars and globular
clusters of the Galactic halo and the Galaxy's dwarf galaxy
satellites.  Further, if one accepts the Galactic
  age-metallicity relationship presented, for example, by
  \citet{freemanbh02}, this restricts discussion to star formation and
  associated chemical enrichment that occurred during the first
  $\sim$4\,Gyr following the Big Bang.  Our abundance restriction
also includes part of the so-called ``metal-weak thick disk (MWTD)''
(see \citealt{chiba&beers00}) and the Galactic bulge, neither of which
will be discussed here.  Currently, no major works have been carried
out that attempt to elucidate differences between halo and MWTD
abundance patterns.  The bulge, on the other hand, is believed to be
the site of some of the very first star formation, the result of which
is seen today admixed with overwhelming later generations at the
Galaxy's center, precluding insight into its metal-poor population.
This is work for the future.

In Section 2, the search for metal-poor stars will be briefly
outlined, and what has been discovered so far.  Section 3 is concerned
with the manner in which chemical abundances are determined (and their
reliability). An overview of metallicity distribution functions (MDF)
of globular clusters, field stars, and satellite dwarf galaxies is
also given. A major focus of this chapter is an introduction to the
interpretation of the relative abundances, [X/Fe], and the
corresponding chemical patterns observed in metal-poor stars.  Against
this background, in Section 4, the body of metal-poor stellar
abundances is presented, and the general abundance trends are
discussed in light of expectations set by models of stellar evolution
and galactic chemical evolution (GCE).  In Section 5, age
determination in a small class of extremely metal-poor stars which
have huge r-process element enhancement is described.  In Section 6,
the implications of deduced abundances for the cosmogony of the early
Universe and the Milky Way system are considered.  Finally, in Section
7, the possibilities and challenges of the future are outlined.

\section{DISCOVERY -- THE SEARCH FOR NEEDLES IN THE HAYSTACK}

\subsection{Historical Perspective} \label{Sec:historical}

\begin{figure*}[!t]
\begin{center}
\includegraphics[width=10.0cm,angle=-90]{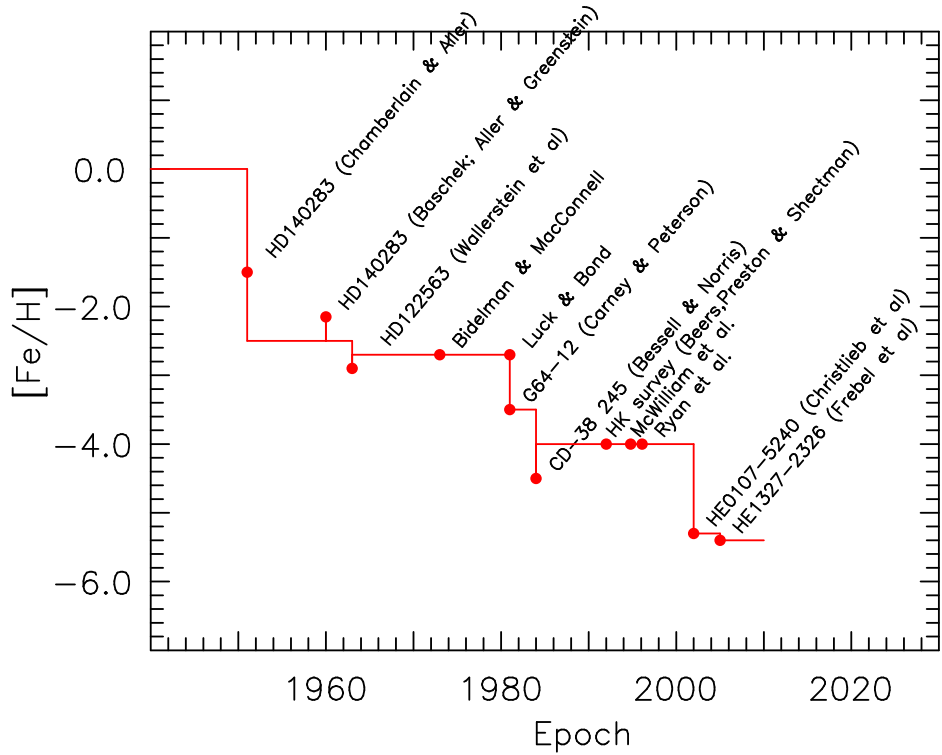}
\caption{\label{Fig:Pasadena}\small [Fe/H] for the most metal-poor
  star then known as a function of epoch.  The symbols denote the
  abundance determined by the authors, while the horizontal lines
  refer, approximately, to currently accepted values. (The abundances
  are based on one-dimensional, Local Thermodynamic Equilibrium model
  atmosphere analysis. See
  Section~\ref{Sec:abundance_determination}.)}
\end{center}
\end{figure*}

Chemical abundance (along with spatial distribution, kinematics, and
age) is one of the basic parameters that define stellar populations
(see e.g., \citealt{sandage86}).  In the middle of the twentieth
century, however, as noted by Sandage, ``There had grown up a general
opinion, shared by nearly all spectroscopists, that there was a single
universal curve of all the elements, and that the Sun and all the
stars shared ... precisely ... this property of identical ratios of
any element A to hydrogen''.  Subtle spectroscopic differences that
had been documented at that time were thought to result from
differences in physical conditions in the atmospheres of stars rather
than in chemical composition.  \citet{chamberlain51} profoundly
changed this concept with their chemical abundance analysis of the
``A-type subdwarfs'' HD~19445 and HD~140243, for which they reported
[Fe/H] ([Ca/H]) = --0.8 (--1.4) and --1.0 (--1.6), respectively. Their
work clearly established the existence of stars with elemental
abundances (relative to that of hydrogen) lower than in the Sun, and
that these lower abundances played a critical role in determining the
strength of their spectral features. (It should be noted in passing
that these early values are believed to have been overestimates, with
the currently accepted values for Fe being [Fe/H] $\sim-2.0$ and
--2.5, respectively.  See \citet[Footnote 2]{sandage86} for an
interesting sociological comment on the differences between the
earlier and current values.)  Soon after that work, \citet{b2fh57}
reviewed the case for the nucleosynthesis of almost all of the
chemical elements within stars.  In the decades that followed,
exhaustive searches for, and analysis of ``metal-poor'' stars -- as
illustrated in Figure~\ref{Fig:Pasadena} -- have led to the discovery
of stars with lower and lower values of [Fe/H] until, at time of
writing, two objects with [Fe/H] $\sim-5.5$ are known.

Two major developments, relevant to the present discussion, occurred
in parallel with the early chemical abundance analyses of stars.  The
first was the wide acceptance of the ``Big Bang'' paradigm as the most
likely description of the Universe.  The second was the demonstration
by \citet{wagoneretal67}, for example, that at the era of decoupling
of radiation and matter, some minutes after the singularity, no
elements beyond lithium had been produced (if isotropy and homogeneity
were assumed).

One might then enquire how best observationally to examine the manner
in which the chemical enrichment of the Universe proceeded.  A first
approach would be to investigate objects at high redshift, such as
galaxies and the {\lya} clouds seen in the spectra of quasars.
\citet{songaila01} and \citet{ryanweberetal09} report that for
redshifts z = 5.0 and 6.0, measures of Si IV and C IV (Songaila) and C
IV (Ryan-Weber et al.)  observed column densities imply intergalactic
metallicity $Z_{\rm IGM}$ $\ga$ 3.5$\times$10$^{-4}Z_{\odot}$ and (9
$\pm$ 5)$\times$10$^{-5}Z_{\odot}$, respectively. Assuming solar
abundance ratios, these intergalactic values correspond to [Fe/H]
$\ga$ --3.4 and --4.0.  It is also important to note in
  this context the recent analyses of very metal-poor Damped
  Lyman--$\alpha$ systems (see \citealt{cooke11}) that are currently
  observed out to redshifts z~$\sim$2 -- 3, and which report
  abundances of $\sim$6 elements, down to [Fe/H] $\sim-3.0$.
Far-field cosmological measurements thus currently reach to abundance
limits 30 times larger than those observed in the most metal-poor
stars in the Milky Way.  Further, while to date only C and Si are
observed at high redshift (z $\ga 5$), some 8 -- 9 elements are
measurable in Galactic stars observed to have [Fe/H] = --5.5
\citep{{HE0107_Nature}, {HE1327_Nature}}. That is to say, it seems
reasonable to suggest that the most metal-poor stars have the
potential to serve as the best cosmological probes of chemical
enrichment at the earliest times.

\vspace{0.2cm}
\subsection{Search Techniques} \label{Sec:search_techniques}

Metal-poor field stars are rare.  To begin with, the proportion of
stars in the solar neighborhood that belong to the halo population is
only $\sim$10$^{-3}$ (see e.g., \citealt{bahcall&soneira80}).
Further, as a rule of thumb, the simple chemical enrichment model of
the halo of Hartwick (1976; see Section~\ref{Sec:mdf_fieldstars}
below) suggests that the number of stars should decrease by a factor
of ten for each factor of ten decrease in abundance.  For example, the
number of stars with [Fe/H] $<$ --3.5 should be smaller by a factor
100 than the number with [Fe/H] $<$ --1.5. (For observational support
for this suggestion, down to [Fe/H] $\sim-4.0$, below which it breaks
down, see \citet{norris99}). Roughly speaking, given that the
stellar halo MDF peaks at [Fe/H] = --1.5, in the solar neighborhood
one might expect to find $\sim$1 in 200,000 stars with [Fe/H] $<$
--3.5.

One thus needs to filter out disk stars if one wishes to find
metal-poor stars.  While important bright extremely metal-poor stars
have been discovered somewhat serendipitously (e.g., the red giant
{\cd} with [Fe/H] = --4.0 and V = 12.8; \citealt{bessell&norris84}),
for stars brighter than B $\sim16$ this has to date been
systematically achieved in one of two ways.  The first uses the fact
that the halo does not share the rotation of the Galactic disk, and a
large fraction of its members have relatively high proper motions.
The first star with [Fe/H] $<$ --3.0 (G64-12;
\citealt{carney&peterson81}) was discovered in this way.  The major
surveys to date that utilized this technique are those of
\citet{ryan&norris91a} and \citet{carneyetal96}, whose samples each
comprise a few hundred halo main-sequence dwarfs with [Fe/H] $<$
--1.0, and who together report $\sim$10 stars having [Fe/H] $<$ --3.0.

The second method has been more prolific and utilizes objective-prism
spectroscopy with Schmidt telescopes, which permit one to
simultaneously obtain low-resolution spectra (resolving power R (=
$\lambda/\Delta\lambda$) $\sim400$) of many stars over several square
degrees.  Examination of the strength of the Ca II K line at
3933.6\,{\AA} with respect to that of nearby hydrogen lines or an
estimate of the color of the star permits one to obtain a first
estimate of whether the star is metal-weak or not.  Candidate
metal-poor stars are then observed at intermediate resolution (R
$\sim2000$) to obtain a measurement of the metal abundance of the
star. The techniques are described in detail by
\citet{beers&christlieb05}, who also document important surveys that
have obtained first abundance estimates for some tens of thousands of
stars brighter than B $\sim16.5$ with [Fe/H] $<$ --1.0.

The most important Schmidt surveys to date have been the HK survey
\citep{bps92} and the Hamburg/ESO Survey (HES)
\citep{christliebetal08}.  In order to give the reader an appreciation
of the scope and many steps involved in the process, here is a brief
description of the HES.  According to N. Christlieb, the HES consists
of some 12 million stars in the magnitude range $10 < B < 18$. In an
effective survey area of some 6700 square degrees, $\sim$21,000
candidate metal-poor stars were selected, for which, at the time of
writing, follow-up spectroscopy has been obtained of $\sim$5200.
Preliminary metal-poor candidates were selected in several steps to
arrive at candidate lists for which medium-resolution spectroscopy was
sought.  Due to limitations of telescope time and target faintness, it
was common that not all stars could be observed.  In the original
candidate list in the magnitude range 13.0 $\la$ $B$ $\la$ 17.5 there
were $\sim$3700 red giants of which about 1700 were observed at
medium-resolution \citep{schoerck}, together with $\sim$3400
near-main-sequence-turnoff stars, of which $\sim$700 have follow-up
spectroscopy \citep{lietal10}.  There is also a bright sample of
$\sim$1800 stars having $B <14.5$, for all of which medium resolution
spectra were obtained by \citet{frebel_bmps}.  From these samples, the
most metal-poor candidates were selected for high-resolution
observation.  Various considerations determined whether a star was
ultimately observed.  These include telescope time allocations,
observability and weather conditions during observing time, target
brightness, reliability of the medium-resolution result, science
questions to be addressed, and of course the preliminary metallicity
of the star.  Given these limitations, fainter stars remain unobserved
on the target lists due to time constraints.

\begin{figure*}[!tbp]
\begin{center}
\includegraphics[width=13.0cm,angle=0]{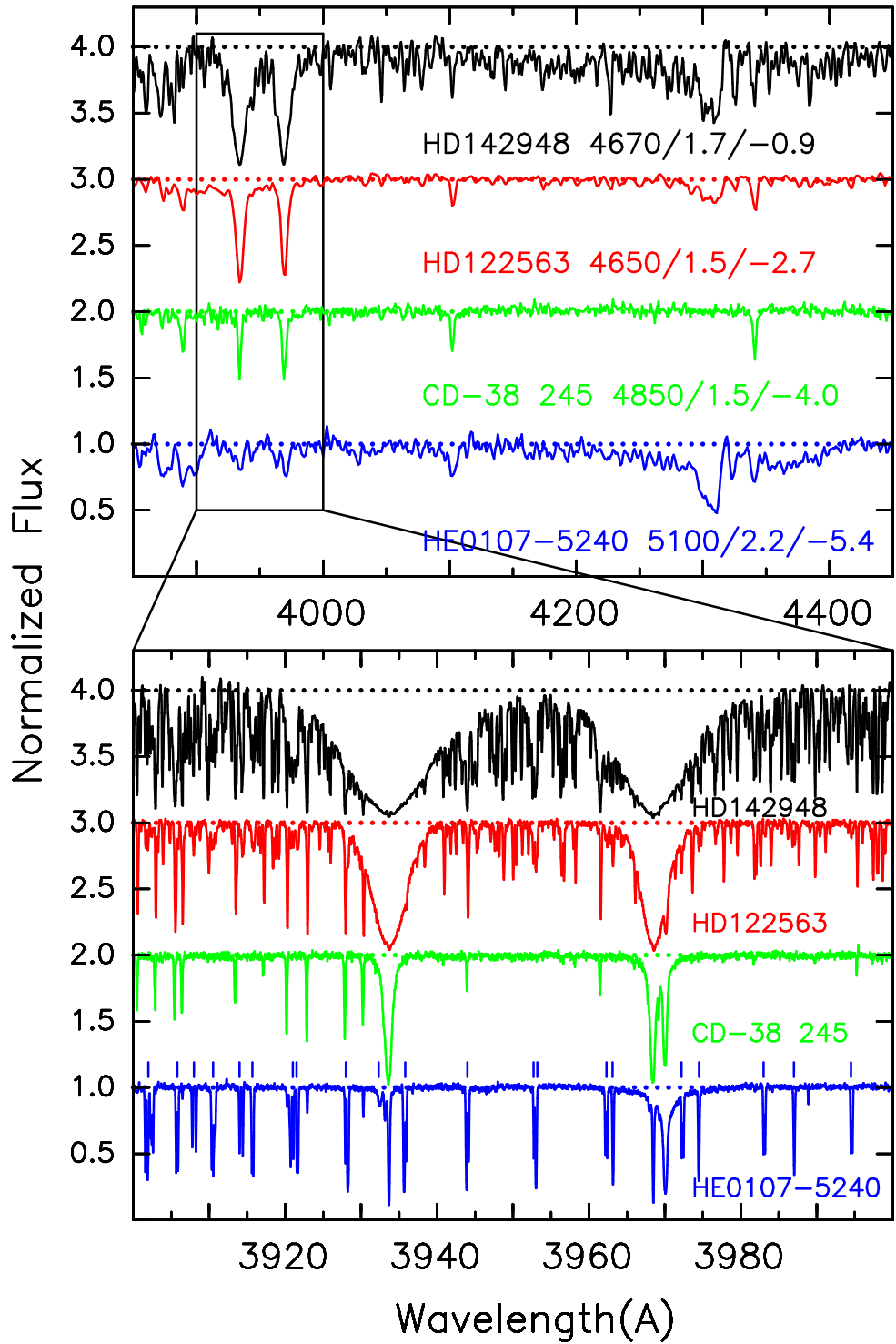}

\caption{\label{Fig:Hires}\small (Upper panel) Spectra at intermediate
  resolution (R $\sim$ 1600) of metal-poor red giants over the range
  --5.4 $<$ [Fe/H] $<$ --0.9.  Note the strong decrease in the
  strengths of the Ca II H \& K lines at 3933.6 and 3968.4\,{\AA}. The
  numbers in the panel represent {\teff}/{\logg}/[Fe/H].  (Lower
  panel) Spectra of the same stars at R $\sim$ 40000 on the range 3900
  -- 4000\,{\AA}.  (Note that while the Ca\,II\,H\,\&\,K lines are
  very weak in the most metal-poor giant, {\hen}, many more lines have
  appeared.  These are features of CH (the positions of which are
  indicated immediately above the spectrum) resulting from an
  extremely large overabundance of carbon relative to iron in this
  object.}
\end{center}
\end{figure*}

To this point the discussion has been confined to surveys that have
concentrated on discovering candidate metal-poor stars with $B$ $\la$
17.5, with follow-up medium-resolution spectroscopy complete in most
cases to only somewhat brighter limits.  Surveys that reach to
considerably fainter limits are the Sloan Digital Sky Survey (SDSS)
and the subsequent SEGUE-I and II surveys (see http://www.sdss.org),
which have obtained spectra with resolving power R $\sim2000$, and are
also proving to be a prolific source of metal-poor stars.  In a sample
of some 400,000 stars, SDSS/SEGUE has discovered 26,000 stars with
spectra having $S/N$ $>$ 10, and [Fe/H] $<$ --2.0 (based on these
intermediate-resolution spectra), while some 400 have [Fe/H] $<$
--3.0.

The search for metal-poor stars remains a very active field, with
several exciting projects coming to completion, currently in progress,
and planned.  This matter will be further discussed in
Section~\ref{Sec:conclusion}.

\subsection{High-Resolution, High $\mathbf{S/N}$ Follow-Up Spectroscopy}

The final observational step in the discovery process is spectroscopy
of the most significant objects (e.g., most metal-poor, or most
chemically peculiar) at very high resolving power (R $\sim$~10$^4$ --
10$^5$) and $S/N$ $\ga$ 100, in order to reveal the fine detail
required for the determination of parameters such as accurate chemical
abundances, isotope ratios, and in some cases stellar ages.  This is
best achieved with 6 -- 10\,m telescope/\'{e}chelle spectrograph
combinations -- currently HET/HRS, Keck/HIRES, Magellan/MIKE,
Subaru/HDS, and VLT/UVES.

In order to give the reader a feeling for both the role of increased
resolution and the manner in which decreasing metallicity affects the
observed flux, Figure~\ref{Fig:Hires} shows the increase in
spectroscopic detail between intermediate (R $\sim1600$) and high (R
$\sim40000$) resolving power for four metal-poor red giants of similar
effective temperature ({\teff}) and surface gravity ({\logg}) as metal
abundance decreases from [Fe/H] = --0.9 to --5.4 (for {\hen}, the most
metal-poor giant currently known).

\subsection{Census of the Most Metal-Poor Stars} \label{Sec:census}

\begin{figure}[!tbp]
\begin{center}
\includegraphics[clip=true,width=9.0cm,bbllx=67, bblly=270,
   bburx=500, bbury=576]{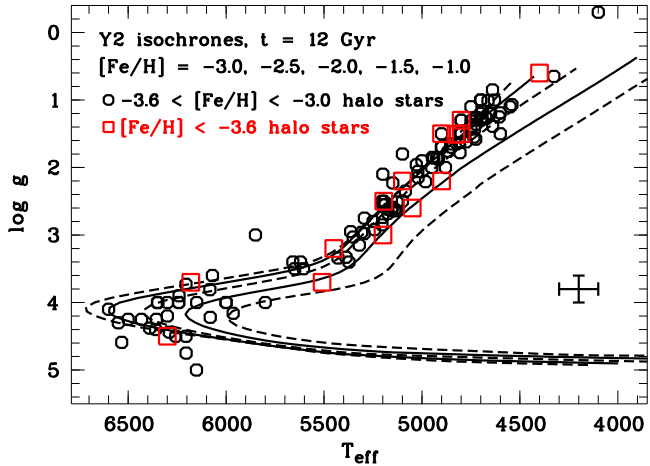}

\caption {\label{Fig:hr_diagram}\small Hertzsprung-Russell diagram of
  the $\sim$130 known metal-poor stars with $\mbox{[Fe/H]}<-3.0$
  studied at high-resolution (from the compilation of
  \citealt{frebel10}). 14 stars with $\mbox{[Fe/H]}<-3.6$ are marked
  with open squares. Typical error bars are indicated at bottom
  right. Several 12\,Gyr isochrone tracks (from
  http://www.astro.yale.edu/demarque/yyiso.html) for different
  metallicities are overplotted for illustration. As can be seen, the
  main-sequence turnoff shifts significantly to hotter temperatures at
  [Fe/H] $<$ --2.0, whereas the giant branch is less affected.}

\end{center}
\end{figure}

This section presents a census of stars having [Fe/H]
$<$ --3.0 and for which detailed high-resolution, high $S/N$,
published abundance analyses are available.  The data set comprises
some 130 objects, and may be found in the compilation of
\citet{frebel10}.  Figure~\ref{Fig:hr_diagram} shows the distribution
of the stellar parameters effective temperature, {\teff}, and surface
gravity, {\logg}, of these objects, in comparison with several 12\,Gyr
isochrones of different metallicities, [Fe/H], in the
Hertzsprung-Russell diagram. The vast majority of the stars are
luminous red giants (4000\,K $<$ {\teff} $<$ 5500\,K, 0.0 $<$ {\logg}
$<$ 3.5), but about 25 main-sequence stars near the turnoff (5800\,K
$<$ {\teff} $<$ 6600\,K, 3.5 $<$ {\logg} $<$ 4.5) are also known.  A
similar ratio is maintained for metallicities below
$\mbox{[Fe/H]}=-3.6$.  (The atmospheric parameters {\teff}, {\logg},
and [Fe/H] are the essential stellar parameters that determine the
structure of a star's outer layers and the details of its emergent
flux, as will be discussed in
Section~\ref{Sec:abundance_determination}.)

Figure~\ref{Fig:lowfe_MDF} presents the histogram of [Fe/H] for the
same group of objects.  Two things are worth noting from this diagram.
First, the number of stars decreases precipitously as one moves
towards lowest abundance, and second, the proportion of carbon-rich
objects increases dramatically. The implications of this remarkable
behavior are discussed further in Section~\ref{Sec:early_uni}.  More
generally, roughly 10\% of the objects in this sample (often those
with enhanced carbon) reveal a chemical nature that is different from
that of ``normal'' halo stars. These objects are of particular
interest as they can be used to address a variety of astrophysically
important questions.  Arguably, the most interesting stars in the
sample are those with metallicities $\mbox{[Fe/H]}\lesssim-3.5$.  It
is probably reasonable to say that all stars known to have
$\mbox{[Fe/H]}\lesssim-3.5$ from medium-resolution spectroscopy and B
$<$ 16.5 are included in this diagram, given their potential for
insight into the early Universe.

There are eight stars {known to have $\mbox{[Fe/H]}\sim-4.0$ or
  less. One of them is {\cd} \citep{bessell&norris84,Norrisetal:2001},
  the first star with $\mbox{[Fe/H]}\sim-4.0$. It was discovered some
  30 years ago, and was a long-standing record holder for the most
  iron-poor object in the Milky Way. It has been observed and analyzed
  many times by most research groups working in the field.  Only four
  stars in this very small sample have $\mbox{[Fe/H]}<-4.3$, with two
  having $\mbox{[Fe/H]}<-5.0$. In 2001, the first star with
  $\mbox{[Fe/H]}<-5.0$ was discovered. Until then, it had not been
  clear whether objects with metallicities lower than that of {\cd}
  existed.  This object, HE~0107$-$5240, is a faint ($V=15.2$) red
  giant with $\mbox{[Fe/H]}=-5.3$ \citep{HE0107_Nature}. In 2004, the
  bright ($V=13.5$) subgiant HE~1327$-$2326 was identified and shown
  to have $\mbox{[Fe/H]}=-5.4$ \citep{HE1327_Nature}, corresponding to
  $\sim1/250000$ of the solar iron abundance. This small stellar Fe
  number density translates to an actual iron mass that is about 100
  times less than that of the earth's iron core!  Both stars were
  found in the Hamburg/ESO survey. Since then, no further objects with
  such record-low Fe values have been discovered. As outlined in
  Section~\ref{Sec:conclusion} new surveys will provide additional
  chances to uncover more of these rare stars.

  The paucity of stars with $-5.3~\la\mbox{[Fe/H]}~\la~-4.3$ sparked
  considerable interest among theorists, with some suggesting that
  there may be a physical reason for this apparent gap in the
  metallicity distribution function (e.g., \citealt{shigeyama}). The
  discovery, however, of the red giant {\hej} \citep{he0557} and the
  dwarf star {\sdss} \citep{caffau11} both with
  $\mbox{[Fe/H]}\sim-4.7$ (adopting 1D Fe abundances) confirmed
  that extremely limited discovery statistics below
  $\mbox{[Fe/H]}\sim-4.3$, driven by only four stars, are most likely
  the cause of the apparent gap.

In summary, as of mid-2010, numerous stars with $\mbox{[Fe/H]}<-3.0$
have been discovered and many ($\sim$130) have been analyzed with
high-resolution spectroscopy.  Stars with $\mbox{[Fe/H]}<-3.5$ are
much rarer, but most likely all known examples ($\sim$25) of them have
high-resolution spectroscopic analyses. Only four stars with
$\mbox{[Fe/H]}<-4.3$ are known, of which two have $\mbox{[Fe/H]}<
-5.0$.

\begin{figure*}[!tbp]
\begin{center}

\includegraphics[width=16.0cm,angle=0]{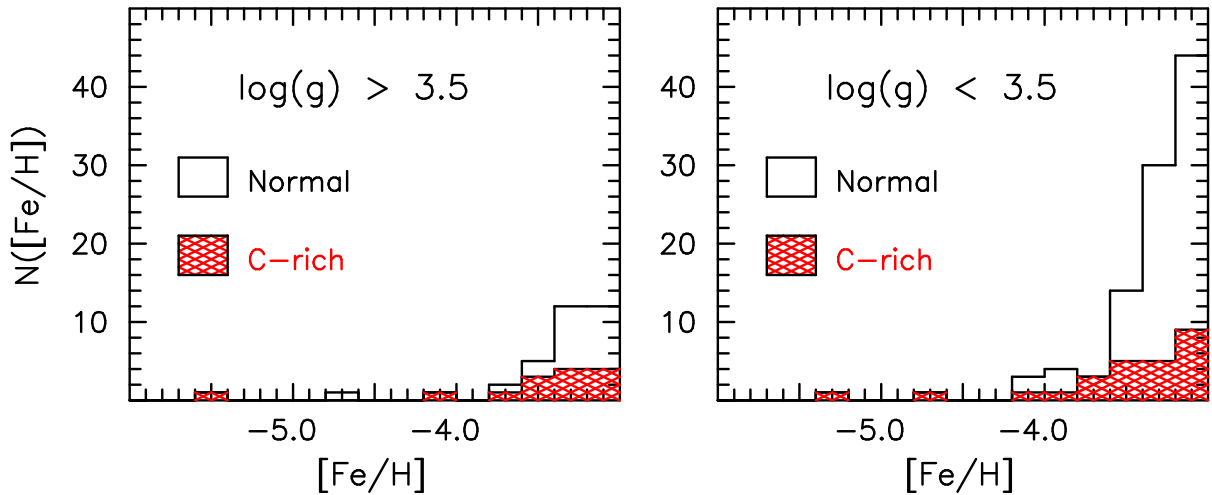}

\caption{\label{Fig:lowfe_MDF} \small [Fe/H] histogram for stars having
  high-resolution, high $S/N$, abundance analyses and [Fe/H] $<$
  --3.0, from the compilation of \citet{frebel10}. On the left are
  results for main-sequence and subgiant stars, while the right
  presents data for red giants.  The shaded regions refer to C-rich
  stars.  Note the rapid decline in the number of stars as [Fe/H]
  decreases, accompanied by an increase in the proportion of
  carbon-rich objects.}

\end{center}
\end{figure*}

\subsection{The Lowest Observable Metallicity}

What is the lowest abundance one might be able to observe
  in the Galactic halo?  From a practical point of view, a useful
limit is set by the abundance corresponding to a measured Ca\,II\,K
line strength of 20\,m{\AA} (roughly 2 -- 4 times the strength on the
weakest lines measurable in high-resolution, high $S/N$, spectra such
as those shown in Figure~\ref{Fig:Hires}) in a cool red giant with
{\teff} = 4500\,K and {\logg} = 1.5.  (Ca\,II K is the strongest
atomic feature in metal-poor stars, and its strength is greater in red
giants than near-main-sequence dwarfs (due to the lower effective
temperatures of the former).  Also, the abundances of red giants are
much less modified by accretion from the Galactic Interstellar Medium
(ISM) than those of main-sequence stars, because of the deep outer
convective regions in giants.)

Adopting a 1D LTE model atmosphere (see Section~\ref{Sec:1D}) with
these parameters and [Fe/H] = --4.0 (the lowest available abundance in
many grids, and which should be adequate for the task), a line
strength of 20\,m{\AA} corresponds to [Ca/H]$_{min}$ = --9.4.  (For
Fe\,I\,3859.9\,{\AA}, the intrinsically strongest Fe\,I line in the
optical spectrum, a line strength of 20\,m{\AA} results in a less
stringent limiting abundance of [Fe/H] = --7.2.)  If one were to
assume that this hypothetical star had [Ca/Fe] = 0.4, similar to that
found in the most metal-poor stars, its iron abundance would be
[Fe/H]$_{min}$ = --9.8.  This can be taken as a rough estimate of the
lowest metallicity practicably detectable.

Even, however, if such a star existed, one should not automatically
interpret the above minimum abundance as the value with which it
formed, given the possibility of accretion of material from the
interstellar medium (ISM) during its $\sim$13\,Gyr lifetime.  Using
calculations described by \citet{poll}, who compute the amount of
material likely to have been accreted onto each of some 470 observed
halo main-sequence stars, it was found that during its time on the RGB
the average amount of material accreted onto a star would have
increased an initial zero heavy element abundance to an observed
atmospheric value of [Fe/H] = --8.6, with a dispersion of 0.8\,dex
(Here the large dispersion is driven by an extremely strong dependence
of the accretion process on the relative velocity of the star with
respect to the ISM.)  From this information, it follows that a star
that formed with [Fe/H]$_{min}$ = --9.8, and experienced average ISM
accretion would be observed during its RGB evolutionary phase as an
object with [Fe/H] $\sim$ --8.6.  Alternatively, given the dispersion
in possible accretion histories, one might also say that the
probability of finding a star that initially had zero heavy-element
abundance (i.e., Population\,III) and observed today during its RGB
phase would have an ``accreted'' abundance of [Fe/H] = --9.8 or
smaller, is $\sim$0.07.

Having assessed the technical feasibility of finding
near-zero-metallicity, low-mass (M $<$ 1\,M$_{\odot}$) stars, one
needs also to consider potential physical processes that may have
played a role in the formation of the most metal-poor stars, and which
lead to abundances between the current lowest observed level of
$\mbox{[Fe/H]} \sim-5.5$ and the potentially detectable $\mbox{[Fe/H]}
=-9.8$.  As will be discussed in Section~\ref{Sec:early_uni} the
critical factor is the cooling mechanisms that determine the
contraction and fragmentation of existing gas clouds.  Two potentially
important cooling mechanisms are noted here, as well as the abundance
limits they impose, following \citet{poll}.  The first is C\,II and
O\,II fine-structure line cooling which leads to [Fe/H]$_{min}= -7.3$.
The second is the major cooling due to dust grains, for which the
limit might be 1 -- 2 orders of magnitude lower, e.g.,
[Fe/H]$_{min}=-8.0$ to $-9.0$.  While more detailed knowledge on
cooling mechanisms may well change these values, the above discussion
shows that one should not be surprised to find stars with
metallicities much lower than those of the most-metal-poor stars
currently known.

\section{DERIVED CHEMICAL ABUNDANCES}

\subsection{Abundance Determination} \label{Sec:abundance_determination}

\subsubsection{One-Dimensional Model Atmosphere Analyses} \label{Sec:1D}

Most chemical abundance determinations are based on one-dimensional
(1D) model stellar atmosphere analyses that assume hydrostatic
equilibrium, flux constancy, Local Thermodynamic Equilibrium (LTE),
and treat convection in terms of a rudimentary mixing length theory.
(In most cases the configurations are plane parallel, but when
necessary spherical symmetry is adopted for giants.) To first order,
the basic atmospheric parameters that define the model are effective
temperature ({\teff}), surface gravity ({\logg}), and chemical
composition.  Given these, one may construct a model atmosphere and
compute the emergent flux for comparison with observations.  Then, on
the assumption that the model well-represents the observed star, when
one obtains a good fit between the model emergent flux (in particular
the strengths of the atomic and molecular features) and the observed
flux, one assumes the chemical abundances of the model correspond to
those of the observed star.  The student should consult \citet{gray05}
and \citet{gustafsson08} for the concepts associated with the
process.

For completeness, it should be briefly  noted that
{\teff} and {\logg} are sometimes derived from atomic and molecular
transitions (excitation temperature and ionization equilibrium,
respectively) and sometimes from measurements of continuum colors, and
the strengths of hydrogen Balmer lines and the Balmer Jump.  Surface
gravity is also often derived from the star's luminosity, {\teff}, and
mass.  A basic shortcoming of 1D modeling is that an artificial,
second-order, extra line-broadening called ``microturbulence'', over
and above the thermal broadening of the models, is always introduced
into the formalism to satisfy the requirement that atomic lines of
different strength yield the same abundance.  This will not be
discussed  here further, except to say that the need for this
``fudge factor'' has proved to be unnecessary in the more physically
realistic 3D modeling.  Finally, best analysis involves an iterative
process that demands the adopted {\teff}, {\logg}, abundances, and
microturbulence are consistent with both the adopted model atmosphere
and all of the details in the star's spectrum that are sensitive to
these parameters.

As discussed in Section~\ref{Sec:definitions}, the analysis produces
stellar atmospheric abundances $\epsilon$(X) for species X relative to
hydrogen, expressed as log$_{10}{\epsilon}$(X)= log$_{10}(N_{\rm
  X}/N_{\rm H}$) + 12; in most cases values are published using the
bracket notation \mbox{[X/H]}$ = \log_{10}(N_{\rm X}/N_{\rm H})_\star
- \log_{10}(N_{\rm X}/N_{\rm H})_\odot$, which expresses the results
relative to solar values.  For completeness it should be noted that
the \textit{elemental} abundances derived in this way represent the
contribution of all isotopes; additional isotope ratios can only be
determined in a few cases (e.g., C). This contrasts nucleosynthesis
models, which yield abundances of each individual calculated isotope
abundance.  (Publicly available model atmospheres and associated
atomic and molecular data may be found, for example, at
http://kurucz.harvard.edu, http://vald.astro.univie.ac.at, and
http://www.physics.nist.gov/PhysRefData/ASD/lines{\_}form.html, while
codes for the computation of emergent fluxes and determination of
chemical abundances may be found, for example, at
http://www.as.utexas.edu/$\sim$chris/moog.html.)  Given the power of
modern computers, this is now a mature and straight-forward process,
and 1D/LTE abundances, [Fe/H] and [X/Fe], based on high resolution,
high S/N, data are currently available for a large number of
metal-poor stars -- for example, for resolving power $>$ 20,000 and
[Fe/H] $<$ --2.0, data exist for some 600 objects.  Two comprehensive
compilations of published material are those of \citet{suda08} and
\citet{frebel10}, the latter of which will be used in what follows.
The precisions of these results are high, typically of order 0.10 dex
(26\%), and in some cases $\sim$0.03 dex (7\%).

\subsubsection{Three-Dimensional Model Atmospheres} \label{Sec:3D}

The question that remains to be answered is: how accurate are these 1D
abundances?  The issues have been addressed by \citet{asplund05}, to
whom the reader is referred.  Three-dimensional (3D) hydrodynamical
models reveal temperature inhomogeneities, which lead to different
temperature structures between 3D and 1D models.  As metallicity
decreases, the inhomogeneities become larger, resulting in significant
negative abundance corrections.  Figures 1 and 2 of Asplund illustrate
the effect: at {\teff} = 5800\,K, {\logg} = 4.4, one finds that while
at [Fe/H] = 0.0 the {\it average} temperature of the 3D model agrees
reasonably well with that of its 1D counterpart, the situation is very
different at [Fe/H] = --3.0, where the 3D model has temperatures lower
by several hundred degrees in its upper layers.  These in turn lead to
significant differences between 1D and 3D abundances as a function of
the metallicity of the star and the excitation potential of the
observed line transition.  Figure 8 of Asplund shows that for
resonance lines of typical elements 1D abundances are too high by 0.1
-- 0.6 dex, with the difference being smaller for lines of higher
excitation potential.  The strength of molecular features is very
sensitive to the temperatures in the outer layers, resulting in
dramatically lower abundances compared with those obtained in 1D
analyses.

Abundance analysis utilizing 3D models is a computationally intensive
exercise, and results are currently available for only relatively few
stars selected for their astrophysical significance.  Examples of this
are the two most metal-poor stars {\hen} and {\hea},  for
  which Figure~\ref{Fig:1D3D} presents abundance differences
[X/Fe](3D) -- [X/Fe](1D) versus atomic number from the work of
\citet{collet06} and \citet{he1327_uves}.  Note the extremely large
differences for C, N, and O -- for which the cited results are
determined from analysis of CH, NH, and OH, respectively.  One must
bear this in mind when seeking to interpret 1D chemical abundances.

\begin{figure}[!tbp]
\begin{center}
\includegraphics[width=0.48\textwidth]{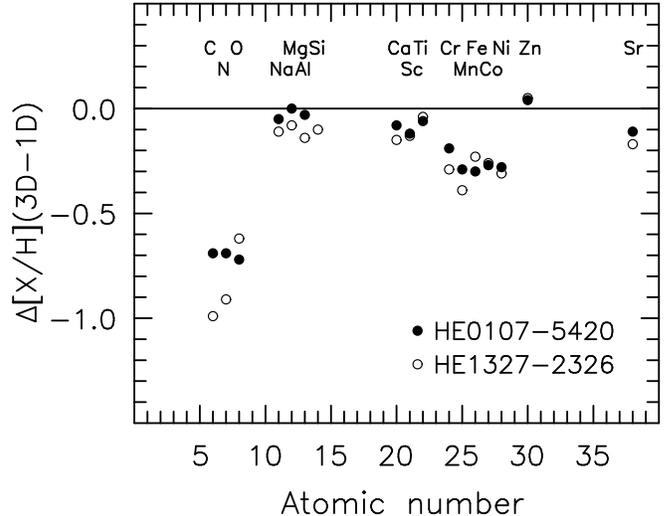}

\caption{\label{Fig:1D3D}\small The difference in abundance,
  [X/Fe]$_{\rm 3D}$ -- [X/Fe]$_{\rm 1D}$, vs. atomic number, deduced
  from analyses based on three-dimensional and one-dimensional model
  atmospheres, for the two most metal-poor stars -- the subgiant
  {\hea} and the red-giant {\hen}, both with [Fe/H]$_{\rm 1D}
  \sim-5.5$.  These stars show the so-far most extreme abundance
  differences between 1D and 3D analyses.  (Data from
  \citealt{collet06} and \citealt{he1327_uves}).}

\end{center}
\end{figure}

\subsubsection {Departures from Thermodynamic Equilibrium (Non-LTE)}

In order to determine chemical abundances one needs to derive the
populations of atomic and molecular energy levels, which depend on
details of the radiative and collisional effects in the regions of
line formation in the stellar atmosphere.  The reader is once again
referred to \citet{asplund05} for a thorough discussion of this
matter.  The proper solution to the problem is sufficiently
computationally intensive that most investigations to date have made
the assumption of LTE.  This approach assumes that collisional effects
dominate over radiative ones, from which it follows that the required
populations can be determined by the Maxwell, Saha, and Boltzmann
distributions, which involve only the local physical parameters
temperature and electron pressure.  To quote Asplund: ``In LTE the
strength of a line can be straightforwardly predicted from a few
properties of the line and the species once the model atmosphere and
continuous opacity are known.  In non-LTE, in principle everything
depends on everything else, everywhere else.'' (For completeness, it
should also be noted that a remaining uncertainty in current non-LTE
analyses is the treatment of inelastic collisions with hydrogen atoms;
see Asplund 2005, his Section 2.1.)

Given the time-consuming nature of non-LTE computations, the large
majority of abundances analyses to date assume LTE.  The advice of
Asplund should, however, be recalled. ``It is always
appropriate to provide the LTE results for comparison purposes, but it
is unwise to ignore the available non-LTE calculations when providing
the final abundance values.''  The present work follows
  this advice where possible, and considers (non-LTE -- LTE)
differences further in Section~\ref{Sec:relative_abundances}.  

\subsubsection{\it Caveat Emptor}

Two caveats are offered in conclusion.  The first is that essentially
all of the abundances presented here have been determined using 1D/LTE
model atmosphere analyses.  In some cases, when 3D and/or non-LTE data
are available, comments are included on the resulting differences
between the two formalisms.  The second point is that for a
comprehensive improvement over 1D/LTE results, one needs to use both
3D and non-LTE and not just one of them: in the case of lithium, for
example, and as will be discussed in Section~\ref{Sec:light_elem}, the
3D and non-LTE corrections are both large, but of opposite sign, and
fortuitously largely cancel to give the 1D/LTE result.
 
\subsubsection{Post-Astration Abundance Modification}

The final question one must address is whether the abundances obtained
from these exhaustive model atmosphere analyses are indeed the values
in the protocloud from which the star formed.  Here, very briefly,
with source material pertinent to metal-poor stars, are important
examples of processes that can modify the original abundance patterns
in the observed surface layers.

\begin{itemize}

\item

Accretion from the interstellar medium over the lifetime of a star
(e.g., \citealt{poll})

\item

Radiative and gravitational diffusion in the stellar surface layers
(e.g., \citealt{behr99})

\item

Macroscopic mixing of nucleosynthesis products from stellar interiors
into their surface layers (e.g., \citealt{gratton2000})

\item

Post-asymptotic-giant-branch evolution, during which the sequence of
element fractionation onto circumstellar grains,
radiation-pressure-driven grain/gas separation, and the formation of a
stellar atmosphere containing the remaining gas produce an Fe-poor,
modified abundance pattern determined by the physics of gas/grain
condensation (e.g., \citealt{giridhar05})

\item

Transfer of material across a multiple stellar system during
post-main-sequence evolution (e.g., \citealt{beers&christlieb05})

\end{itemize}

\subsection{Abundance Patterns}

\subsubsection{Metallicity Distribution Functions (MDF)} \label{Sec:mdf_clusters}

\begin{center}{\it The Galactic Globular Cluster System}\\ \end{center} \vspace{-2mm}

With very few exceptions (which will be considered in
Section~\ref{Sec:gc_evol}) the Milky Way's globular clusters are
individually chemically homogeneous with respect to iron.  The
collective MDF of the cluster system is bimodal, as first definitively
shown by \citet{zinn85} and presented here in
Figure~\ref{Fig:carretta09} (based on the more recent abundance
compilation of \citealt{carrettaetal09}).  The two components,
initially designated ``halo'' and ``disk'' have mean metallicities
[Fe/H] $\sim-1.5$ and --0.4.  This terminology, however, appears to be
an oversimplification: some clusters with abundances as low as [Fe/H]
$\sim-1.5$ have disk-like kinematics; some of the inner ``disk''
sub-population have been suggested to be members of the Galactic
bulge; and consideration of the horizontal branch morphologies of
globular clusters first led to the suggestion of old and young
subgroups in the halo sub-population \citep{zinn93}.  Clearly, the
situation is a very complicated one.

\begin{figure}[!tbp]
\begin{center}

\includegraphics[width=8.7cm,angle=0]{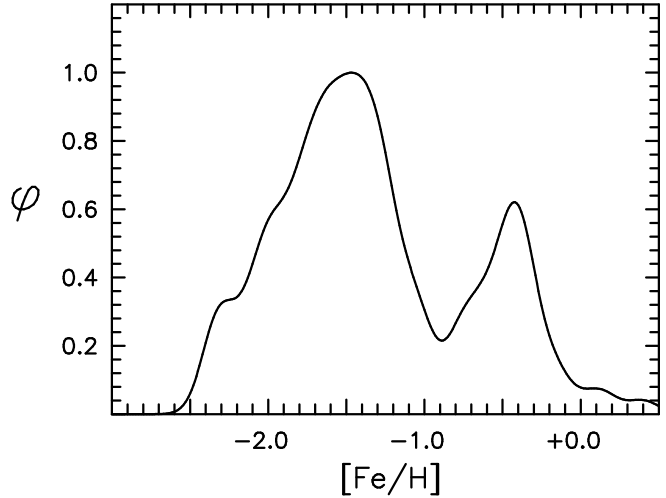}

\caption{\label{Fig:carretta09}\small MDF of the Galactic globular
  clusters (data of \citealt{carrettaetal09}).  Note the clearly
  bimodal distribution, with peaks at [Fe/H] = --1.5 and --0.4,
  corresponding predominantly to halo and disk/bulge material,
  respectively.  (The histogram was generated with a gaussian kernel
  having $\sigma$ = 0.10\,dex.)}

\end{center}
\end{figure}

\begin{center}{\it Field Stars}\label{Sec:mdf_fieldstars} \\ \end{center} \vspace{-2mm}

MDFs are also available for local metal-poor samples of both
kinematically selected main-sequence dwarfs \citep{ryan&norris91b} and
\citep{carneyetal96}, and spectroscopically selected giants
\citep{schoerck} and dwarfs \citep{lietal10}.

The field star distributions differ from that of the globular clusters
in one important aspect: all halo field star samples contain objects
having abundances considerably lower ([Fe/H] = --4.0 to --3.0) than
that of the most metal-poor globular cluster ([Fe/H] $\sim$~--2.5).
According to \citet{carneyetal96} the difference between halo clusters
and kinematically selected dwarfs is highly significant (at the 93 --
99.9\% level); this effect is shown here in Figure~\ref{Fig:carney96},
based on more recent data.  For spectroscopically selected samples, on
the other hand (which by definition have a strong abundance selection
bias towards more metal-poor stars, not present in kinematically
selected samples), the significance is less clear.  According to
\citet{schoerck}: ``A comparison of the MDF of Galactic globular
clusters ... shows qualitative agreement with the halo [field star]
MDF, derived from the HES, once the selection function of the latter
is included.  However, statistical tests show that the differences
between these are still highly significant.''  The problem with the
spectroscopically chosen HK and HES samples is that the corrections
that must be applied to compensate for the selection function are very
large, as clearly shown in Figure~\ref{Fig:tsujimoto99} below (see
also Sch{\"o}rck et al.\ 2009, their Figure~17).

\begin{figure}[!t]
\begin{center}

\includegraphics[width=8.7cm,angle=0]{fn_fig7.eps}

\caption{\label{Fig:carney96} \small Comparison of the MDFs of the
  halo globular clusters (thin line) and kinematically chosen halo
  field main-sequence dwarfs (thick line). The selection of the halo
  samples follows Section 3.5 of \citet{carneyetal96}: for the
  clusters only objects more than 8 kpc from the Galactic center are
  included (with abundances from \citealt{carrettaetal09}), while for
  the dwarfs the data are from \citet{carneyetal96}. (The histograms
  were generated with a gaussian kernel having $\sigma$ = 0.15\,dex.)}

\end{center}
\end{figure}

\begin{figure*}[!tbp]
\begin{center}
\hspace{-5mm}
\includegraphics[clip=true,width=0.485\textwidth,bbllx=45, bblly=375,
   bburx=570, bbury=730]{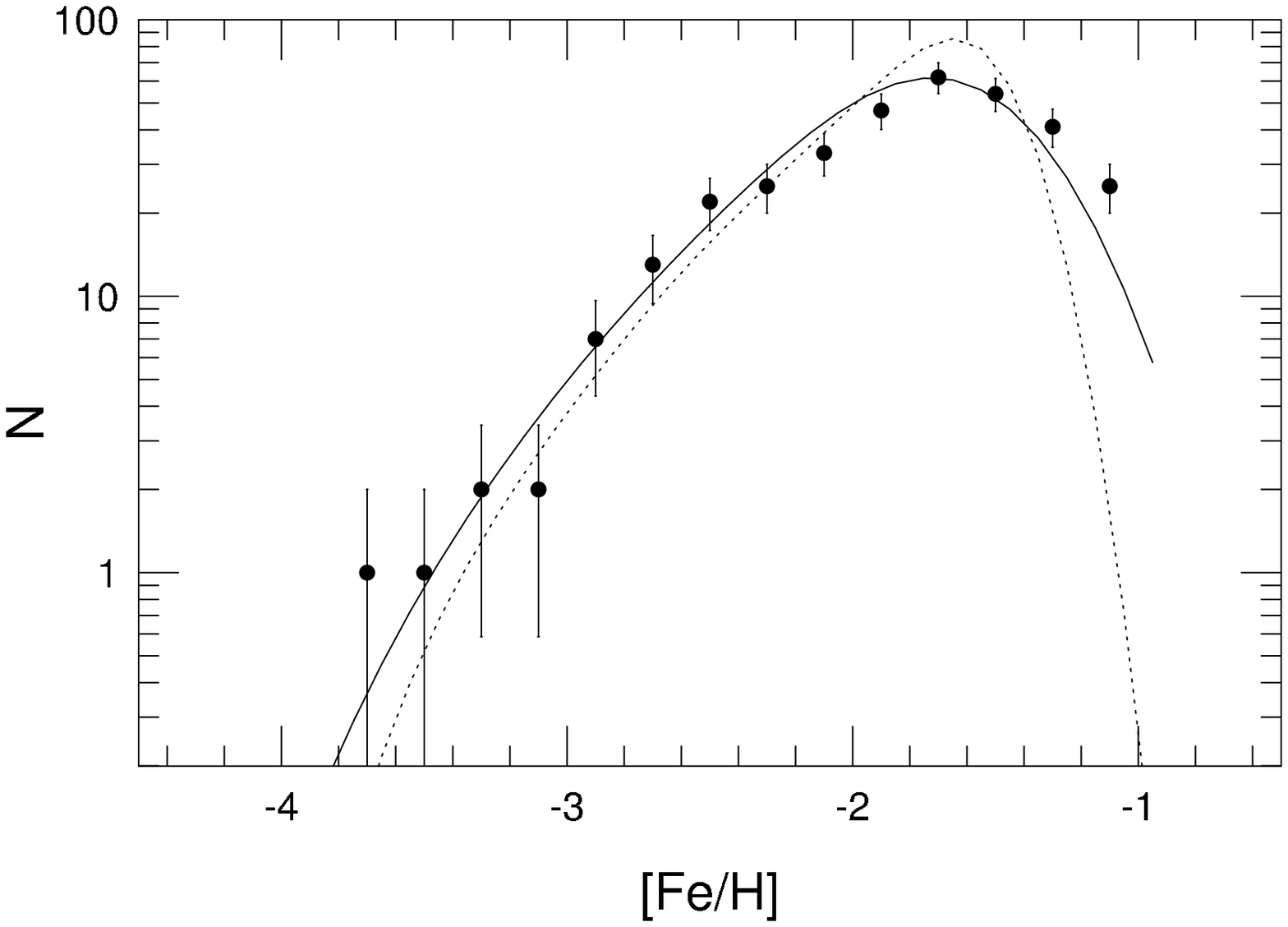}
\includegraphics[clip=true,width=0.50\textwidth,bbllx=82,bblly=420,%415,
   bburx=490, bbury=690]{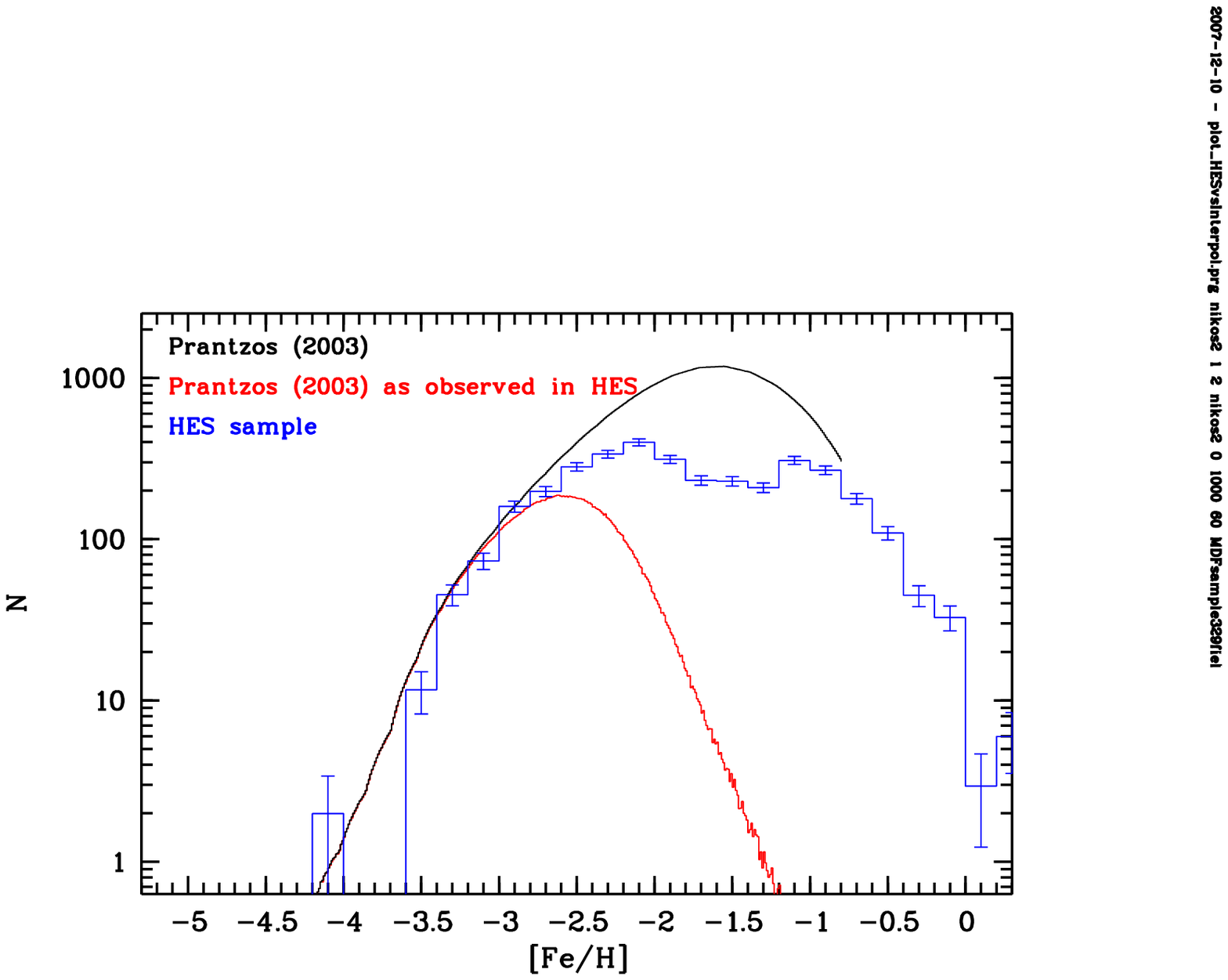}

\caption{\label{Fig:tsujimoto99} \small Left: comparison of the MDF of
  the kinematically selected halo main-sequence dwarf sample of
  \citet{ryan&norris91b} with the simple model (dotted line) of
  \citet{hartwick1976} and the more realistic supernova-induced
  star-formation model (solid line) of \citet{tsujimotoetal99}.  (The
  figure comes from \citealt{tsujimotoetal99}.) Right: MDF of the
  spectroscopically chosen halo giant sample of \citet{schoerck} in
  comparison with the GCE model of \citet{prantzos03}.  Note the large
  correction necessary to modify the model (the upper continuous line)
  for the abundance incompleteness caused by the spectroscopic
  selection function (the lower continuous line) (from
  \citealt{schoerck}). }

\end{center}
\end{figure*}

The fundamental importance of metallicity distribution functions is
that they provide essential constraints on galactic chemical
enrichment (GCE) models.   The starting point adopted for
the present discussion is the simple model of halo chemical
enrichment of \citet{hartwick1976}, who assumed that initially the
halo contained zero heavy elements, and was chemically enriched by the
ejecta of massive stars on timescales short compared with those of the
halo's dynamical evolution (instantaneous recycling).  He also assumed
that the initial mass function was constant with time, and in order to
reproduce the MDF of the halo globular clusters, postulated that gas
was removed from the system at a rate proportional to that of star
formation.  The left panel of Figure~\ref{Fig:tsujimoto99} shows a
comparison of this simple model (dotted line) with the observations of
halo field dwarfs by \citet{ryan&norris91b}.  The solid line in the
figure, which somewhat better fits the data, represents a model of
\citet{tsujimotoetal99} which involves star formation on shells swept
up by the ejecta of the supernova explosions of massive stars.
  
A point worth reiterating from Section~\ref{Sec:search_techniques} is
that the simple Hartwick model predicts the number of metal-poor stars
should decrease by a factor of ten for each factor of ten decrease in
abundance: \citet{norris99} and \citet{schoerck} report that this
appears to be the case down to [Fe/H] = --4.0 and --3.6, respectively,
below which there is a large dearth of stars.  Recall also from
Section~\ref{Sec:census} that only four stars with [Fe/H] $\la$ --4.3
are currently known.

Several other GCE models have been proposed which modify the basic
assumptions of the Hartwick model.  As an example, the right panel of
Figure~\ref{Fig:tsujimoto99} shows the comparison between the
spectroscopically selected halo giant sample of \citet{schoerck} and
the model of \citet{prantzos03} (which investigates improvement of the
instantaneous recycling approximation and possible gaseous
infall). The reader is referred to \citet{schoerck} and
\citet{lietal10} for comparison of the observations with other GCE
models from T.~Karlsson (delayed chemical enrichment at the earliest
times), S.~Salvadori and co-workers ($\Lambda$CDM framework with a
critical metallicity for low-mass star formation), and N. Prantzos
(semi-analytical model within the hierarchical merging paradigm).

This section is concluded with a caveat concerning the above
comparison of MDFs.  There has been growing evidence over some two
decades, beginning with the seminal works of \citet{hartwick87} and
\citet{zinn93} that the Galactic halo comprises more than one
component, with different properties as a function of Galactocentric
distance; see \citet{carollo10} and \citet{morrisonetal09}, and
references therein for details.  The multiplicity of the Galactic halo
will be discussed in Section~\ref{Sec:mw}.  Suffice it here to say it
makes little sense to compare the MDFs of samples (observational
and/or theoretical) that have different properties (except to test the
null hypothesis).  It is essential to match the underlying
characteristics of the theoretical models and observed samples that
are being compared.

\begin{center}{\it Dwarf Spheroidal Galaxies (dSph) } \label{Sec:mdf_dsph} \\ \end{center} \vspace{-2mm}

In stark contradistinction to the Milky Way's globular clusters, its
dwarf spheroidal galaxy satellites all show large internal spreads in
the abundance of iron.  The MDFs for five of the $\sim$25 currently
known systems are shown in Figure~\ref{Fig:mdf_dsph}.  (Also shown in
the figure, for comparison purposes, is the MDF of $\omega$~Cen, the
most massive Milky Way globular cluster, and one of only $\sim$5 halo
clusters known to exhibit a dispersion in iron greater than
$\sigma$[Fe/H] $\sim$~0.03\,dex.)  In each panel the abundance dispersion
$\sigma$[Fe/H] and integrated absolute visual magnitude,
M$_{V,\,\rm total}$, are also shown.

It has been known for some time that the metallicities of elliptical
galaxies and the more luminous dSphs collectively decrease as
luminosity decreases \citep{mateo98}.  As reported by \citet{kirby08},
the mean [Fe/H] of dSphs continues to decrease with decreasing
luminosity over the range of the ultra-faint systems as well.  As
shown here in Figure~\ref{Fig:kirby} the relationship holds over the
range 3.5 $\la$ log(L$_{\rm tot}$/L$_{\odot}$) $\la$~7.5.  This is a
clear signal that the dwarf galaxies have undergone internal chemical
evolution.  Examination of Figure~\ref{Fig:mdf_dsph} also shows that
in the faintest of the dwarf systems (M$_{\rm V,\,total} \la$ --7)
there is a large fraction of stars with [Fe/H] $<$ --3.0, suggesting a
relationship between ultra-faint dwarf galaxies and the most
metal-poor stars in the Milky Way halo.  This topic will be further
addressed in Section~\ref{Sec:dg_evol}, but it is worth noting here
that an essential difference between the Milky Way's globular clusters
and dSph systems is that (for objects with integrated magnitudes
M$_{\rm V,\,total} \ga -10$) the dSphs are embedded in dark matter
halos (with M/L$_{\rm V}$ $\sim$ 10 -- 10$^{4}$ in solar units), while
almost all clusters contain relatively little or no dark matter
(M/L$_{\rm V} \la 5$).  This is almost certainly the essential
difference behind the large [Fe/H] dispersions observed in the dSph
systems, but absent from the globular clusters.

\begin{figure}[!tbp]
\begin{center}

\includegraphics[width=8.7cm,angle=0]{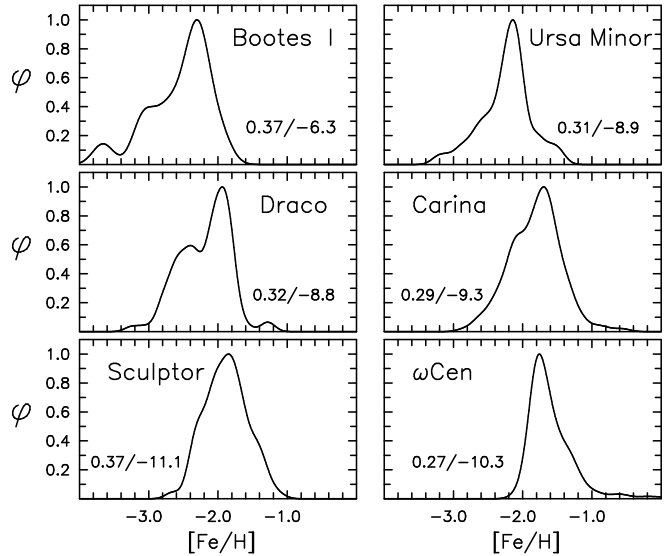}
  
\caption{\label{Fig:mdf_dsph}\small The Metallicity Distribution
  Functions for five Milky Way dwarf galaxies and the globular cluster
  $\omega$ Centauri. Also shown in each panel are
  $\sigma$[Fe/H]/M$_{\rm V,\,total}$ (the dispersion in [Fe/H] and the
  integrated absolute visual magnitude of the system).  See
  \citet{norris10booseg} for source material. (The histograms were
  generated with gaussian kernels having $\sigma$ = 0.10 --
  0.15\,dex.)}

\end{center}
\end{figure}

\begin{figure}[!t]
\begin{center}
%\vspace{-0.5 cm}
\includegraphics[clip=true,width=8.7cm,bbllx=30, bblly=10,
   bburx=410, bbury=370]{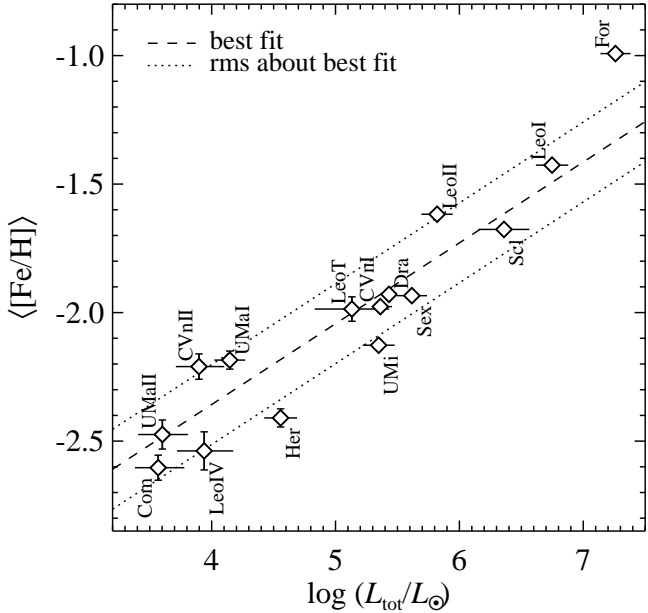}

\caption{\label{Fig:kirby}\small Mean [Fe/H] vs. luminosity for the
  Milky Way's dwarf spheroidal galaxies.  Systems with L$_{\rm V} \leq
  10^{5}\,$L$_{\odot}$ are designated ``ultra-faint'' dwarf galaxies, since
  they are fainter than the long-known ``classical'' dwarf
  galaxies. It is very likely, however, that there exists a continuous
  transition between the physical properties of the two groups. (Prepared
  by E. N. Kirby using data from Kirby et al.\ 2008, and references
  therein.)}

\end{center}
\end{figure}

\subsubsection{Relative Abundances} \label{Sec:relative_abundances}

\begin{figure*}[!h]
\begin{center}

\includegraphics[width=1.00\textwidth]{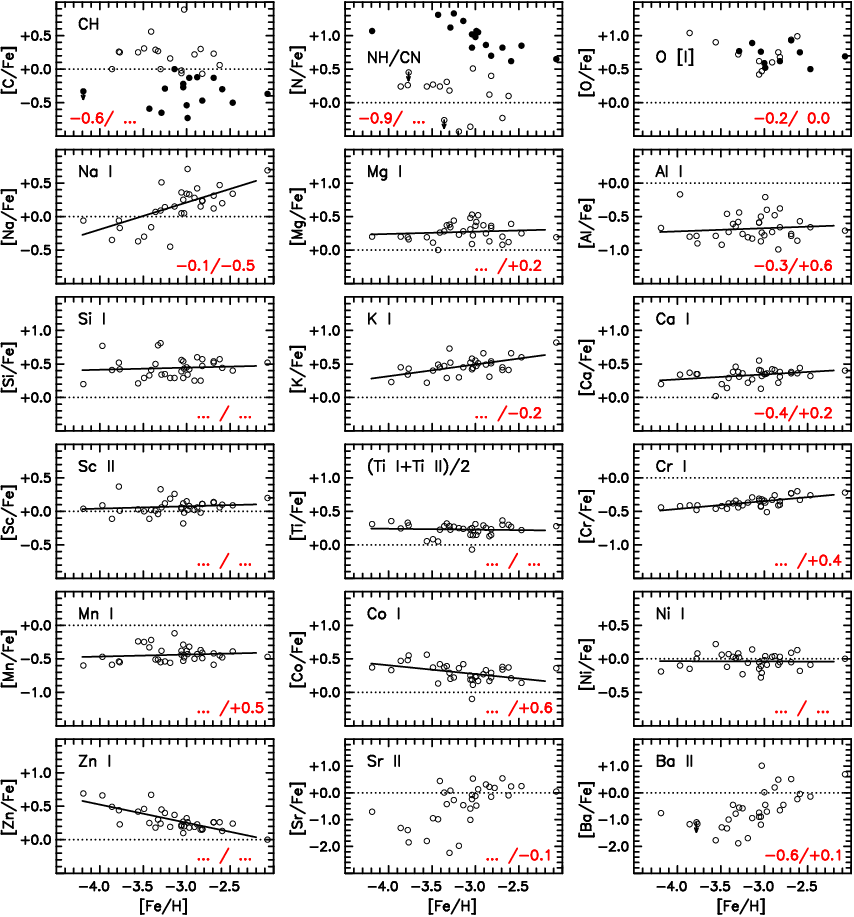}

\caption{\label{Fig:cayrel04} \small 1D/LTE relative abundances
  ([X/Fe]) vs. [Fe/H] for metal-poor halo red giants from the work of
  \citet{cayrel2004}, \citet{spite05}, and \citet{francois07}. In the
  top row, filled and open circles refer to ``mixed'' and ``unmixed''
  stars, respectively, as defined by Spite et al. (see
  Section~\ref{Sec:carbon}).  Also shown at the bottom of each panel
  are indicative (3D -- 1D)/(non-LTE -- LTE), abundance differences as
  discussed in Section 3.1.2 (``...'' indicates a potential
  incompleteness in our literature search or the absence of relevant
  information).}

\end{center}
\end{figure*}

Just as [Fe/H] is adopted as proxy for a star's overall metallicity,
the abundances of the other elements are most often expressed relative
to Fe, i.e., as [X/Fe] for element X.  (This is a somewhat arbitrary
definition, driven by the practicality of the richness of the Fe\,I
spectrum, and from time-to-time the implications of adopting an
alternative element as reference is investigated.  For an example of
this see \citet{cayrel2004}.)  Element abundances are thus directly
related to the element that represents the end stage of stellar
evolution and provides a good indicator of core-collapse SN
nucleosynthesis.  Note that all such (relative) abundances are relative
not only to Fe, but also to the abundances measured for the Sun (the
bracket notation). This should be kept in mind when ``reading'' the
chemical relative abundance trends in metal-poor stars in terms of Galactic
chemical evolution.

To give the reader a feeling for the scope of the observed trends,
Figure~\ref{Fig:cayrel04} shows the 1D/LTE relative abundances for
metal-poor Galactic halo red giants from the work of
\citet{cayrel2004}, \citet{spite05}, and \citet{francois07}, which
covers the range --4.5 $<$ [Fe/H] $<$ --2.0, and is regarded by many
as the ``gold standard'' of the state-of-the-art for this type of
work. The reader should note that the scale in 16 of the
18 panels of Figure~\ref{Fig:cayrel04} is the same, with a range in
[X/Fe] of 2\,dex.  For the remaining two cases ([Sr/Fe] and [Ba/Fe])
this is insufficient to cover the range in the early Universe, and for
these the relevant panel range is 5\,dex!  The dotted lines in the
figure correspond to the solar value.  The solid lines in the panels
of sodium through zinc represent the regression lines of
\citet{cayrel2004}; for these elements these authors report errors of
measurement $\sigma\sim$~0.05 -- 0.10 dex, and dispersions about the
regressions of $1 - 2\sigma$. Also shown, at the bottom
right of each panel, are indicative 3D/non-LTE corrections for stars with
[Fe/H] $\sim$~--3.0 that have been gleaned from \citet{asplund05} and
other sources in the literature (such as the works of
S.~M.~Andrievsky, D.~Baum\"{u}ller, M.~Bergemann, D.~V.~Ivanova,
L.~Mashonkina, and co-workers).

By way of introduction to what follows, several aspects
of the figure are highlighted.

\begin{figure*}[!htp]
\begin{center}
\includegraphics[width=17.0cm,angle=0]{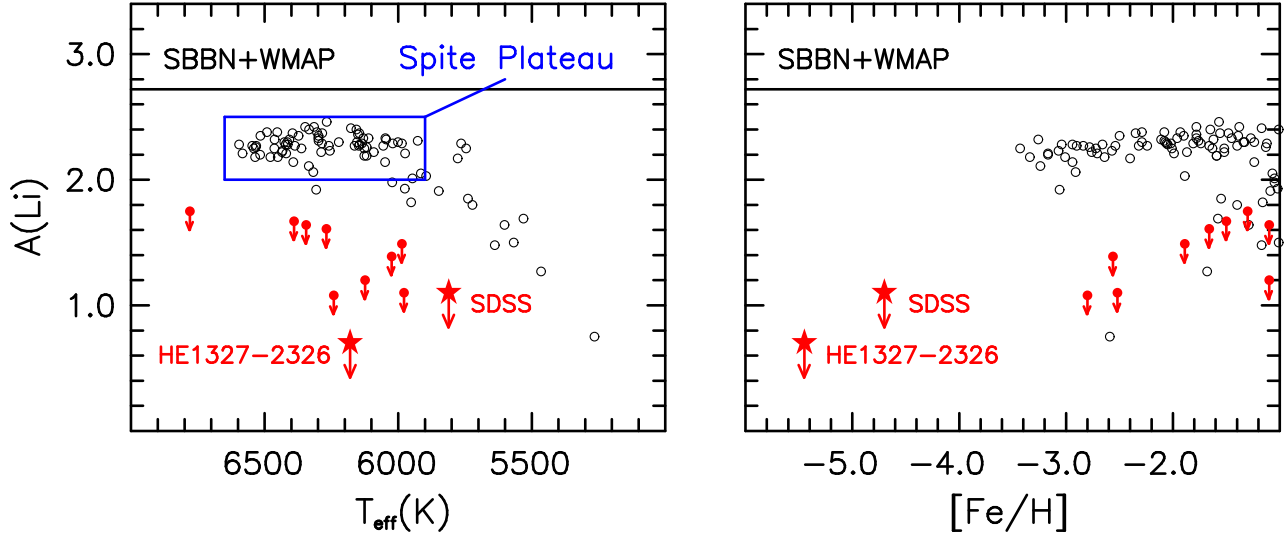}
\caption{\label{Fig:liall} \small 1D/LTE lithium abundance, A(Li), as
  a function of {\teff} (left) and [Fe/H] (right), from the work of
  Frebel et al.\ (2008, {\hea}) and Caffau et al.\ (2011, {\sdss})
  (large filled stars), Mel\'{e}ndez et al.\ (2010, open circles), and
  Ryan et al.\ (2001, filled circles).  The horizontal line in each
  panel is the value predicted from the observations by WMAP
  interpreted in terms of SBBN \citep{cyburt08}.}
\end{center}
\end{figure*}

\begin{itemize}

\item

The large spreads in C, N, Sr, and Ba are real and tell us much about
internal mixing during red giant evolution (C and N:
Section~\ref{Sec:carbon}), and the processes that produce the heavy
neutron-capture elements (Sr and Ba: Section~\ref{Sec:ncap}).

\item

Systematic enhancements of the $\alpha$-elements Mg, Si, Ca, and
(partially) Ti, lead to an explanation involving SNe of Type Ia and
II, operating at different times (Sections~\ref{Sec:alphas} and
\ref{Sec:dg_evol}).

\item

Tight solar-like correlations, such of those of the iron-peak elements
Sc and Ni, suggest a close relationship between the production
mechanisms of some of the iron-peak elements, indicative of processes
similar to those responsible for the enrichment of the Sun.

\item

In contrast to the previous point, the trends shown by iron-peak
elements such as Cr, Mn, and Co cast doubt on the previous suggestion.
This ``contradiction'' is indicative perhaps of differences related to
the location of the mass-cut radius within the progenitor of the SN
explosion (above which all material is expelled) or to non-LTE effects
(Section~\ref{Sec:ironpeak}).

\item

Large corrections to some of the 1D/LTE abundances are clearly
necessary to take into account 3D effects and a more realistic
treatment of non-LTE before they may be closely and reliably compared
with the prediction of stellar evolution and GCE computations.

\end{itemize}

\section{THE CHEMICAL EVOLUTION OF THE UNIVERSE}

\subsection{Relics of the Big Bang}\label{Sec:light_elem}

According to Standard Big Bang Nucleosynthesis (SBBN), some minutes
after the singularity at the era of decoupling of radiation and
matter, the only chemical elements in the Universe were hydrogen,
helium and lithium.  With the additional constraint of the results of
the Wilkinson Microwave Anisotropy Probe (WMAP), the predicted
relative mass densities of these elements are 0.75, 0.25,
2.3$\times$10$^{-9}$, respectively \citep{WMAP}.  All other elements
have been produced subsequently.

\subsubsection{Helium}

No reliable spectroscopic determinations exist of the abundance of
helium in the atmospheres of stars having [Fe/H] $<$ --1.0.  Most are
too cool ({\teff} $<$ 7000\,K) for lines of neutral helium to be
currently useful for abundance analysis (see e.g., \citealt{dupree11});
and in metal-poor stars hot enough for the test to be made (the
so-called blue-horizontal-branch stars), strong diffusive processes
are clearly at work in their outer layers and preclude determination
of the chemical abundances in the material from which they formed
(\citealt{behr99}).  The best estimates of primordial helium abundance
based on spectral features of helium come from the analysis of helium
lines in gaseous nebulae, for which \citet{WMAP} report a primordial
helium abundance $Y_{p}$ = 0.232 -- 0.258.

\subsubsection{Lithium} \label{Sec:lithium}

\citet{spite82} first demonstrated that the Li abundance of
metal-poor, near main-sequence-turnoff stars appears constant in the
temperature range {\teff} = 5500 -- 6250\,K, and concluded: ``the
abundance of lithium at the beginning of the Galaxy was: {\it N}$_{\rm
  Li}$ = 11.2 ($\pm$3.8) 10$^{-11}${\it N}$_{\rm H}$'', i.e., A(Li) =
2.05 $\pm$ 0.16.  Astronomers today discuss this fundamental discovery
not so much in respect of the Galaxy, to which it is certainly
pertinent, but rather in terms of the Li abundance that emerged from
the Big Bang.

The effect is shown in Figure~\ref{Fig:liall}, based on more recent
observational data, where 1D/LTE values of A(Li) are presented as a
function of {\teff} and [Fe/H]. (It is noted here  for
  completeness that the accepted temperature scale for metal-poor
main-sequence stars has become some 300\,K hotter since the work of
\citet{spite82}.)  One sees that the so-called Spite Plateau remains
clearly defined, and as appreciated by \citet{spite82}, for {\teff}
$<$ 5900\,K on the new scale, lithium is destroyed by strong
convective circulation that brings it into deeper and hotter regions.
During extensive expansions on the original sample, some stars were
found in which Li could not be detected.  It has been suggested that
these stars, which comprise only a small fraction of their parent
population, have ultra-low lithium abundances as the result of
phenomena related to binarity and blue stragglers, during which Li is
converted into other elements at the high temperatures experienced
during convective mixing in their outer layers (see \citealt{ryan01}).

There are, however, two extremely important further points to be taken
from Figure~\ref{Fig:liall}.  The first is that, ignoring the obvious
outliers, the mean abundance of the Plateau, A(Li) = 2.28 $\pm$ 0.01,
lies some 0.4 -- 0.5\,dex below the value that has been predicted from
the results of WMAP, interpreted in terms of the predictions of SBBN
of A(Li) = 2.72$_{-0.06}^{+0.05}$ \citep{cyburt08}. (Here the error in
the mean abundance of the Plateau admits no slope or step as a
function of {\teff}, both of which have been claimed in the
literature.)

The second point is that for the most metal-poor
near-main-sequence-turnoff stars, {\hea} and {\sdss} with [Fe/H]$_{\rm
  1D, LTE}$ = --5.4 and --4.7, lithium is not detected, leading to the
extremely puzzling limits of A(Li) $<$ 0.7 and $<$ 1.1. respectively.
Given that these objects have {\teff} = 6180\,K and 5810\,K, one would
have expected them to lie on the Spite Plateau. (Note also that there
is no evidence yet for binarity, or any other (non-abundance)
peculiarity, for these stars.)  This question will be addressed
further in Section~\ref{Sec:early_uni}, where they are discussed in
more detail.

Before proceeding, it should be noted that available 3D/non-LTE
computations appear to be in agreement with those based on the 1D/LTE
assumptions.  The reader is referred to \citet{asplund03}, who report
that for two stars (with {\teff}/{\logg}/[Fe/H] = 5690/1.67/--2.50 and
6330/2.04/--2.25) the 3D and non-LTE corrections are both large, with
absolute values of $\sim$0.3 dex, but of opposite sign, which
essentially cancel to yield a total correction of only $\sim$0.05 dex.
That is to say, 1D/LTE Li abundances are fortuitously valid.

Given the accuracy of the WMAP/SBBN prediction of the primordial Li
abundance, the most widely held view seems to be that the abundance
obtained from the analysis of observed Li line strengths in
near-main-sequence-turnoff metal-poor stars is not the primordial
value, and that an explanation of the difference will lead to a deeper
understanding of the astrophysics of stars and galaxies. The
reader should consult \citet{korn07}, \citet{lind09}, and
\citet{melendez10} for recent examples of this approach, based on
lithium abundances of field (Mel\'{e}ndez et al.) and globular cluster
(Korn et al.\ and Lind et al.) near-main-sequence-turnoff stars.
\citet{melendez10} report ``Models including atomic diffusion and
turbulent mixing seem to reproduce the observed Li depletion ... which
agrees well with current predictions from ... standard Big Bang
nucleosynthesis'', while \citet{lind09} state ``We confirm previous
findings that some turbulence, with strict limits to its efficiency,
is necessary for explaining the observations''.  Both, on the other
hand, issue a {\it caveat emptor}: ``We caution however that although
encouraging, our results should not be viewed as proof of the
... models until the free parameters required for the stellar modeling
are better understood from physical
principles.''(\citealt{melendez10}); and ``However, these models fail
to reproduce the behavior of Li abundances along the plateau,
suggesting that a detailed understanding of the physics responsible
for depletion is still lacking''(\citealt{lind09}).

\subsection{The Milky Way Halo} \label{Sec:MW_chem_evol}

The evolution of the chemical elements began shortly after the Big
Bang, and is an ongoing process.  It can be traced in detail in the
Milky Way with stars of different metallicities, ranging from the most
metal-deficient to the most metal-rich.  Iron abundance serves as
proxy not only for the overall metallicity of a star, but also for the
evolutionary timescales it took to enrich the gas from which stars
formed.  It is not possible in most cases, however, to determine the
ages of individual field stars, and what is known of Milky Way halo
ages is derived from the fitting of globular cluster and field star
near-main-sequence-turnoff color-magnitude diagrams to stellar
evolution modeling, and nucleo-chronometry of metal-poor field stars.
(This topic is further addressed in Section~\ref{Sec:ages}.)  As noted
in Section~\ref{Sec:plan_attack} the present discussion is restricted
principally to stars of the Galactic halo having [Fe/H] $<$ --1.0.
The Galactic age-metallicity relationship suggests that it took of
order $\sim$4\,Gyr to reach this abundance (see e.g.,
\citealt{freemanbh02}).  For comparison, the (one zone) galactic
chemical enrichment model of \citet{kobayashietal06} also takes
$\sim$4\,Gyr to reach [Fe/H] = --1.0.  The abundance trends discussed
in the following thus describe the first $\sim$5\,Gyr of the evolution
of the Milky Way -- which, for the present discussion, is taken as a
first approximation to the timescale for a similar enrichment of the
Universe.

To understand the production of the elements and the observed trends
found for metal-poor stars as a function of overall metallicity, most
subsections below begin with a description of the relevant
nucleosynthesis processes.  \citet{arnett96} and \citet{Wallerstein97}
provide general introductions to this topic.
\citet{woosley_weaver_1995}, among others, have carried out extensive
core-collapse SN yield calculations to investigate the synthesis of
the different isotopes during stellar evolution and subsequent
supernova explosion. Progenitor masses of 11 -- 40\,M$_{\odot}$ and
different metallicities were considered. Since the details of the
explosion mechanism of SNe remain largely unknown, a piston
approximation (for the sudden injection of energy -- the
``explosion'') is employed so that the post-SN nucleosynthesis can be
calculated.  Fortunately, relatively few isotopes appear to be
significantly affected by this uncertainty.  The overall explosion
energy and the ``mass cut'' (a specific radius above which material is
ejected, rather than falling back onto the nascent black hole or
neutron star) thus have significant impact on the final abundance
distribution. 

Mainly intermediate mass elements, with $Z\le30$, are produced and
ejected by core-collapse supernovae. Traces of the so-called
``neutron-capture'' elements ($Z\ga30$) are believed to be produced by
SNe and also during the asymptotic giant branch (AGB) phase of
evolution of low and intermediate mass ($\sim$1 -- 8 M$_{\odot}$)
stars. These heavier elements are about one million times less
abundant than the lighter ones. Irrespective of their quantities,
however, all elements play an important role in our understanding of
galactic chemical evolution since each reflects the interplay of all
the astrophysical processes and sites that produced the elements as they are 
known today.

In what follows, an unfortunately somewhat incomplete discussion is
presented of many of the elements that are observed in metal-poor
stars. Reasons for the incompleteness range from simple space
limitations of this chapter to the fact that not all chemical elements
can be observed in the relatively cool main-sequence and giant stars
reviewed here.

\subsubsection{The Evolution of Carbon through Zinc}\label{Sec:carbon}

\begin{center}{\it Carbon, Nitrogen, Oxygen} \\ \end{center} \vspace{-2mm}

C, N, and O are synthesized during quiescent stellar evolution, and in
Type II (core-collapse) SN explosions. Carbon is produced in the
triple-$\alpha$ process during advanced evolutionary stages such as
the AGB phase. It is released into the interstellar medium during
supernova explosions if the star is massive enough, or through stellar
winds, if significant. Whenever the triple-$\alpha$ process is at
work, some oxygen is created as a by-product in the
$\alpha$-process. Hence, O can be regarded as an ``$\alpha$-element''
(see below), and abundance studies have shown that O does indeed
exhibit this behavior in metal-poor stars (see
Figure~\ref{Fig:cayrel04}, top right, and the discussion below).
Nitrogen is produced during H-burning in the CNO-cycle, during which
C, N, and O act as catalysts.  In this process, C and O abundances
decrease while N increases, as the CNO-cycle approaches
equilibrium. At the same time the $^{12}$C/$^{13}$C ratio is driven to
small values ($\sim$4).  The production of nitrogen can be increased
by stellar rotation: fast rotating, massive Population\,III stars
\citep{meynet2005} may, for example, have been significant producers
of the first enrichments in CNO elements.

\begin{figure}[!tbp]
\begin{center}
\includegraphics[clip=true,width=8.7cm,bbllx=45, bblly=230,
   bburx=445, bbury=770]{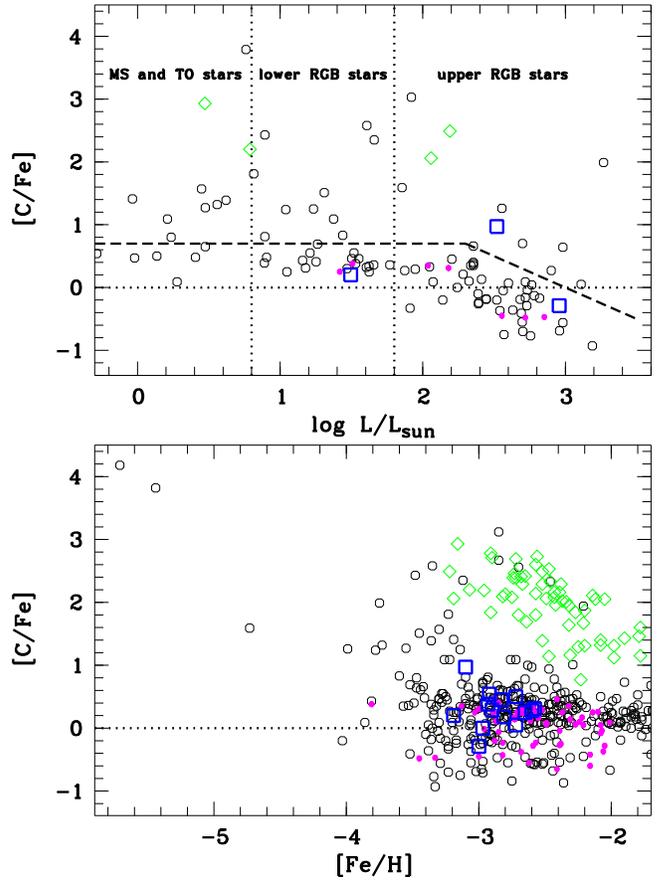}

\caption{\label{Fig:CFe} \small 1D/LTE [C/Fe] abundances as a function
  of [Fe/H] and luminosity (data from the \citealt{frebel10}
  compilation). In the top panel only stars with $\mbox{[Fe/H]}<-3.0$
  are included. The definition for C-rich objects of
  $\mbox{[C/Fe]}>0.7$ but with a luminosity dependent decline
  reflecting internal mixing processes is shown as a dashed line (see
  also Table~\ref{Tab:definitions}). Open diamonds are used for s- and
  r+s-process-rich metal-poor stars, open squares refers to r-II and
  small filled circles to r-I r-process-rich objects, which are
  further discussed in Section~\ref{Sec:s-proc}.  }

\end{center}
\end{figure}

In metal-poor stars, the abundances of each of C, N, and O can be
determined from observations of their hydrides -- the G-band of CH at
$\sim$4300\,{\AA}, the near-UV NH feature at 3360\,{\AA}, and the UV
features of OH at 3100\,{\AA}. CN and/or C$_{2}$ bands can also
provide constraints at optical wavelengths.  The point that must be
repeated here is that (as noted in
Section~\ref{Sec:relative_abundances}) 1D/LTE abundances determined
from CH and NH may overestimate C and N abundances by up to
$\sim$0.7\,dex.  Atomic features of C\,I at $\sim$9070\,{\AA},
together with the forbidden O\,I line at 6300\,{\AA} and the O\,I
triplet at $\sim$7770\,{\AA}, provide other important constraints on
the abundances of these elements. The three diagnostics involving
oxygen yield different 1D/LTE abundances, driven by 3D and non-LTE
effects.  Again, the reader should consult \citet{asplund05} for a
thorough discussion of the problem: suffice it here to say that only
for the forbidden O\,I line are the 1D/LTE abundances relatively
unaffected.  Recent results, taking into account various abundance
corrections (e.g., \citealt{fabbian09}), indicate relatively small
variation of [O/Fe] as a function of [Fe/H] in metal-poor stars.

Evolutionary mixing effects also modify initial surface abundances.
Dredge-up events and mixing bring nuclei from interior layers to the
surface, including CNO-processed material. The surface abundances of
heavier elements are not affected by these mixing processes and their
relative fractions remain unchanged.  The effect is shown in
Figure~\ref{Fig:cayrel04} where one sees a clear anticorrelation
between C and N in ``mixed'' stars (filled circles; [C/Fe]
  $<$ 0.0 and [N/Fe] $>$ +0.5) and ``unmixed'' stars (open circles;
  [C/Fe] $\geq$ 0.0 and [N/Fe] $<$ +0.5), as defined by
  \citet{spite05}.  These mixing effects are observed in metal-poor
stars with increasing luminosity on the upper RGB, as illustrated in
the upper panel of Figure~\ref{Fig:CFe}, where only stars with [Fe/H]
$<$ --3.0 are shown. A downturn of the [C/Fe] ratio can be clearly
seen at luminosity $\log$\,L/L$_{\odot}>2$.  When studying the CNO
group, stellar evolutionary status must therefore be taken into
account.

Many metal-poor stars show C abundances well in excess of the general
carbon trend set by most Population\,II stars (and unrelated to
abundance changes due to mixing) at all metallicities. Such ``extra''
C must have come either from additional sources that enriched the
material from which the star formed or from enriched material that was
added to the star at a later time. Carbon measurements in metal-poor
stars thus provide important information on the various previous
enrichment events and the nature of the first stars.  The lower panel
of Figure~\ref{Fig:CFe} shows [C/Fe] as a function of [Fe/H] for a
large sample of metal-poor stars. The separation of stars with large
carbon overabundances from those with more ``normal'' values is
clearly seen.  These {\bf C}arbon {\bf E}nhanced {\bf M}etal {\bf
  P}oor stars are referred to as CEMP stars, a significant fraction of
which have [Fe/H] $>$ --3.0 and inherited their carbon from a binary
companion star through a mass transfer event. This type of enrichment
is further discussed in Section~\ref{Sec:s-proc}.  The
origin of the carbon richness for the majority of stars with [Fe/H]
$<$ --3.0 remains the topic of intense interest, which is also
described in Section~\ref{Sec:early_uni}.

\begin{center}{\it $\alpha$-Elements} \label{Sec:alphas} \\ \end{center} \vspace{-2mm}

The $\alpha$-elements (Mg, Ca, Si, Ti) are built from multiples of He
nuclei via $\alpha$-capture during various stages of stellar evolution
(carbon burning, neon burning, complete and incomplete Si
burning). Although Ti ($Z=22$) is not a true $\alpha$-element, in
metal-poor stars the dominant isotope is $^{48}$Ti, which behaves like
one.  Produced in massive stars, these $\alpha$-element nuclei are
then dispersed during subsequent SN explosions. Abundance studies have
shown that the majority of metal-poor stars with $\mbox{[Fe/H]}<-1.0$
show an enhanced $\mbox{[$\alpha$/Fe]}$ ratio compared with the solar
value. Figure~\ref{Fig:alphas} illustrates that below this value
$\mbox{[$\alpha$/Fe]}$ plateaus at $\sim$0.4 due to the correlated
production and release of $\alpha$-elements and Fe.  This
characteristic overabundance in halo stars reflects an enrichment by
core-collapse SNe in the early Universe. At later times (roughly
1\,Gyr after the Big Bang), once the first lower-mass stars reached
the end of their lifetimes, SN\,Ia explosions began to dominate the
production of Fe. The main yield of SNe\,Ia is C, O, and Fe-peak
elements. This change in Fe producers can be seen in the abundance
trends of metal-poor stars.  Above metallicities of
$\mbox{[Fe/H]}\sim-1.0$, the onset of SNe\,Ia and their Fe
contribution to the chemical evolution of the Milky Way manifests
itself in a pronounced decrease of the stellar $\mbox{[$\alpha$/Fe]}$
values (e.g., \citealt{ryan96}) until $\mbox{[$\alpha$/Fe]}=0.0$ is
reached at $\mbox{[Fe/H]}=0.0$.

\begin{figure}[!tbp]
\begin{center}

\includegraphics[clip=true, width=8.70cm,bbllx=45, bblly=25,
  bburx=545, bbury=780]{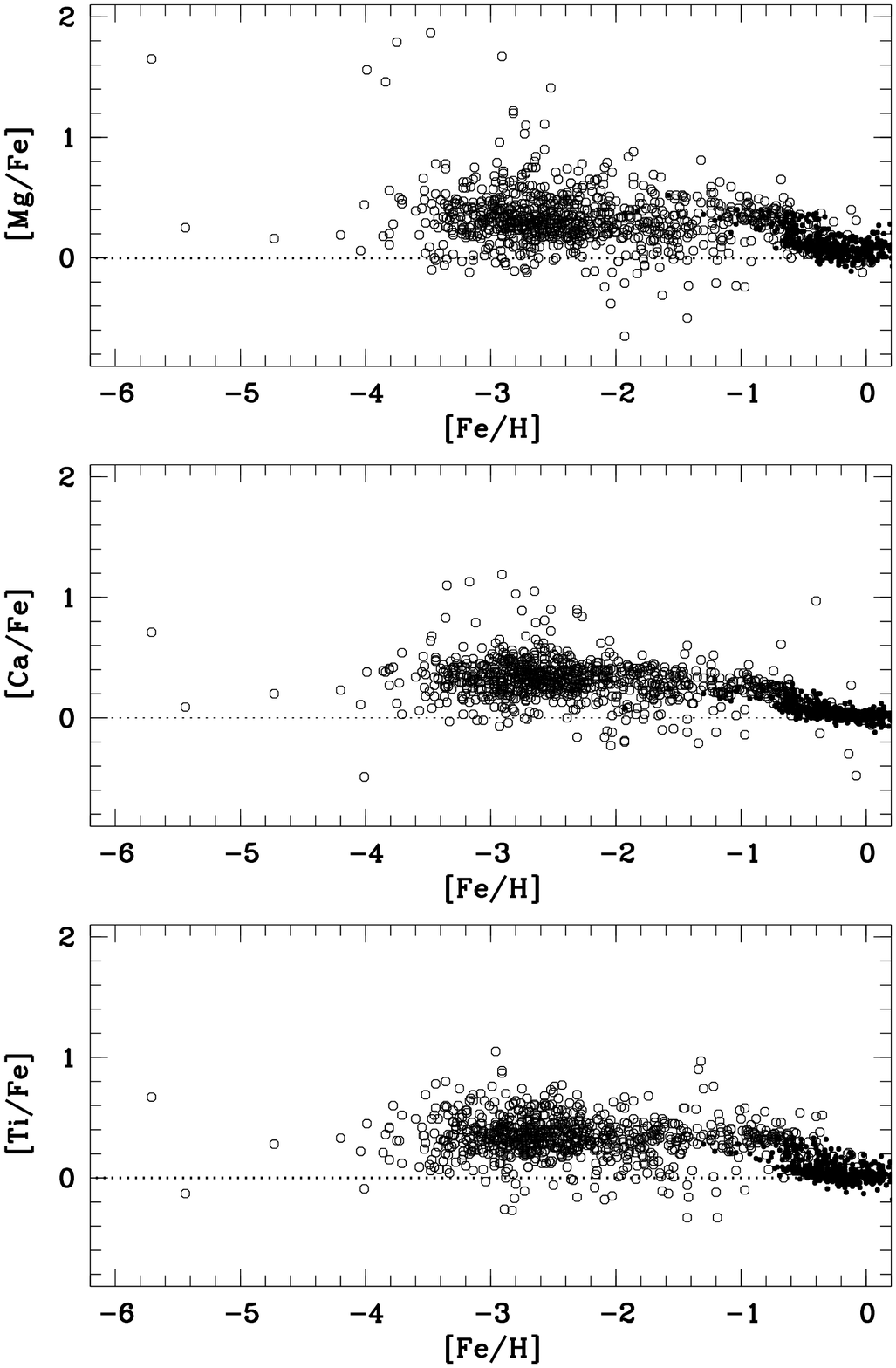}

\caption{\label{Fig:alphas} \small 1D/LTE $\alpha$-element ratios
  [Mg/Fe], [Ca/Fe], and [Ti/Fe] as a function of [Fe/H]. Open circles
  denote halo stars from the compilation of \citet{frebel10}.  Small
  dots are more metal-rich thin and thick disk stars from the
  \citet{venn04} compilation. At $\mbox{[Fe/H]}>-1.0$, all three
  element ratios decrease with respect to the average halo value of
  $\mbox{[$\alpha$/Fe]}\sim0.4$. This is due to the onset of SNe\,Ia,
  which contribute relatively high Fe to the chemical enrichment of
  the Galaxy. By $\mbox{[Fe/H]}\sim~0.0$, the solar ratio of
  $\mbox{[$\alpha$/Fe]}$ is reached.}

\end{center}
\end{figure}
There are important exceptions to this generalization.  Some
metal-poor stars show large Mg and Si abundances possibly due to
unusual supernova explosions and associated nucleosynthesis processes
\citep{aoki_mg, HE1327_Nature}.  Others, as will be discussed in
Sections~\ref{Sec:dg_evol} and \ref{Sec:mw}, exhibit lower values,
driven by the different star formation rates and different relative
$\alpha$/Fe contributions from Type II and Type Ia supernovae.

The $\alpha$-elements also serve to highlight a further potentially
important role of relative abundances as a function of [Fe/H].
Because the abundance of a given element contains the history of all
the SNe that have contributed to the cloud from which a star forms,
the dispersion of observed relative abundances contains potentially
strong constrains on the relative yields of SNe, the stellar mass
function, and the efficiency with which the ejecta of SNe have been
mixed with the existing ISM.  Several authors have emphasized the
small values of the dispersion in [Mg/Fe] ($\sim$0.06 -- 0.10 dex) in
homogeneously selected and analyzed halo samples, which lead to
interesting restrictions on the above possibilities.

\begin{center}{\it Iron-peak Elements}\\ \label{Sec:ironpeak} \end{center} \vspace{-2mm}

In the early Universe, the iron-peak elements (V to Zn; $23\le Z
\le30$) were synthesized during the final evolution of massive stars,
in a host of different nucleosynthetic processes before and during
core-collapse SN explosions.  These include direct synthesis in
explosive burning stages (explosive oxygen and neon burning, and
complete and incomplete explosive Si burning), radioactive decay of
heavier nuclei or neutron-capture onto lower-mass Fe-peak elements
during helium and later burning stages, and $\alpha$-rich freeze-out
processes.

\begin{figure}[!t]
\begin{center}

\includegraphics[clip=true, width=8.7cm,bbllx=45, bblly=25,
  bburx=545, bbury=780]{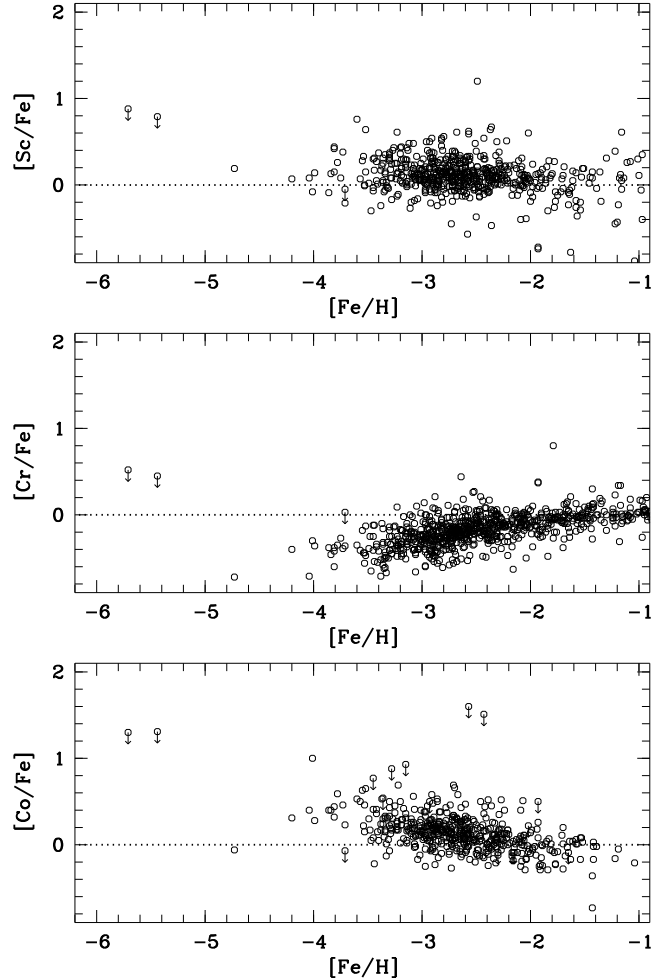}

\caption{\label{Fig:fepeak} \small 1D/LTE Fe-peak relative abundances
  [Sc/Fe], [Cr/Fe], and [Co/Fe] vs. [Fe/H] (data from the compilation
  of \citealt{frebel10}).  See text for discussion.}

\end{center}
\end{figure}

In Figure~\ref{Fig:fepeak}, 1D/LTE relative abundances, [Sc/Fe],
[Cr/Fe], and [Co/Fe] are presented as a function of [Fe/H], which
demonstrate quite different behavior as [Fe/H] increases.
Figure~\ref{Fig:cayrel04} also presents data for other Fe-peak
elements (Mn, Ni, and Zn) on the range $-4.0 \la \mbox{[Fe/H]}\la-2.5$.
The abundance trends of Cr and Mn have a pronounced positive slope:
their abundances at the lowest metallicities are subsolar
($\mbox{[Cr,Mn/Fe]}\sim-0.5$ at [Fe/H] $\sim$ --3.5), becoming
solar-like at [Fe/H] $\sim$ --1.0. In contradistinction, the Co and Zn
abundance trends are in the opposite sense. Their abundances decrease
from $\mbox{[Co,Zn/Fe]}\sim+0.5$ at [Fe/H] $\sim$ --3.5, to roughly
solar values at higher metallcities ([Fe/H] $\sim$ --1.0).  Finally,
Sc and Ni remain relatively unchanged with respect to [Fe/H].

Investigations of these very different behaviors have involved two
essentially different approaches.  The first was to consider whether
they could be explained in terms of the explosion energy and position
of the mass-cut of core-collapse SN models (see
\citealt{umeda&nomoto05} and references therein), which has been only
partially successful, and to which the reader is referred.  Second, in
the context of non-LTE effects, Bergemann and co-workers
(e.g., \citealt{bergemann10a}) have reported that for each of Cr\,I,
Mn\,I, and Co\,I, abundance differences (in the sense
$\Delta$[X/Fe](non-LTE -- LTE), are small at high abundance and
increase to $\sim$+0.4 to +0.6 at [Fe/H] $\sim\,-3.0$.  Consideration
of Figures~\ref{Fig:cayrel04} and \ref{Fig:fepeak} shows that while
this acts to remove the downward trends for [Cr I/Fe] and [Mn/Fe] (and
consistency then with [Cr II/Fe] results), it exacerbates the upward
behavior observed for [Co/Fe], leading to an excess of 1 dex above the
solar value at [Fe/H] = --3.5 -- providing an even larger challenge
for an understanding of this abundance ratio.

\subsubsection{The Evolution of Neutron-Capture Elements}\label{Sec:ncap}

Only elements up to zinc can be synthesized via nuclear fusion.  Most
heavier elements are built up by the slow neutron capture process, the
$s$-process, and the rapid neutron capture process, the $r$-process
(e.g., Meyer 1994 and references therein). The $s$-process operates
under conditions of relatively low neutron densities ($n_{n} \sim
10^{7}$ neutrons cm$^{-3}$). In this regime the timescale for neutron
capture is slower, in general, than the $\beta$-decay rate of unstable
isotopes. In contradistinction, when an extremely strong neutron flux is
present, the $r$-process occurs on timescales of only a few seconds so
that neutron-capture takes place within the $\beta$-decay rates of the
newly created unstable isotopes.  The majority of elements with $Z >
30$ can be produced by either the $s$- or $r$-process and it is not
trivial to disentangle the different production mechanisms.
Metal-poor stars offer an opportunity to obtain ``clean''
nucleosynthetic signatures of each process, as will be described in
this section.  This opportunity provides unparalleled insight into the
details of nucleosynthesis in the early Universe and the onset of
chemical evolution of the heaviest elements.

\begin{center}{\it s-process} \label{Sec:s-proc} \\ \end{center} \vspace{-2mm}

$s$-process nuclei are produced in the interiors of low and
intermediate-mass AGB stars and in the He- and C-burning phases of
massive stars.  On timescales longer than that of a $\beta$-decay,
neutrons are added to a seed nucleus (i.e., Fe) to build up heavier,
stable nuclei.  When neutron-capture creates a radioactive isotope, it
will in general decay to its stable daughter isotope before capturing
another neutron. In this way, nuclei along the ``valley of
$\beta$-stability'' are created. The overall extent to which heavier
and heavier isotopes are made is determined by the strength of the
neutron flux and the timescale over which it operates. This is known
as the time-integrated neutron-flux or neutron-exposure. As a
consequence, the s-process is more efficient in low metallicity AGB
stars due to a relatively larger ratio of neutrons to Fe seeds owing
to the primary nature of the neutron source.  In massive stars,
however, the efficiency of the s-process strongly depends on whether
the neutron source is of primary or secondary nature, and may depend
on stellar rotation.

About half of the isotopes of the elements heavier than iron can be
created through the s-process.  Nuclei with atomic numbers that are
equivalent to the magic numbers of neutrons, $A = 90$ ($N = 50$),
$A=140$ ($N = 82$), $A=208$ ($N = 126$), are produced in larger
quantities owing to their small neutron-capture cross-sections.  (Here
and in what follows, A refers to the mass number, Z to the atomic
number, and N to the neutron number of a nucleus.)  This results in
three so-called ``s-process peaks'' that make up a distinct
neutron-capture abundance signature.  The first peak is located at Sr,
Y, and Zr, the second at Ba, La, and Ce, and the third occurs at the
end point of the s-process, Pb and Bi. For the Sun, the s-process
component can be calculated, and subtracted from the total
neutron-capture pattern to obtain the r-process
contribution. Figure~\ref{Fig:r-s-proc} shows the neutron-capture
abundance distributions of both the s- and r-process. The s-process
peaks are clearly seen, as well as the relative contributions of each
process to a given element.

Overall, the s-process is rather well understood theoretically, even
though there remain uncertainties with regard to the modeling of the
amount of $^{13}$C that acts as a major neutron source and other
reaction rates associated with it (e.g., \citealt{arlandini1999,
  snedenetal08}).  The ``main'' component of the s-process occurs in
the helium shells of thermally pulsing lower mass AGB stars and is
believed to account for elements with $Z \ge 40$.  Examples of
s-process yields obtained from models of AGB stars with masses of 1 to
6\,M$_{\odot}$ and different metallicities can be found in
\citet{karakas07}.

The main neutron sources are $\alpha$-captures on $^{13}$C and
$^{22}$Ne nuclei. The former creates a low neutron density of $n_{n}
\sim 10^{7}$ neutrons cm$^{-3}$, whereas the latter can provide a
burst of neutrons with fluxes up to $n_{n} \sim 10^{13}$ neutrons
cm$^{-3}$ during convective thermal pulses.  The concentration of
$^{13}$C and the low reaction rate at the temperatures under which the
$^{13}$C($\alpha$,n)$^{16}$O reaction occurs in the He-shell maintain
the s-process for thousands of years. Moreover, the repeated exposure
of the He-shell to neutron fluxes is important for forming the
heaviest elements in AGB stars. On the contrary, the
$^{22}$Ne($\alpha$,n)$^{25}$Mg source has a timescale of only $\sim10$
years.  During the final stages of AGB evolution, s-process material is
dispersed by stellar winds.  

The so-called ``weak'' component of the s-process occurs in the He-
and C-burning cores of more massive stars of roughly solar
metallicity, and preferentially produces elements around $Z
\sim$~40. These stars are just massive enough (perhaps around
$8$\,M$_{\odot}$) to eventually explode as core-collapse SNe during
which the s-process material is ejected into the ISM. Regardless of
the mass range, the AGB phase includes a series of dredge-up episodes
that transport the newly created material to the surface. Through
stellar winds, the ISM is immediately enriched with s-process
elements, making AGB stars significant contributors to Galactic
chemical evolution.

\begin{figure}[!tbp]
\begin{center}

\includegraphics[clip=true, width=8.7cm,bbllx=41, bblly=405,
  bburx=530, bbury=750]{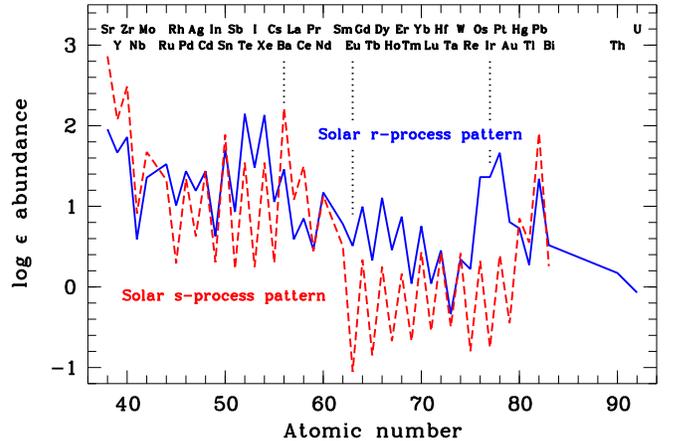}
  
\caption{\label{Fig:r-s-proc} \small Solar s- and r-process patterns
  (dashed and full lines, respectively) (data from
  \citealt{2000burris}).  After billions of years of chemical
  evolution, the different contributions of the s- and r-processes are
  clearly seen. Sr and Ba are predominantly produced in the s-process,
  whereas Eu, Os, Ir and Pt originate mainly in the r-process.  (The
  vertical dotted lines have been added to facilitate identification
  of Ba, Eu, and Ir.).  }

\end{center}
\end{figure}

Given that many stars occur in binary systems, a common scenario is
mass transfer during which s-process elements are transferred to a
lower mass companion. This fortuitously provides an indirect method of
studying a clean AGB nucleosynthesis signature.  The process occurs
not only in the early Universe, but also among higher metallicity
stars. The so-called ``Ba-stars'' are the ``receiver'' stars within
Population\,I binaries and the ``CH stars'' those within mild
Population\,II systems.  The characteristic s-process signature seen
above in Figure~\ref{Fig:r-s-proc} has been observed in many
metal-poor stars as the result of a more massive companion going
through the AGB phase and transferring some material onto its
companion (e.g., \citealt{2001aokisprocess}). In
Figure~\ref{Fig:s-proc}, abundances for several s-process-enhanced
stars are shown in comparison with the scaled solar s-process pattern.
(One should recall here that an abundance definition that classifies
stars as s-process-rich is given in Table~\ref{Tab:definitions}.)  As
may be seen in the figure, there is good agreement between the scaled
solar pattern and the stellar abundances. (It should be remarked in
passing that the relatively poor agreement for Pb results from a
significant underproduction of Pb in earlier s-process solar models,
such as the one presented in Figure~\ref{Fig:s-proc} (R. Gallino,
private communication). Correspondingly, the scaled-solar r-process Pb
predictions are too high. There remain disagreements, however, between
the observed Pb abundances and the model predictions, suggesting that
either our understanding of these processes is still rather limited or
that there are systematic uncertainties in the abundance
determinations, or both.)

This agreement is remarkable given the fact that the
solar neutron-capture material is a product of $\sim8$\,Gyr of
integrated chemical evolution, whereas the halo stars received these
elements directly from one of the AGB stars that made s-process
elements early in the Universe.  Overall, the abundance match
indicates a solid theoretical understanding of the s-process. This is
also demonstrated by a small number of s-process stars that show
extremely large enhancements of Pb, as predicted for third
peak-elements (see \citealt{snedenetal08}).  Additionally, since
carbon is also produced during the AGB phase, the mass transfer
usually includes large amounts of carbon. (s-process-enhanced stars
are marked as such in Figure~\ref{Fig:CFe}, with diamond symbols, to
illustrate this point.)  Most importantly, it should be kept in mind
that the carbon excess in these stars is dominated by an extrinsic
source and not representative of the intrinsic carbon abundance of the
stars' birth cloud.  Finally, note that despite the mass transfer, the
s-process-enhanced metal-poor stars exhibit lighter element ($Z<30$)
abundance patterns that are the same as those of other normal
metal-poor halo stars. One exception is fluorine which, if
significantly enhanced, is a signature of low-mass AGB pollution
together with usual s-process-element and carbon-enhancement.

\begin{figure}[!tbp]
\begin{center}
\includegraphics[clip=true, width=8.7cm,bbllx=26, bblly=200,
   bburx=528, bbury=630]{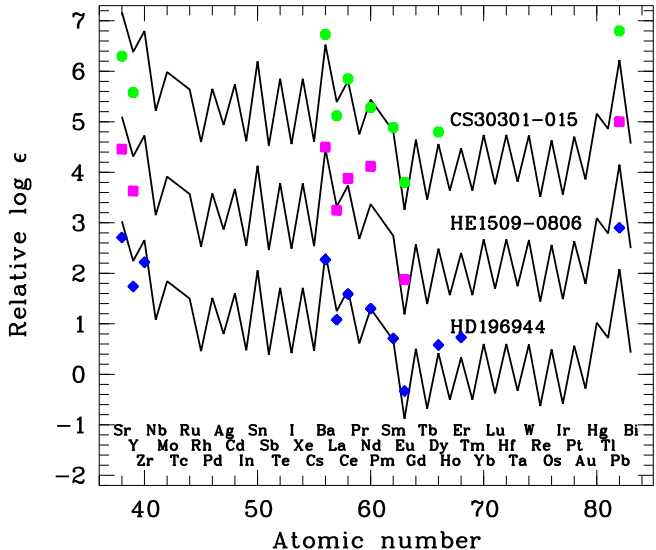}

\caption{\label{Fig:s-proc} \small 1D/LTE abundances of
  s-process-enhanced metal-poor stars compared with the scaled solar
  s-process pattern. See text for discussion.}

\end{center}
\end{figure}

The evolution of representative neutron-capture elements as a function
of [Fe/H] is shown in Figure~\ref{ncap}. Sr and Ba are predominantly
produced in the s-process (89\% and 85\% in the Sun, respectively); see
\citet{2000burris} for details.  Since the first lower mass stars in
the Universe reached their AGB phase $\sim$1\,Gyr after the Big Bang,
s-process enrichment occurs with some delay with respect to
core-collapse SN enrichment. This is indeed as observed: at a
metallicity of $\mbox{[Fe/H]}\sim-2.6$, the s-process is in full
operation, including significant neutron-capture element ``pollution''
of the Galaxy by AGB stars \citep{simmerer2004}, as can be seen in the
top panels of Figure~\ref{ncap}.  Metal-poor stars with an obvious
s-process signature from a mass transfer event are also
carbon-rich. At these and higher metallicities, all stars thus formed
from gas that was intrinsically enriched in s-process elements,
irrespective of whether or not they received extra s-process material
from a companion. As can be seen in the bottom panels of the figure,
there is a main branch in the [Ba/Fe] vs. [Fe/H] plane.  Above
$\mbox{[Fe/H]}\sim-2.6$, it is dominated by stars formed from AGB
enriched gas.

There are some exceptions with respect to clean s-process signatures
in metal-poor stars. A handful of objects display a mixed abundance
signature originating from both the s- and the r-process (see
e.g., \citealt{jonsell06} for an extensive discussion on the origin of
s-/r-mixtures observed in metal-poor stars). This includes some stars
with $\mbox{[Fe/H]}<-2.6$, and their unusual chemical patterns are
perhaps due to earlier more massive stars expelling some s-process
elements when they exploded as core-collapse SNe. Several different
scenarios have been invoked to explain the combination of the two
neutron-capture processes originating at two very different
astrophysical sites. No completely satisfactory explanation, however,
has yet been found.

\begin{figure*}  [!ht]
\begin{center}
%\hspace{+12mm}

\includegraphics[width=13.0cm,clip=true,bbllx=45,bblly=202,bburx=505,bbury=720]{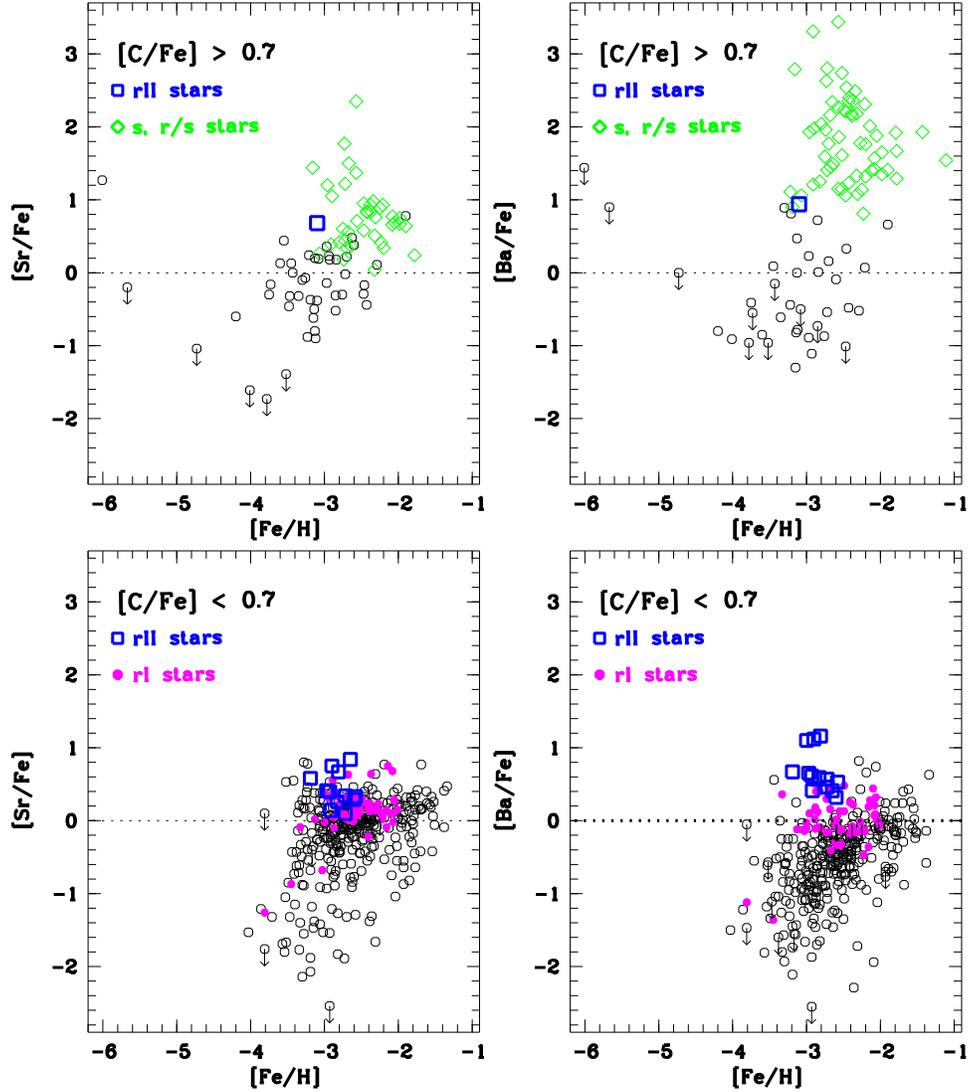}

\caption{\label{ncap} {\small 1D/LTE neutron-capture element
    abundances ratios [Sr/Fe] and [Ba/Fe] as a function of [Fe/H] for
    carbon-enhanced objects ($\mbox{[C/Fe]}\ge0.7$; top panels) and other halo
    stars (bottom panels). The range in [Sr/Ba] and [Ba/Fe] is much
    larger than uncertainties or systemic differences between
    individual studies, indicating a cosmic origin.  Below
    $\mbox{[Fe/H]}\sim-3.0$, the evolution is dominated by r-process
    enrichment. For $\mbox{[Fe/H]}\ga -2.6$, the s-process
    significantly contributes neutron-capture material (see
    \citealt{simmerer2004}). Arrows indicate upper limits, while the
    solar ratio is indicated by dotted lines.}}

\end{center}
\end{figure*}

\begin{center}{\it r-process} \label{Sec:rprocess} \\ \end{center} \vspace{-2mm}

Heavy elements are also produced in the rapid (r-) process, which
takes place over just a few seconds.  Seed nuclei (e.g., C or Fe) are
bombarded with neutrons ($\sim$10$^{22}$ neutrons cm$^{-2}$
sec$^{-1}$) to quickly form large radioactive nuclei far from
stability.  After the strong neutron flux ceases, the nuclei decay to
form stable, neutron-rich isotopes.  The r-process does not, however,
produce infinitely large nuclei because of a significant decrease in
the cross sections of neutron-capture nuclei with closed neutron
shells. Other unfavorable reaction rates and problems with nuclear
stability in the heavy-isotope region also play a role. These factors
eventually terminate the r-process at nuclei around $A=270$, far in
the trans-uranium regime. Those nuclei all decay to eventually become
Pb. Approximately half of the neutron-capture isotopes heavier than
iron are produced in this way, including the heaviest, long-lived
radioactive elements thorium and uranium.

The r-process also manifests itself in a characteristic abundance
pattern, showing three large peaks at elements with $A\sim$\,80
($Z\sim$\,33; Se-Br-Kr), $A\sim$\,130 ($Z\sim$\,52; Te-I-Xe) and
$A\sim$\,195 ($Z\sim$\,77; Os-Ir-Pt), similar to the s-process
peaks. The latter two of these may be seen in
Figure~\ref{Fig:r-s-proc}. The peaks form because nuclei with closed
neutron shells only reluctantly capture any neutrons (i.e., they have
extremely small cross-sections). With their long $\beta$-decay
lifetimes, they act as bottlenecks to additional neutron-captures
creating even heavier nuclei. Hence, nuclei with atomic masses at or
just below the closed-shell nuclei pile up during the process.

Unlike the situation for the s-process, the astrophysical site(s) that
provide the extreme neutron fluxes required for the r-process have not
yet been identified.  Neutron-star mergers have been considered, but
their long evolutionary timescale prior to merging argues against them
being the primary r-process site in the early Galaxy.  Neutrino-driven
winds emerging from the formation of a neutron-star during a
core-collapse SN explosion are more promising locations
\citep{qian_wasserburg03}.  Since massive SNe dominate chemical
enrichment in the early Universe (e.g., as documented through the
$\alpha$-element enhancement found in halo stars), the neutrino-driven
wind model agrees naturally with such an early SN enrichment mode.

\begin{figure*}[!tbp]
\begin{center}

\includegraphics[clip=true, width=12.0cm,bbllx=27, bblly=205,
   bburx=555, bbury=690]{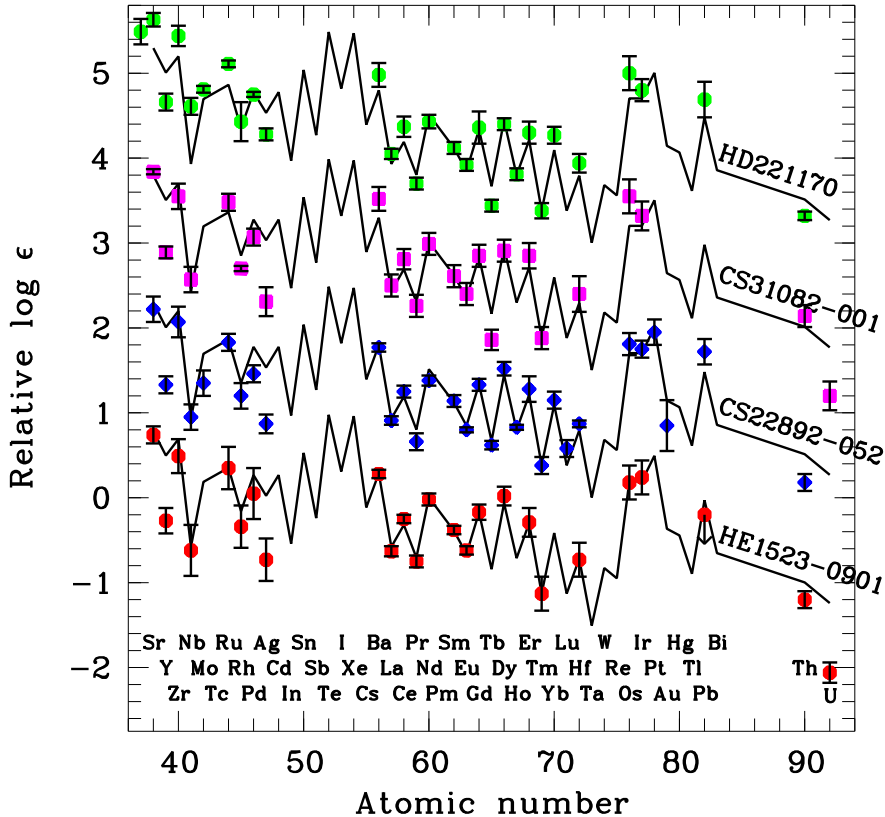}

\caption{\label{Fig:r-proc} {\small 1D/LTE abundances in
    r-process-enhanced metal-poor stars compared with the scaled solar
    r-process pattern. Note the remarkable agreement for elements
    heavier than Ba ($Z\ge56$).}}

\end{center}
\end{figure*}

In order to learn about the details of the r-process and its site, it
is of great importance to obtain actual data of a ``clean'' r-process
signature. The best candidates for this are the r-II stars (see
Table~\ref{Tab:definitions} for definitions), the most strongly
r-process-enhanced objects, which comprise about $5\%$ of stars with
$\mbox{[Fe/H]}\lesssim-2.5$ (see \citealt{heresII}).  All but one of
the r-II stars have metallicities close to $\mbox{[Fe/H]}=-3.0$, with
the outlier having an even lower [Fe/H]. The metallicities are thus
distinctly lower than the value of $\mbox{[Fe/H]}=-2.6$ discussed
above as corresponding to the onset for AGB s-process enrichment.  (It
should be recalled that mildly enriched r-I stars are found up to
metallicities of $\mbox{[Fe/H]}\sim-2.0$; while at higher values, the
signature becomes less clean since the more metal-rich star would have
formed from material already significantly enriched in r-process
elements.) This suggests that the r-process enhancement comes from
stars slightly more massive than those that experience the s-process
during AGB evolution.

The ``main'' r-process operates in the full range of neutron-capture
elements, up to $Z = 92$.  Model calculations have shown that it
probably only occurs in a specific, yet unidentified type of
core-collapse SN, or perhaps only in a particular mass range ($\sim$8
-- 10\,M$_{\odot}$; \citealt{qian_wasserburg03}). Examination of the
ratios of the heavy ($Z>56$) neutron-capture abundances in r-process
enhanced stars (e.g., \citealt{Snedenetal:1996, Hilletal:2002,
  heresII, he1523}) shows that the abundance distribution of each
closely matches that of the scaled solar r-process pattern (e.g.,
\citealt{2000burris}). Figure~\ref{Fig:r-proc} shows data for four
well-studied r-II stars. Given that the Sun was born $\sim$8\,Gyr
later than these otherwise ordinary metal-poor stars, this is a
remarkable finding. Assuming that the r-process takes place only in
core-collapse SNe, the match of the stellar and solar patterns
suggests that the r-process is universal: that is, no matter when and
where is happens, it always produces its elements with the same
proportions. Otherwise, the integrated pattern observed in the Sun
would not resemble the individual pattern found in a $\sim$13\,Gyr old
star.

While there is excellent agreement with the scaled solar r-process
pattern for elements heavier than Ba, deviations have been found among
the lighter neutron-capture species.  This indicates that the origin
of the lighter elements is more complex, with perhaps both the ``main''
and ``weak'' r-processes contributing in different mass ranges (see,
e.g., \citealt{travaglio}).  The ``weak'' r-process is thought to
produce mainly the lighter neutron-capture elements ($Z<56$) and
little or no heavier material, such as Ba. Possibly, this occurs
mainly in massive ($\gtrsim20$\,M$_{\odot}$) core-collapse SNe (see
e.g., \citealt{wanajo_ishimaru}). A candidate for an observed ``weak''
r-process signature is provided by the r-process-poor, metal-poor star
HD~122563 \citep{honda06}, which displays a depleted, exclusively
light neutron-capture-element pattern.  The [Sr/Ba] ratio in this and
other stars can be employed to learn about the relative contributions
of the two r-processes, and potentially the origin of the overall
abundance pattern. In this scenario, the ``main'' r-process would
produce lower [Sr/Ba] ratios than the ``weak'' one. 

Overall, neutron-capture elements are produced in limited amounts.
Their abundances in solar system material is about 6 orders of
magnitude less than those of the Fe-peak elements. Nevertheless, they
provide invaluable insight into various early nucleosynthesis
processes. The enormous scatter of neutron-capture abundances (e.g.,
[Ba/Fe]), as a function of [Fe/H], suggests that the production of
neutron-capture elements is completely decoupled from that of Fe and
other elements. As described earlier, and displayed in
Figures~\ref{Fig:alphas} and \ref{Fig:fepeak}, the $\alpha$- and
Fe-peak element abundances show very little scatter, probably because
the ISM was already relatively well-mixed by
$\mbox{[Fe/H]}\sim-4.0$. As can be seen in Figure~\ref{ncap},
especially at $\mbox{[Fe/H]}\sim-3.0$, there is a range of $\sim$6\,dex
in neutron-capture abundances. This must reflect strongly varying
degrees of neutron-capture yields at the earliest times, and probably
also different processes contributing different groups of these
elements in various amounts (for example more Sr than Ba at the very
lowest metallicities).  Only at somewhat higher metallicities, when
the s-process begins to dominate the neutron-capture inventory, does
the bulk of the stellar abundance ratios become more solar-like.

\subsection{The Milky Way Globular Clusters and Dwarf Galaxies}

\subsubsection{Globular Clusters} \label{Sec:gc_evol}

The internal relative abundance patterns of the Galactic globular
clusters are distinctly different in many respects from those of the
Galactic halo, and require an explanation involving poorly understood
intracluster self-enrichment processes.  While this is a
  very important field of endeavor, insofar as it lies beyond the
scope of the present chapter, a comprehesive description of this topic
can not be provided here.  Instead, some of the main differences
between the globular cluster and halo field stars are briefly
highlighted.

\begin{itemize}

\item Most clusters are chemically homogeneous with respect to iron,
  at the $\sim$0.03 dex level.  The clear exceptions among the halo
  clusters are $\omega$ Centauri, M22, and M54, where ranges of
  $\Delta$[Fe/H]~$\sim$~0.3 to 1.5\,dex have been observed.  In
  Figure~\ref{Fig:mdf_dsph} the reader can see the MDF of $\omega$
  Cen, the cluster with the largest spread; a large number ($\sim$5)
  of sub-populations, with distinctly different mean [Fe/H] have been
  identified in this system.

\item All globular clusters so far studied have been found to be
  chemically inhomogeneous in a number of light elements that are
  produced or modified in nuleosynthetic $\mbox{(p, $\gamma)$}$
  reactions.  Beginning in $\sim$1970, observations and analyses of
  cluster members have over time added the following elements to the
  list -- C, N, O, Mg, Na, and Al.  Strong correlations and
  anti-correlations exist among the abundances of these elements; and
  within a given cluster, sub-populations have been identified based
  on abundance patterns involving them.  Lithium and
  heavy-neutron-capture element variations have also been reported
  that correlate with those of the above elements in some clusters.
  Intermediate mass AGB stars are most commonly identified as the
  nucleosynthesis sites responsible for these variations, together in
  some cases with internal mixing high on the present-day RGB.

\item Finally, some (but not all) of the most massive clusters
  ($\omega$ Cen, NGC 2808) show multiple main sequences in the
  color-magnitude diagram, for which the only empirically consistent
  explanation yet proposed is that there are sub-populations within
  these systems that have distinct helium abundances in the
  astoundingly large range Y $\sim$~0.23 -- 0.35 (Y is the
  helium fraction by mass).  No completely satisfactory explanation
  has yet been given, although several authors identify massive AGB
  stars in an early generation of cluster stars as the prime
  candidate.

\end{itemize}

Various models have been proposed that are unique to the globular
cluster environment and which involve a number of stellar generations
that chemically enrich the material from which subsequent generations
form.  An example of such a model, which also provides references to
the observational material described above is provided by \citet{conroy11}.

\subsubsection{Dwarf Galaxies} \label{Sec:dg_evol}

One of the important unsolved problems in cosmology is understanding
the formation of galaxies.  Studying the compositions of stars in
dwarf galaxies provides information on the chemical evolution of these
systems.  The Milky Way's dwarf galaxy satellites, with a large range
in masses and luminosities very different from those of the Milky Way
itself, permit a comparison of their chemical evolution histories,
which in turn provides clues to the origin and overall evolution of
different types of galaxies.  Specifically, the connection between the
surviving dwarf systems and those believed to have been captured and
dissolved to form the Milky Way halo is best addressed by examining in
detail the stellar chemical abundances of present-day dwarf galaxies
(see also Section~\ref{Sec:mw} on this topic).  The most metal-poor
(and hence oldest) stars in a given system permit unique insight into
the earliest phases of star formation.  Stars born at later times (and
thus with higher metallicities) contain the integrated effects of
internal chemical evolution in their atmospheric compositions. (See
\citealt{kirby11} for an overview of the history and current state of
simple chemical evolution models.)

\begin{figure*}[!t]
\begin{center}

\includegraphics[clip=true, width=12.0cm,bbllx=30, bblly=100,
   bburx=480, bbury=730]{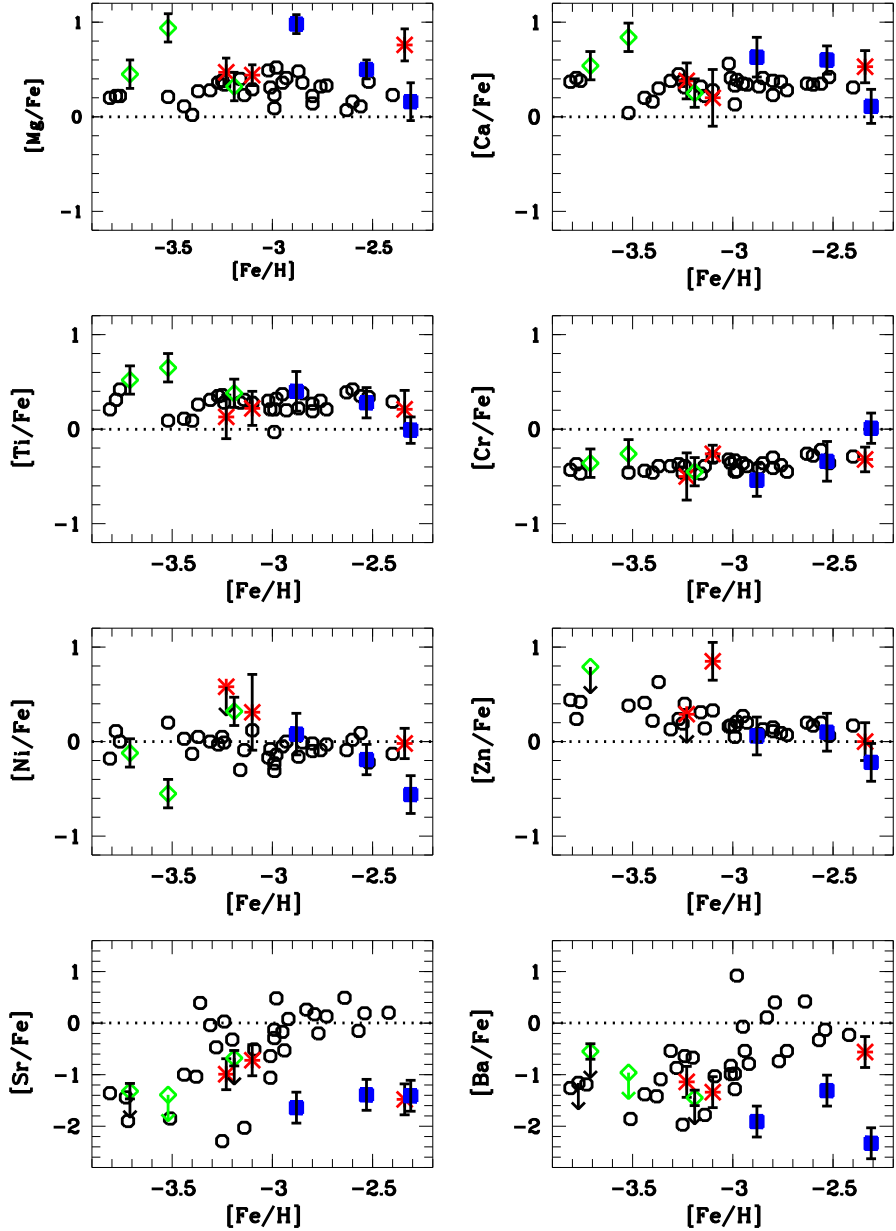}

\caption{\label{Fig:abund_comp}\small Comparison of stars in the
  Galactic halo (circles: \citealt{cayrel2004,francois07}) and dwarf
  galaxies (asterisks: Ursa Major\,II, filled squares: Coma Berenices,
  diamonds: Segue\,1, Bootes\,I and Leo IV) in the 1D/LTE relative
  abundances ([X/Fe]) vs [Fe/H] diagram.  While the light element
  abundances agree very well, dwarf galaxy stars have relatively low
  neutron-capture abundances, albeit still within the range of the halo
  stars.}

\end{center}
\end{figure*}

Dwarf spheroidal galaxies are relatively simple systems that allow us
to study, both observationally and theoretically, the basic processes
that led to their origin and evolution. They are generally old,
metal-poor, have no gas, and thus no longer support star formation.  On
the other hand, a large fraction of their mass comprises dark matter,
with the least luminous of them having mass-to-light ratios of order
10$^{3}$ (in solar units). Some 25 such systems are currently known
orbiting the Galaxy today (see \citealt{tolstoy_araa} for a
review). The $\sim$10 recently discovered ``ultra-faint'' dwarf
galaxies ($L_{\rm V} \leq 10^{5}\,$L$_{\odot}$; \citealt{martin08}) are
some orders of magnitude fainter than the more luminous,
``classical'', Milky Way dwarf spheroidal galaxies.  As has been
outlined in Section~\ref{Sec:mdf_dsph}, all of these dwarf systems
follow a (metallicity, luminosity)--relationship, with the classical
dwarfs being on average more metal-rich and containing more stars than
their less luminous ultra-faint siblings.

With $\mbox{[Fe/H]}\gtrsim-2.0$, stars in the classical dwarf galaxies
were found to have abundance ratios different from halo stars at the
same metallicity (e.g., \citealt{shetrone03, geisler05}). Most
strikingly, the $\alpha$-element abundances are not enhanced to the
SN\,II enrichment level of $\mbox{[$\alpha$/Fe]}\sim0.4$.  This
indicates different enrichment mechanisms and longer timescales in the
dwarf galaxies; due to a slower evolution, the Fe contribution from
SN\,Ia occurred ``earlier'', at a time when the entire system had not
yet reached a metallicity of $\mbox{[Fe/H]}\sim-1.0$, the turn-down
point of $[\alpha$/Fe] vs. [Fe/H] in the Milky Way (see
Figure~\ref{Fig:alphas}).

Only very recently, a handful of stars with metallicities of
$\mbox{[Fe/H]}<-3.0$ was discovered in the classical dwarf galaxies,
with some of them having $\mbox{[Fe/H]}\sim-4.0$ \citep{scl}. While
these dwarfs have been studied for many decades, problems with earlier
search techniques had prevented the discovery of extremely metal-poor
stars (\citealt{starkenburg10}). The existence of such objects shows
that a metallicity range of $\sim$3\,dex is present, at least in the
Sculptor and Fornax dSphs.  At $\mbox{[Fe/H]}<-3.0$, the chemical
abundances, obtained from high-resolution spectra, are remarkably
similar to those of Galactic halo stars at similar metallicities. This
is in contrast to the deviations at higher [Fe/H], and provides
evidence for a change in the dominant enrichment mechanisms. For these
types of dwarf galaxies, the transition from halo-typical abundance
ratios (as a result of SN\,II enrichment) to more solar-like values
(SN\,Ia-dominated Fe production) appears to take place around
$\mbox{[Fe/H]}= -3.0$ \citep{cohen09, aoki09}.  As a consequence,
chemical evolution may be a universal process, at least at the
earliest times, the very regime that is probed by the most metal-poor
stars.

The first extremely metal-poor stars not belonging to the Galactic
halo field population were found in some of the ultra-faint dwarf
galaxies, even before such stars were discovered in the classical
dwarfs \citep{kirby08}. Due to their distance and low stellar density
these systems contain few stars brighter than $V=19$, making the
collection of spectroscopic data a challenge.  Nevertheless,
high-resolution spectra of a handful of individual metal-poor stars in
Ursa Major\,II, Coma Berenices, Bootes\,I, Segue\,1, and Leo\,IV
(\citealt{frebel_umacom10,norris10,norris10seg,simon10}) have been
obtained. A large fraction of them are extremely metal-poor (i.e.,
[Fe/H] $<$ --3.0). With one exception, all of their light element
($Z<30$) abundances show the halo-typical core-collapse SNe signature,
resembling those of similarly metal-poor Galactic halo stars.  This
may be clearly seen in the upper six panels of
Figure~\ref{Fig:abund_comp}, where the relative abundances of the
metal-poor halo red giant sample of \citet{cayrel2004} and
\citet{francois07} (presented above in
Section~\ref{Sec:relative_abundances}) are compared with those
available for red giants in the ultra-faint dwarf galaxy stars.  The
exception is the CEMP-no star (see Table~\ref{Tab:definitions})
Segue~1--7, a radial-velocity member or Segue~1, which has [Fe/H] =
--3.52, [C/Fe] = +2.3, and [Ba/Fe] $<$ --1.0 \citep{norris10seg}.  The
200-fold overabundance of carbon relative to iron in this extremely
metal-poor star is quite remarkable.  This shows that the CEMP-no
phenomenon is not restricted to the Milky Way halo and may provide
important clues to the origin of these stars.

Some abundances, however, indicate that the chemical evolution in
these small systems may have been moderately inhomogeneous (see
Section~\ref{Sec:mdf_dsph}, Figure~\ref{Fig:mdf_dsph}), with some
stars perhaps reflecting enrichment by massive Population\,III
stars. The chemical similarity to halo stars is also found at higher
metallicity, up to $\mbox{[Fe/H]}\sim-2.0$, in contrast to what has
been found for the classical dwarfs.  This remarkable similarity
between the abundance profiles of the halo and the dwarf galaxies
supports the view that chemical evolution is independent of galaxy
host mass in this metallicity regime.  Moreover, this (together with
the existence noted above of a CEMP-no star in the ultra-faint
Segue~1) renews support for a scenario in which the metal-poor end of
the Milky Way halo population was built up from destroyed dwarf
galaxies (see Section~\ref{Sec:mw}).

Finally, neutron-capture abundances should be mentioned. These are
extremely low in the ultra-faint systems, and as may be seen in the
two bottom panels of Figure~\ref{Fig:abund_comp} in the range $-3.0 <$
[Fe/H] $< -2.0$, the observed Sr and Ba values lie well below those
found in typical Milky Way halo stars.  A more general statement is
that the mean values of [Sr/Fe] and [Ba/Fe] are significantly smaller
in the ultra-faint dwarfs than in the halo.  Comparably low values for
Sr and Ba are also found in the more luminous dwarfs Hercules
\citep{koch_her} and Draco \citep{fulbright_rich} despite their
sometimes relatively high Fe values of $\mbox{[Fe/H]}\sim-2.0$.

\section{COSMO-CHRONOMETRY}\label{Sec:ages}

Because of their low metallicity, metal-poor stars are usually
regarded as having been formed at the earliest times, when the first
elements heavier than helium were being synthesized.  The most
metal-poor stars are thus regarded as being almost as old as the
Universe.  Age determinations for field stars are, however, difficult,
since they do not belong to a distinct single-age population such as a
globular cluster.  Cluster ages are based on fitting isochrones to
their color-magnitude diagrams.  The age dating of globular clusters
will not be discussed here, and the reader is referred to
\citet{vandenberg96} for details, and to \citet{marinfranch09} for
more recent results which are addressed further in
Section~\ref{Sec:mw}.  Suffice it to say that although the clusters
are not as metal-poor as the most metal-poor field stars, the ages of
the older of them range from 10 to 14\,Gyr, placing them among the
oldest objects in the Universe. The main point of focus in this
chapter is dating techniques for individual r-process-enhanced
Galactic halo field stars.

\subsection{Nucleo-Chronometry of Metal-Poor Field Stars}

A fundamental way to determine the age of a {\it single} star is
through radioactive decay dating. Elements suitable for this procedure
are not, however, present in sufficient quantities in ordinary stars.
There is also the problem of finding stars that have experienced
enrichment from a single source so that the decay tracks the time from
just one production event until the time of measurement. Fortunately,
in strongly r-process-enhanced metal-poor stars, radioactive age
dating is possible through abundance measurements of Th ($^{232}$Th,
half-life $14$\,Gyr) and/or U ($^{238}$U, half-life $4.5$\,Gyr).
These half-lives are sufficiently long for measurements of cosmic
timescales, and stellar ages can be determined based on radioactive
decay laws that lead to simple equations for different chronometer
ratios involving Th, U, and stable r-process elements.  Observed
abundances in r-processed-enhanced stars provide determinations of
their remaining radioactive material, e.g., $\rm
\log\epsilon(Th/r)_{now}$, with r being a stable element such as Eu,
Os, and Ir, for which the following relationships (derived from
radioactive decay laws in combination with known nuclear physics)
obtain.

\begin{itemize}
\item[1.] \mbox{$\Delta t = 46.78[\log{\rm (Th/r)_{initial}} - {\rm
      \log\epsilon(Th/r)_{now}}]$ Gyr}
 
\item[2.] \mbox{$\Delta t = 14.84[\log{\rm (U/r)_{initial}} - {\rm
      \log\epsilon(U/r)_{now}}]$ Gyr}

\item[3.] \mbox{$\Delta t = 21.76[\log{\rm (U/Th)_{initial}} - {\rm
      \log\epsilon(U/Th)_{now}}]$ Gyr} 
\end{itemize}

Only theoretical r-process calculations can provide the initial
production ratios ($\log{\rm (Th/r)_{initial}}$ and $\log{\rm
  (U/r)_{initial}}$) that describe how much r-process material,
including Th and U, was made in the production event, i.e., the SN
explosion. This implies that, technically, the SN is dated rather than
the star. The time span, however, of the formation of the star after
the SN is regarded as negligibly short compared to the star's age.
Currently, the astrophysical site of the r-process remains unclear,
and the associated initial conditions are not known, making yield
predictions difficult. Nevertheless, some calculations involving
various approximations are available (e.g.,
\citealt{schatz_chronometers}).  It should be kept in mind that the
universality of the r-process, noted in Section~\ref{Sec:rprocess},
(at least for $Z\ge56$) is a major ingredient in predicting the
relative elemental ratios, such as Th/r.

\begin{figure*}[!t]
\begin{center}

\includegraphics[clip=true, width=17.5cm,bbllx=30, bblly=308,
   bburx=525, bbury=515]{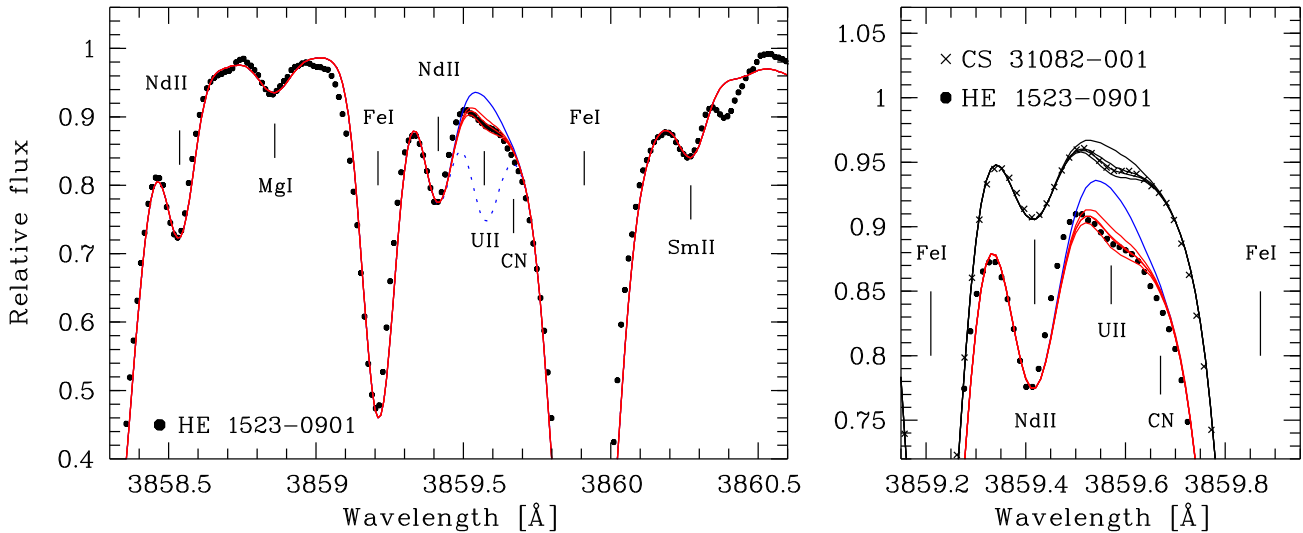}
  
\caption{\label{Fig:U_region} \small Spectrum synthesis of the U line
  region at 3860\,{\AA} in HE~1523$-$0901 (left panel, whole region;
  right panel, detailed view of just the line) and also CS~31082-001
  (right panel only). Dots indicate the observed spectrum and
  continuous lines present synthetic spectra computed with a range of
  U abundances for comparison with the observed one.  The latter are
  best illustrated in the right panel where the lowest three lines
  correspond to log $\epsilon$(U) = --1.96, --2.06, and --2.16, and
  the uppermost line includes no U. The dotted line in the left panel
  represents a synthetic spectrum with an estimated U abundance if U
  was not radioactive and had not decayed over the past
  $\sim$13\,Gyr. From \citet{he1523}. }

\end{center}
\end{figure*}

The first r-II star, CS~22892-052, was discovered more than a decade
ago in the HK survey \citep{McWilliametal}. Its Th/Eu ratio yielded an
age of $14$\,Gyr \citep{sneden03}.  A second object, CS~31082-001, in
which both Th and U were measurable, was then also shown to be 14\,Gyr
old, based on its U/Th abundance ratio \citep{Hilletal:2002}. A large
campaign was then initiated to observe metal-poor candidate stars from
the Hamburg/ESO survey to discover such objects.  Identifying stars
with a strong Eu II line at 4129\,{\AA}, the main r-process indicator
in stellar spectra, led to the discovery of several strongly
r-process-enhanced (r-II) stars (see \citealt{heresII}, and references
therein) and dozens of mildly enriched (r-I) objects.  With the
exception of CS~31082-001, many of these r-process-enhanced stars
could be dated with only the Th/Eu chronometer.

In CS~31082-001, chronometer ratios involving any stable elements
(e.g., Th/Eu) yielded {\it negative} ages. This is due to unusually
high Th and U abundances compared with values expected for these
elements from their overall r-process pattern (scaled to the
Sun). Since only the elements heavier than the 3$^{\rm rd}$
$r$-process peak are (equally) affected \citep{roederer09}, the U/Th
ratio still gives a reasonable age for this star. The behavior was
termed an ``actinide-boost'' \citep{schatz_chronometers} and indicates
an origin different from ``normal'' r-process-enhanced stars and/or
multiple r-process sites \citep{Hilletal:2002}. Since then, three more
r-process-enhanced stars with such high Th/Eu ratios ($\sim$20\% of
r-process stars) have been found \citep{honda04,lai2008}.  The
underlying physical process(es) leading to the large fraction of the
actinide-boost stars will need to be thoroughly investigated over the
next few years. It is crucial to assess whether the apparent
universality of the r-process of elements with $Z\ge56$ seen in
``regular'' r-process-enhanced stars remains truly universal, or if it
is simply an artefact of our limited understanding of the r-process
and/or insufficient numbers of such stars.

For only one r-II star has it so far been possible to determine ages
from more than just one chronometer ratio.  The bright giant
HE~1523$-$0901 ($V = 11.1$) has the strongest enhancement in r-process
elements so far observed, $\mbox{[r/Fe]}=1.8$
\citep{frebel_bmps,he1523}, and among the measured neutron-capture
elements are Os, Ir, Th, and U.  Its average stellar age of
$\sim$13\,Gyr is based on seven chronometers Th/r, U/r and U/Th
involving combinations of Eu, Os, Ir, Th and U.  Only in cool r-II red
giants can the many weak and often partially blended neutron-capture
features be measured. The two most challenging examples are the
extremely weak U\,II line at 3859\,{\AA} and the even weaker Pb\,I
line at 4057\,{\AA}; these two lines are the strongest optical
transitions of the two elements. (It should be mentioned
that both lines are blended with a strong CH feature.  Hence, U and Pb
can be detected best in stars with subsolar carbon abundances, which
minimizes the blending effect. In CS~22892-052, a carbon-rich
r-process-enhanced star, neither element will ever be measurable.)

Figure~\ref{Fig:U_region} shows the spectral region around the U line
in HE~1523$-$0901. To be useful for age determination, r-II stars should be
as bright as possible (preferably $V<13$) so that very high-resolution
spectra with very high $S/N$ can be collected in reasonable observing
times. A successful U measurement requires a high-resolution spectrum
(R $>$ 60000) with $S/N$ of at least 350 per pixel at 3900\,{\AA}. A
Pb measurement may be attempted in a spectrum with $S/N\sim500$ at
4000\,{\AA}. Only {\it three} stars have had U measurements. They are
HE~1523$-$0901, CS~31082-001, and a somewhat uncertain detection in BD
+17$^{\circ}$ 3248, of which the age of HE~1523$-$0901 is currently
the most reliable.

Compared with Th/Eu, the U/Th ratio is more robust against
uncertainties in the theoretically derived production ratio because Th
and U have similar atomic masses (for which uncertainties largely
cancel out; e.g., \citealt{wanajo2002}). Hence, stars displaying Th {
  \it and} U are the best for age determination. For the same reason,
stable elements of the 3$^{rd}$ r-process peak ($76 \le Z \le 78$) are
best used in combination with Th and U.  Nevertheless, realistic age
uncertainties range from $\sim$2 to $\sim$5\,Gyr depending on the
chronometer ratio (see \citealt{schatz_chronometers}, and
\citealt{he1523} for discussions). In any case, age measurements of
old stars naturally provide an important independent lower limit to
the age of the Universe, currently inferred to be 13.73 $^{+0.16}
_{-0.15}$\,Gyr with WMAP \citep{WMAP}. In the absence of an
age-metallicity relationship for field halo stars, the
nucleo-chronometric ages thus demonstrate that these metal-deficient
stars, with [Fe/H] $\sim-3$, are indeed very ancient, leading to the
corollary that stars of similar [Fe/H], but with no overabundance in
r-process elements, have a similar age.

The r-process-enhanced stars fortuitously bring together astrophysics and
nuclear physics by acting as a ``cosmic laboratory'' for both fields
of study. They provide crucial experimental data on heavy-element
production that is not accessible to nuclear physics experiments.
Since different r-process models often yield different final r-process
abundance distributions, particularly in the heavy mass range,
self-consistency constraints are very valuable. The stellar abundance
triumvirate of Th, U, {\it and} Pb provides such constraints. These
three elements are intimately coupled not only with each other but
also to the conditions (and potentially also the environment) of the
r-process. Pb is the $\beta$- plus $\alpha$-decay end-product of all
decay chains in the mass region between Pb and the onset of dominant
spontaneous fission above Th and U. It is also built up from the decay
of Th and U isotopes. All three measurements thus provide important
constraints on the poorly understood decay channels.  They offer an
opportunity to improve r-process models which, in turn, facilitates
the determination of improved initial production ratios necessary for
the stellar age dating.

%\vspace{1.1cm}
\section{COSMOGONY}

\subsection{The Early Universe}\label{Sec:early_uni} 

Simulations of the hierarchical assembly of galaxies within the Cold
Dark Matter (CDM) paradigm pioneered by \citet{white&rees78} and today
referred to as $\Lambda$CDM (e.g., \citealt{diemand07};
\citealt{springel}) demonstrate that structure formation in the
Universe proceeded hierarchically, with small dark matter halos
merging to form larger ones which eventually led to the build-up of
larger galaxies like the Milky Way. This is further described in
Section~\ref{Sec:mw}. The very first stars (Population III) formed in
small, so-called minihalos of $\sim$10$^{6}$\,M$_{\odot}$ that
collapsed at $z\simeq$ 20 -- 30 \citep{tegmark97} a few hundred
million years after the Big Bang.  Hydrodynamical cosmological
simulations have shown that due to the lack of cooling agents in
primordial gas, significant fragmentation was largely suppressed so
that these first objects were very massive, of order
$\sim$100\,M$_{\odot}$ (a ``top-heavy'' initial mass function; e.g,
\citealt{bromm&larson04} and references therein) and likely fast
rotating.  This is in contrast to low-mass stars ($<$1\,M$_{\odot}$)
dominating today's mass function (often referred to as the Salpeter
mass function).

In this scenario, the massive first generation stars (designated
Population\,III.1) soon exploded as core-collapse SNe leaving remnant
black holes (for progenitor masses of 25\,M$_{\odot}<$ M
$<140$\,M$_{\odot}$ and M~$>~260$\,M$_{\odot}$), or even more
energetic pair-instability SNe (PISNe; 140\,M$_{\odot}$ $<$ M $<$
260\,M$_{\odot}$; \citealt{heger2002}) with complete disruption.  A
specific, predicted ``chemical fingerprint'' of the putative PISN
explosion has not (yet) been identified in any metal-poor star. (Given
that luminous supernovae (having peak M$_{\rm V} < -21$) in external
galaxies have been associated with massive progenitor stars (M $>$ 100
M$_{\odot}$) in low-metallicity regions (\citealt{neill11}) it can not
be excluded, however, that such a signature will be found.)  In their
final stages, all these massive objects provided vast amounts of
ionizing radiation (and some of the first metals) that changed the
conditions of the surrounding material for subsequent star formation,
even in neighboring minihalos. Partially ionized primordial gas
supported the formation of first H$_{2}$ and then HD, which in turn
facilitated more effective cooling than would be possible in neutral
gas. Any metals or dust grains left behind from PISNe would have
similar cooling effects.  Hence, there likely was a second generation
of metal-free stars (Population\,III.2) that, for the first time,
included stars of somewhat smaller masses (M~$\sim10\,$M$_{\odot}$.)
This generation, however, was still top-heavy, in contrast to typical
present-day stars (M~$\sim1\,$M$_{\odot}$).  For a recent review of
this topic, see \citet{bromm09}.  Soon thereafter, the
first low-mass metal-poor stars were born. In their atmospheres they
locked in the chemical fingerprint of the very first supernova
explosions. Investigating the chemical abundances of the most
metal-poor stars is thus the only way to gain detailed information of
the nature and properties of the first stars without going to the very
high redshift Universe.  Even with the James Webb Space Telescope (see
http://www.jwst.nasa.gov), the sensitivity will not be sufficient to
directly observe the first stars. The first galaxies may, however,
just be reachable.

\begin{figure}[!t]
\begin{center}
%\includegraphics[clip=true, width=11.0cm,bbllx=30, bblly=100,
%   bburx=530, bbury=430]{fn_fig22_v2.eps}
\includegraphics[width=8.6cm,angle=0]{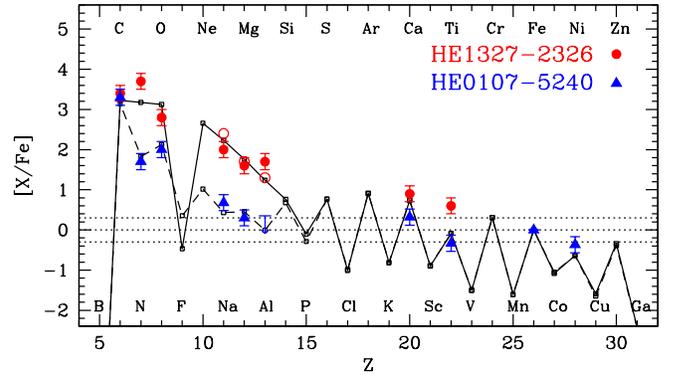}
\caption{\label{Fig:sn} \small Abundance distribution vs. atomic
  number for the two most Fe-poor stars {\hen} and {\hea} (circle and
  triangles, respectively) compared with the best fit models of
  ``mixing and fallback'' core-collapse SNe (from \citealt{nomoto06}).
  The middle dotted line shows the solar abundance ratio.  See text
  for more details on the SNe models.}
\end{center}
\end{figure}

As described in Section~\ref{Sec:census}, four halo stars with the
exceptionally low values of $\mbox{[Fe/H]}<-4.3$ are currently
known. An immediate question arises: do their abundance patterns
reflect the chemical yields of the first stars? Before attempting to
answer this question, their detailed chemical abundances have to be
considered.  The most striking features in both stars with
$\mbox{[Fe/H]}<-5.0$ are the extremely large overabundances of the CNO
elements ($\mbox{[C,N,O/Fe]}\sim+2$ to +4). HE~0557$-$4840 (with
$\mbox{[Fe/H]}=-4.8$) partially shares this signature by also having a
fairly large value of [C/Fe].  SDSS~J102915+172927, however, does
not. This object has an abundance signature that resembles typical
metal-poor halo stars, including its carbon signature, and no
exceptional over- or underabundances.  In contrast, other element
ratios, [X/Fe], are somewhat enhanced in HE~1327$-$2327 with respect
to stars with $-4.0<\mbox{[Fe/H]}<-2.5$, but less so for the giants
HE~0107$-$5240 and HE~0557$-$4840.  No neutron-capture element was
detected in HE~0107$-$5240, HE~0557$-$4840, or SDSS~J102915+172927,
whereas, unexpectedly, a large value of [Sr/Fe] = 1.1 was obtained for
HE~1327$-$2326.  Despite expectations, and as discussed in
Section~\ref{Sec:lithium}, lithium was not detected in either the
relatively unevolved subgiant HE~1327$-$2326 or the dwarf
SDSS~J102915+172927. The lithium abundance upper limits are
$\log\epsilon ({\rm Li})<0.7$ \citep{he1327_uves} and $<1.1$
\citep{caffau11}, respectively.  These results are extremely
surprising.  Given that {\hea} and SDSS~J102915+172927 have {\teff} =
6180\,K and 5810\,K, respectively, one would expect them to lie on the
Spite Plateau, with $\log\epsilon ({\rm Li}) = 2.3$.  Somewhat
unsatisfactory conjectures that might explain the non-detection
include: (1) Li at the epoch of lowest metallicity was below the
abundance of the Spite Plateau due to its destruction early in the
Universe (see e.g., \citealt{piau06} for an interesting scenario), and
(2) Li was destroyed by phenomena associated with (not yet detected)
binarity. Progress will probably only be made when more
near-main-sequence-turnoff stars with [Fe/H] $\la$ -- 4.0 are
discovered which permit clarification of this issue.

\begin{figure*}[!tbp]
\begin{center}
\includegraphics[clip=true, width=15.5cm,bbllx=30, bblly=450,
  bburx=540, bbury=700]{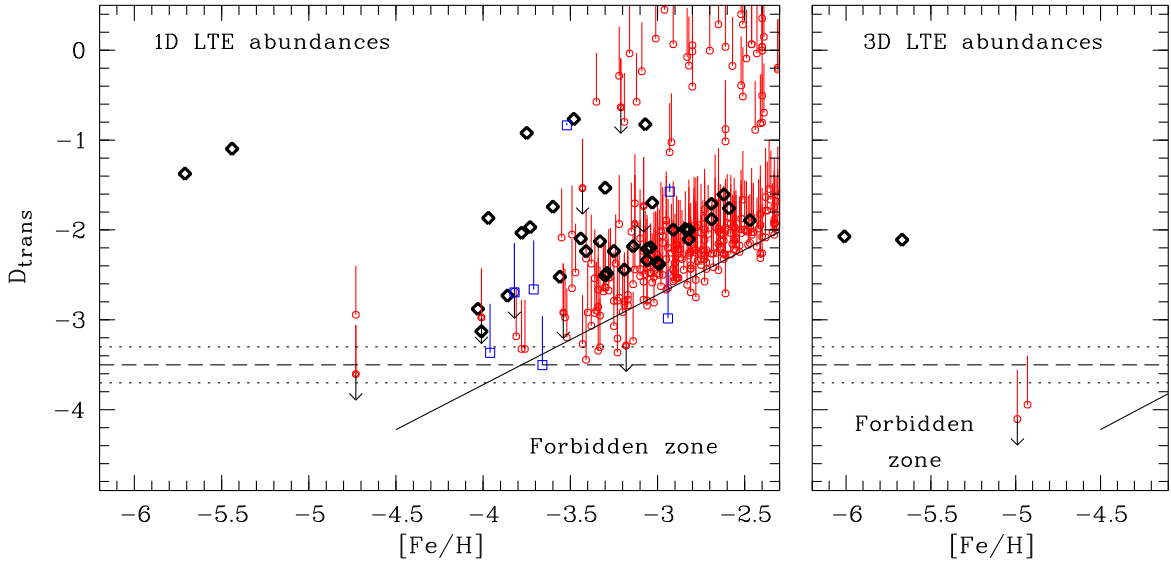}
\caption{\label{Fig:dtrans} \small Left panel: Transition
  discriminant, $D_{\rm trans}$, for Galactic halo (small red circles,
  thick black diamonds) and dwarf galaxy (blue squares) metal-poor
  stars as a function of [Fe/H], based on 1D abundances. Black
  diamonds show stars with $D_{\rm trans}$ values calculated from
  their C and O abundances. Red circles and blue squares depict lower
  $D_{\rm trans}$ limits based on only a known C abundance. The
  corresponding vertical bars show the potential range of $D_{\rm
    trans}$ for a given star assuming O to be tied to C within the
  range $-0.7 < \mbox{[C/O]} < 0.2$. (If an upper limit on O is
  available and less than the maximal assumed O abundance, the bar is
  correspondingly shorter.)  Circles or squares with bars plus
  additional arrows indicate interesting cases were only upper limits
  of C abundances are available and nothing is known about the O
  abundance.  The solid line represents the solar C and O abundances
  scaled down with [Fe/H], while dashed and dotted lines display the
  transition discriminant $D_{\rm trans}=-3.5$ together with
  uncertainties. The ``Forbidden zone'' indicates the region with
  insufficient amounts of C and O for low-mass star formation.  (Based
  on Figure~1 of \citealt{dtrans} with recent additions from the
  literature such as \citealt{caffau11}).  Right panel: Same as left
  panel, but for [Fe/H] $< -4.1$ and using 3D carbon and oxygen
  abundances, for the four most iron-poor stars.}
\end{center}
\end{figure*}

Both HE~0107$-$5240 and HE~1327$-$2326 are benchmark objects with the
potential to constrain various theoretical studies of the early
Universe, such as the formation of the first stars, calculations of
Population\,III SN yields, and the earliest chemical evolution.
Several different scenarios have been offered that seek to explain the
highly individual abundance patterns of both stars as early, extreme
Population\,II, stars that display the ``fingerprint'' of just one
Population\,III SN. These include: (1) ``mixing and fallback'' models
(\citealt{UmedaNomotoNature}; \citealt{iwamoto05}; \citealt{nomoto06})
of a faint (i.e., low energy) 25\,M$_{\odot}$ supernova in which a
large amount of C, N, and O but little Fe is ejected, while a large
fraction of Fe-rich ejecta is postulated to fall back onto the newly
created black hole. (See Figure~\ref{Fig:sn} for comparison of the
observed and predicted abundances for HE~0107$-$5240 and
HE~1327$-$2326); (2) the modeling of \citet{heger_woosley10} who fit
the observed stellar abundances by searching for a match within a
large grid of Population\,III SN yields.  Their best fit involved
typical halo stars with a power-law IMF in the range $M=11 -
15\,$M$_{\odot}$, low explosion energy, and little mixing; and (3) the
investigation of \citet{meynet2005} who explored the influence of
stellar rotation on elemental yields of 60\,M$_{\odot}$
near-zero-metallicity SNe.  Mass loss from rotating massive
Population\,III stars qualitatively reproduces the CNO abundances
observed in HE~1327$-$2326 and other carbon-rich metal-poor stars.  In
a somewhat different model, \citet{suda} proposed a scenario in which
the abundances of HE~0107$-$5240 originated in a Population\,III
binary system that experienced mass transfer of CNO elements from the
more massive companion during its AGB phase, together with subsequent
accretion of heavy elements from the ISM onto the (less massive)
component now being observed.  Along the same lines,
\citet{campbell10} suggested a binary model for HE~1327$-$2326 in
terms of s-process nucleosynthesis and mass transfer via a stellar
wind. Qualitatively, the high C,N,O and Sr stellar abundances could be
explained this way if the star were in a wide binary.  That said,
neither HE~0107$-$5240 nor HE~1327$-$2326 shows radial velocity
variations that would indicate close binarity.

Stars with $\mbox{[Fe/H]}>-4.3$ and ``classical'' halo abundance
signatures have also been reproduced with Population\,III SN
yields. Average abundance patterns of four non-carbon-enriched stars
with $-4.2<\mbox{[Fe/H]}<-3.5$ were modeled with the yields of massive
($\sim$30 -- 50\,M$_{\odot}$), high explosion energy ($\sim$20 --
40$\times\,10^{51}$\,ergs), Population\,III hypernovae
\citep{tominaga07_b} and also fit with integrated yields of a small
number of Population\,III stars \citep{heger_woosley10}. Special types
of SNe or unusual nucleosynthesis yields have been considered for
stars with chemically peculiar abundances, e.g., high Mg. It is,
however, often difficult to explain the entire abundance pattern in
this way. Abundances of additional stars with {Fe/H] $<$ --4.0, as
well as a better understanding of the explosion mechanism and the
effect of the initial conditions on SNe yields are required to arrive
at a more comprehensive picture of the extent to which metal-poor
stars reflect the ejecta of the original Population\,III objects or,
alternatively, those of later generations of SNe.

Some metal-poor stars display abundance ratios of a few elements that
differ by large amounts from the general halo trend (e.g., C, Mg).
The level of chemical diversity also increases towards the lowest
metallicities.  For instance, as discussed in
Section~\ref{Sec:carbon}, a large fraction of the most metal-poor
stars is very carbon-rich (i.e., $\mbox{[C/Fe]}>0.7$).  In the
compilation of \citet{frebel10} the C-rich fraction of stars with
[Fe/H]$<$ --2.0 is $\sim$0.17.  Most significantly, the fraction
increases with decreasing metallicity (see
Figure~\ref{Fig:lowfe_MDF}); and indeed, three of the four stars with
$\mbox{[Fe/H]}<-4.3$ are extremely carbon-rich. Reasons for this
general behavior remain unclear. Could there be a cosmological origin
for the large fraction of carbon-rich stars? Ideas for the required
cooling processes necessary to induce sufficient fragmentation of the
near-primordial gas to enable low-mass star formation include cooling
based on enhanced molecule formation due to ionization of the gas,
cooling through metal enrichment or dust, and complex effects such as
turbulence and magnetic fields \citep{bromm09}.  Fine-structure line
cooling through C\,I and O\,II was suggested as a main cooling agent
\citep{brommnature}.  These elements were likely produced in vast
quantities in Population\,III objects (see Section~\ref{Sec:carbon}),
and may have been responsible for the ISM reaching a critical
metallicity, sufficient for low-mass star formation.

The existence and level of such a ``critical metallicity'' can be
probed with large numbers of carbon and oxygen-{\it poor} metal-poor
stars: if a threshold exists, all of these objects should have a
combination of C and/or O abundances {\it above} the threshold for a
critical metallicity. A transition discriminant was defined by
\citet{dtrans}, which has since been slightly revised to $D_{\rm
  trans}={\rm log} (10^{{\rm [C/H]}} + 0.9 \times 10^{{\rm [O/H]}})$
(V. Bromm 2011, private communication). No low-mass metal-poor stars
should exist below the critical value of $D_{\rm trans}=-3.5$.

As can be seen in Figure~\ref{Fig:dtrans}, at metallicities of
$\mbox{[Fe/H]}\gtrsim-3.5$, most stars have C and/or O abundances that
place them well above the threshold. They simply follow
the solar C and O abundances scaled down to their respective Fe
values. Naturally, this metallicity range is not suitable for directly
probing the very early time. Below $\mbox{[Fe/H]}\sim-3.5$, however,
the observed C and/or O levels must be {\it higher} than the Fe-scaled
solar abundances to be above the critical metallicity.
Indeed, apart from one object, none of the known lowest-metallicity
stars appear to have $D_{\rm trans}$ values or limits below the
critical value, consistent with this cooling theory. The exception is
{\sdss}, which has an upper limit for carbon of only
$\mbox{[C/H]}<-3.8$ (1D) and $<-4.3$ (3D). Assuming the above [C/O]
range, this leads to $D_{\rm trans} < (-3.6$ to $-3.0$) (1D) and $<
(-4.1$ to $-3.5$) (3D).  However, only with a known O abundance can
this low $D_{\rm trans}$ value be conclusively determined. As can be
seen in Figure~\ref{Fig:dtrans}, several other stars also have values
that are close to $D_{\rm trans}=-3.5$, based on their 1D
abundances. Very high $S/N$ spectra suitable for measurements of very
weak CH and OH molecular features will be required to determine
exactly how close the $D_{\rm trans}$ of these objects are to the
critical value.

%Examples are one star in the ultra-faint dwarf galaxy Bo\"otes\,I
%  with $\mbox{[Fe/H]}=-3.7$ and [C/H] = --3.4 (see
%  \citealt{norris10}) has $D_{\rm trans}=-3.0$ (in the absence of a
%  detection of oxygen, [O/C] = +0.3 has been assumed, since it is
%  similar to values found for Galactic halo stars).  Another, in the
%  classical dwarf galaxy Sculptor, with $\mbox{[Fe/H]}=-3.8$ and
%  $\mbox{[C/H]}<-3.6$ \citep{scl} has $D_{\rm trans}<-3.2$, where
%  again [O/C] = 0.3 has been assumed. The latter object is, however,
%  a cool giant, and may have experienced internal mixing that would
%  have lowered its surface [C/H], likely by several tenths (dex).
%  Keeping this in mind, the limit for this star may be very close to
%  the $D_{\rm trans}$ threshold.  

% AF: this para isn't needed anymore since there are several stars
%  that have lower Dtrans limits and the mixing correction is already
%  everywhere taken into account.

% AF: we don't need to mention SDSS here again, after it has been mentioned
%in the same capacity in the previous (now much shorter) paragraph.
  The likely exception of {\sdss} and several interesting ``border
  line'' cases notwithstanding, this cooling theory suggests that at
  low metallicity, carbon excesses are a requirement for the formation
  of most low-mass metal-poor stars. This is qualitatively in line
  with the empirical finding of a large fraction of carbon-rich stars
  and may thus reflect a generic avenue for low-mass star formation.
  Individual objects, of course, could be the result of unusual
  circumstances or different mechanism. For example, {\sdss} could
  have formed from a gas cloud that was primarily cooled by dust
  grains, rather than atoms, made by the first stars. If future data
  show that the \textit{majority} of the most metal-poor stars have
  $D_{\rm trans}<-3.5$, then dust cooling (inducing a much lower
  critical metallicity) would be a dominant cooling mechanism in the
  early Universe.

\begin{figure*}[!t]
\begin{center}
\includegraphics[width=16.0cm,angle=0]{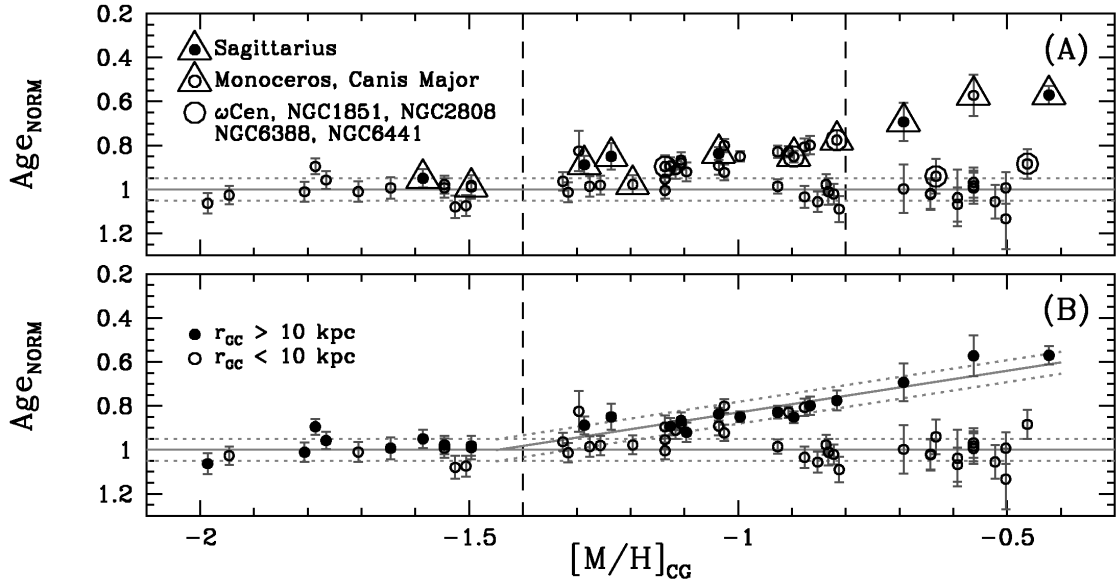}
  
\caption{\label{Fig:Marin-Franch} \small Relative ages, Age$_{\rm
    NORM}$, for the Galactic globular clusters as a function of
  cluster metallicity, [M/H] (where [M/H] = [Fe/H] + log(0.638f + 0.362), and
  log(f) =[$\alpha$/Fe]), from the work of Mar\'{i}n-Franch et
  al.\ (2009, Figure 13).  Approximate absolute ages may be obtained
  as 12.8$\times$Age$_{\rm NORM}$.}

\end{center}
\end{figure*}

\subsection{The Milky Way}\label{Sec:mw}

During the past half century, two basically different observationally
driven paradigms were proposed for the formation of the Galactic halo.
The first was the monolithic collapse model of \citet{els} (hereafter
ELS), and the second the accretion model of \citet{sz78} (hereafter
SZ).  At the same time, \citet{white&rees78} proposed, in a more
general context, their CDM hierarchical clustering paradigm in which
``The entire luminosity content of galaxies ... results from the
cooling and fragmentation of residual gas within the transient
potential wells provided by the dark matter.''  ELS predicted a very
rapid collapse phase (of a few 10$^8$\,yr), and a dependence of
kinematics on abundance together with a radial abundance gradient for
halo material.  SZ, in contradistinction, predicted a longer formation
period of a few 10$^9$\,yr, no dependence of kinematics on abundance,
and no radial abundance gradient.  Not too surprisingly, perhaps,
neither gives a complete explanation of the more complicated reality.
On the one hand, early work revealed no dependence of kinematics on
abundance for [Fe/H] $\la$~--1.7 (see \citealt{chiba&beers00}, and
references therein) , while on the other, globular cluster age
measurements demonstrated that although some clusters were
significantly younger than the majority, the age spread was small for
the bulk of the system.  Figure~\ref{Fig:Marin-Franch}, from the
recent work of \citet{marinfranch09}, presents the relative ages of 64
clusters as a function of metallicity, [M/H], which illustrates this
point.

A turning point in the discussion came with the discovery by
\citet{Ibataetal95} of the Sagittarius dwarf galaxy, which has been
captured by the Milky Way and is currently being torn apart in its
gravitational field.  Some six of the Galactic globular clusters are
believed to have once been part of the Sgr system.
\citet{marinfranch09} comment on similar over-densities in Monoceros
and Canis Major that may contain several other globular clusters and
be associated with similar accretions.  Against this background, it is
then very instructive to consider the detail of
Figure~\ref{Fig:Marin-Franch}.  \citet{marinfranch09} identify two
groups of globular clusters: ``a population of old clusters with an
age dispersion of $\sim$5\% (i.e., $\sim$0.6\,Gyr) and no
age-metallicity relationship, and a group of younger clusters with an
age-metallicity relationship similar to that of the globular clusters
associated with the Sagittarius dwarf galaxy.''  Two thirds of the
sample belong to the old group, one third to the younger.

As noted above in Section~\ref{Sec:mdf_fieldstars}, there has been
growing evidence that the field halo stars of the Milky Way comprise
more than one population.  The reader should consult \citet{carollo10}
for the development of the case that the Galaxy's halo contains an
inner and an outer component.  They report the following essential
differences between the two components, which are dominant interior
and exterior to $\sim$15\,kpc: (1) the inner component is more
flattened than the outer component, with axial ratio (c/a) values of
$\sim$0.6 and 1.0, respectively; (2) the inner component has small
prograde systemic rotation, $\langle$V$_{\phi}\rangle$ = +7 $\pm$
4\,{\kms} (i.e., rotating in the same sense as the Galactic disk),
while the outer has retrograde rotation $\langle$V$_{\phi}\rangle$ =
--80 $\pm$ 13\,{\kms}; and (3) the inner component is more metal-rich,
with peak metallicity [Fe/H] = --1.6, while the outer one has peak
metallicity [Fe/H] = --2.2.

\begin{figure*}[!tbp]
\begin{center}
\hspace{-5mm} 
\includegraphics[width=0.54\textwidth]{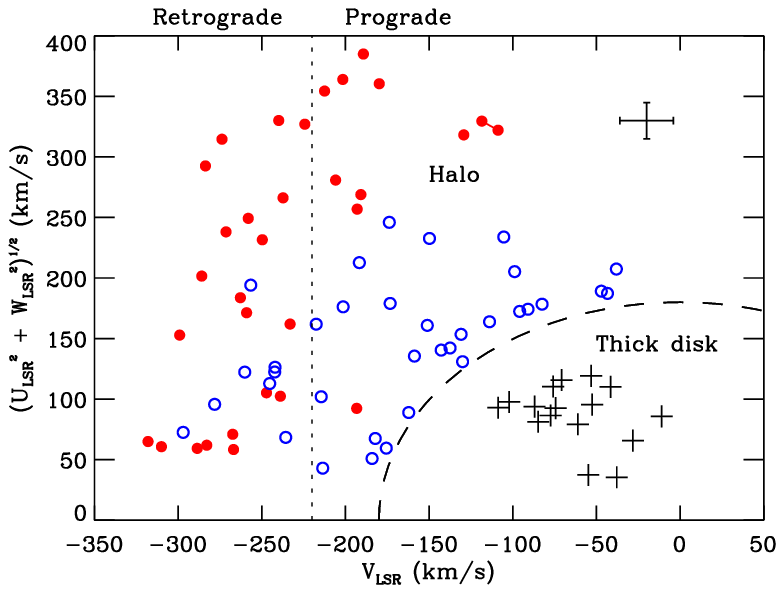}
\includegraphics[width=0.47\textwidth]{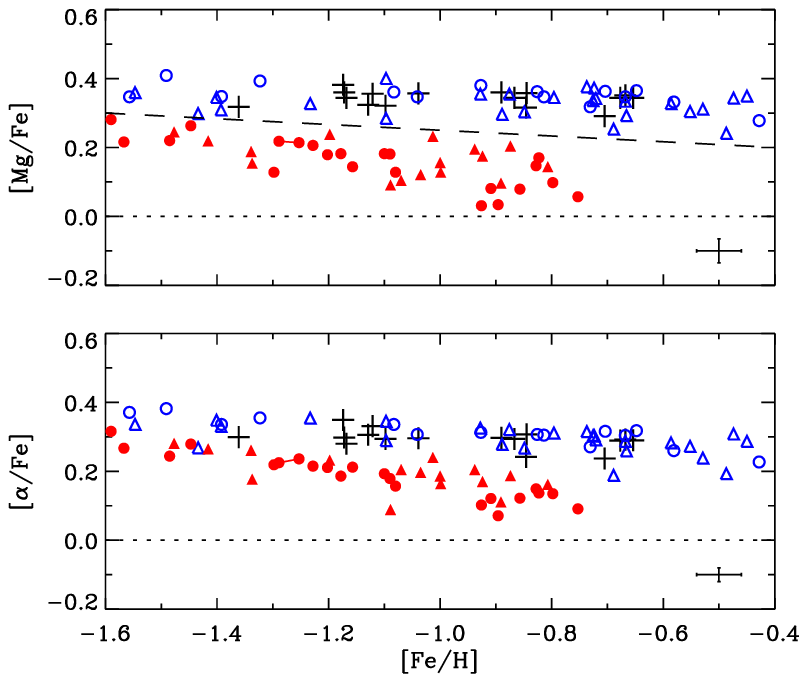}
\caption{\label{Fig:nissen10} \small Left: kinematics (in which U, V,
  W are velocity components in the Galactic frame) and right:
  [$\alpha$/Fe] vs. [Fe/H] for metal-poor Milky Way stars, from the
  work of \citet{nissen10}.  Circles and triangles refer to halo (and
  crosses to thick disk) stars, respectively.  Note the strongly
  correlated relative positions of the filled and open circles in the
  two panels.  See text for discussion.}
\end{center}
\end{figure*}

Against this background, \citet{morrisonetal09} have reported another,
more highly flattened halo component, with c/a $\sim 0.2$, which has
``a small prograde rotation ... supported by velocity anisotropy, and
contains more intermediate-metallicity stars (with --1.5 $<$ [Fe/H]
$<$ --1.0) that the rest of [the] sample''.

While the detailed nature and relationships of these components
remain to be fully understood, it seems likely the answer will be found
within the hierarchical $\Lambda$CDM paradigm reported above. The work
of \citet{zolotov09}, for example, while supporting the SZ paradigm of
halo formation, also produces a dual halo configuration of
``{\it in situ}'' and ``{\it accreted}'' components, not unlike those
envisaged in the ELS and SZ observational paradigms.  Remarkably,
these paradigms were first established on essentially observational
grounds only. They are now being explained in terms of a theoretical
framework based on tracing the dark matter evolution from initial
density fluctuations early in the Universe.

Further support for a two component model comes from recent work of
\citet{nissen10}, who have investigated the abundances of
$\alpha$-elements in the abundance range --1.6 $<$ [Fe/H] $<$ --0.4,
as a function of kinematics, with a view to comparing (in
\citealt{zolotov09} terminology) the ``{\it in situ}'' and ``{\it
  accreted}'' components.  Their results are shown in
Figure~\ref{Fig:nissen10}.  In the right panel one sees a large spread
in [$\alpha$/Fe], at fixed [Fe/H], that correlates strongly with
position in the kinematic (so-called ``Toomre'') diagram on the left.
The simplest and also extremely significant interpretation of this
figure is that stars with [$\alpha$/Fe] $\sim$+0.3 to +0.4, with
prograde kinematics, are part of the ``in situ'' component, while
those with [$\alpha$/Fe] $\la$ +0.3, on principally retrograde orbits,
belong to the ``accreted'' component.  The reader will recall from
Section~\ref{Sec:dg_evol} that low [$\alpha$/Fe] is a key signature of
the Milky Way's dwarf galaxies in the range --1.5 $<$ [Fe/H] $<$ 0.0
(see also Tolstoy et al.\ 2009, their Figure 11).  Said differently,
Figure~\ref{Fig:nissen10} is consistent with the view that dwarf
galaxies have played an important role in the formation of the Milky
Way halo.

A complementary way to study the origin of the Milky Way, its halo,
and similar large galaxies more generally, is through large-scale
$\Lambda$CDM simulation of the growth of structure formation. A
prominent issue with Milky Way-size halos at redshift $z=0$ is the
predicted large number of CDM substructures that surround such a
galaxy.  The number of observed dwarf galaxies surrounding the Milky
Way is, however, much lower and does not agree with such
predictions. This mismatch has been termed the ``missing-satellite''
problem (e.g., \citealt{moore}).

It is thus apparent that many more questions about galaxy assembly and
evolution still need to be resolved. Crucially, it remains to be seen
to what extent small dark halos contained baryonic matter,
subsequently observed as gas and stars, and how they evolved with
time. One way to learn about the luminous content of small sub-halos is to
investigate in detail the surviving dwarf galaxies, in particular the
ultra-faint systems, that orbit the Milky Way.  Studying the onset of
star formation and associated chemical evolution in these satellites
will provide some of the currently missing information for our
understanding of how the observed properties of small, faint systems
relate to the dark matter substructures that built up larger galaxies.

\section {CONCLUSIONS \& FUTURE PROSPECTS} \label{Sec:conclusion}

% ------  CONCLUSIONS ------------
Old metal-poor stars can be employed as tools to learn about the
conditions in the early Universe. The scientific topics that can be
addressed in this way are numerous, and this chapter describes the
most prominent questions to which metal-poor stars can provide unique
insights. These include the origin and evolution of the chemical
elements, the relevant nucleosynthesis processes and sites, and the
overall chemical and dynamical history of the Galaxy. By extension,
the abundance patterns in metal-poor stars provide constraints on the
nature of the first stars and the initial mass function, and the chemical
yields of first/early SNe. Moreover, studying metal-poor stars in
dwarf galaxies opens up ways to learn about early star and early
galaxy formation processes, including the formation of the Galactic
halo through hierarchical assembly.

Our review has highlighted the tension between the approximations
inherent in 1D/LTE model atmosphere abundance analyses, on the one
hand, and the more physically realistic 3D/non-LTE (and more
computationally challenging) formalism, on the other.  Given abundance
differences $\sim$0.5 -- 0.9\,dex between the two formalisms for many
elements, there is an urgent need for comprehensive investment in
self-consistent 3D/non-LTE modeling of the relevant regions of
{\teff}/{\logg}/[Fe/H] space.

% ------  FUTURE PROSPECTS ------------
The most metal-deficient stars are extremely rare, but past surveys
for metal-poor halo stars have shown that they can be systematically
identified through several selection steps. Typically, the metal-poor
halo stars found to date are located no further away than $\sim$10 --
15\,kpc, with $B \la 16$.  This brightness limit ensures that adequate
$S/N$ spectra can be obtained in reasonable observing times with
existing telescope/instrument combinations.  The outer halo beyond
$\sim$15\,kpc, however, is largely unexplored territory, at least in
terms of high-resolution spectroscopy.  Recent work has shown that the
ultra-faint dwarf galaxies orbiting the Galaxy contain larger
fractions of extremely metal-poor stars (i.e., [Fe/H] $<$ --3.0) than
does the Galactic halo. That said, there is a high price to be paid in
order to observe these objects because most of them are extremely
faint. Those that are currently observable with 6 -- 10\,m telescopes
are only the brightest in a given system, and these are usually
located on the upper RGB. At the limit are objects at 19$^{\rm th}$
magnitude that can just be observed at high spectral resolution,
requiring exposure times up to $\sim$10\,h per star in order to reach
the minimum useful $S/N$ ratio in the final spectrum. This is feasible
only for individual stars, not for large-scale investigations.
Objects lower on the RGB or even the main sequence ($>21$\,mag) are
out of reach, even for medium-resolution studies.

Over the next few years all of these brightest dwarf galaxy stars will
have been observed. What then? Either new larger telescopes, or
additional dwarf galaxies that harbor more observable stars, are
required. Addressing both options is currently underway. To chemically
characterize the Galactic halo in detail (including its streams,
substructures and satellites) wide-angle surveys with large volumes
are needed. The Australian Skymapper photometric survey (to begin in
2011) is optimized for stellar work. It will provide a wealth of
metal-poor candidates in need of detailed high-resolution follow up to
determine their abundances.  The footprint of this project will be
some three times larger than that of the HES.  Newly discovered stars
with B $\lesssim16$ will enable an important advance in stellar
archaeology by (hopefully) trebling the number of ``bright'' objects
available for high-resolution abundance studies with existing
facilities.  A significant fraction of SkyMapper candidates will,
however, be too faint for practical and efficient follow-up
observations. In particular, the most metal-poor stars require high
$S/N$ to enable the detection of very weak absorption features.  These
will be the target of the high-resolution spectrographs on the new
generation of 20 -- 30\,m telescopes.  Among these new discoveries, it
is expected that more of the most metal-poor stars (e.g., those with
$\mbox{[Fe/H]}<-5.0$; \citealt{HE0107_Nature} and
\citealt{HE1327_Nature}) will be found.

With SkyMapper, many more faint dwarf galaxies are expected to be
found. Even though the brightest stars in them will still be at the
observational limit for high-resolution spectroscopy, having more of
these dwarf galaxy stars available for detailed studies will provide
new insights into the nature and evolution of these small systems and
their relationship to the building-up process(es) of the Milky
Way. Other photometric surveys such as Pan-Starrs and LSST are also
expected to yield new dwarf galaxies. These surveys, however, will
be useful for the search for metal-poor stars in dwarf galaxies
only if coupled with additional follow-up efforts, due to the lack of
sufficiently metal-sensitive filters.

In addition to these photometric surveys, the Chinese LAMOST
spectroscopic survey will provide numerous metal-poor candidates in
the northern hemisphere, all based on medium-resolution spectra. GAIA
is an astrometric space mission led by ESA, scheduled to begin
observations in 2012.  It will obtain high-precision phase-space
information for one billion stars in the Galaxy, along with the
physical parameters and the chemical composition of many of these
stars. These new data will revolutionize our understanding of the
origin, evolution, structure, and dynamics of the Milky Way as a whole
and of its components. In particular, the kinematic information (e.g.,
proper motions) that will become available for many known metal-poor
stars will enable detailed studies of how the abundances of different
populations depend on kinematics. Furthermore, a precise selection of
low-metallicity candidate stars based on, for example, extreme
kinematic signatures, will become feasible. Since this is currently
beyond reach for most metal-poor halo giants, the GAIA astrometry
should increase the yield of fainter metal-poor stars at larger
distances.

By having the opportunity to access fainter stars in the outer
Galactic halo and dwarf galaxies, the next major frontier in stellar
archaeology and near-field cosmology can be tackled.  High-resolution
follow-up of faint stars may become a reality with the
light-collecting power of the next generation of optical telescopes,
the Giant Magellan Telescope, the Thirty Meter Telescope and the
European Extremely Large Telescope. All three telescopes are currently
in the planning and design phase with completions scheduled around
2020.  When equipped with high-resolution spectrographs, such
facilities would not only permit in-depth analysis of new metal-poor
stars in the Galaxy's outer halo and dwarf galaxies, but also make it
feasible to obtain very high-$S/N$ data of somewhat brighter stars, to
permit investigation, for example, of isotopic ratios such as
$^{6}$Li/$^{7}$Li and r-process-enhanced stars, to provide crucial
empirical constraints on the nature of the site and details of
critical nucleosynthesis processes.  This is currently possible only
for the very brightest stars.  At the faintest magnitudes, individual
stars in the Magellanic Clouds and perhaps even the brightest objects
in Andromeda could be observed with high-resolution spectroscopy.
Studying massive dwarf galaxies and another spiral system that
resembles the Milky Way would provide unprecedented new insight into
the chemical evolution of large systems and their formation
process(es).

All of these new observations will be accompanied by an increased
theoretical understanding of the first stars and galaxies, SN
nucleosynthesis, the mixing of metals into the existing gaseous
medium, and feedback effects in the early Universe, as well as cosmic
chemical evolution.  New generations of hydrodynamical,
high-resolution cosmological simulations will enable a direct
investigation of chemical evolution by including more than one SN and
corresponding feedback(s), for example, in first-galaxy simulations.
These will be sufficient for tracing the corresponding metal
production and spatial distributions and enable direct comparisons
with abundance measurements in the stars of dwarf galaxies.  This in
turn will shed new light on the question of whether the least-luminous
dwarf galaxies resemble the first galaxies, and if they are early
analogs of the building blocks of the Galactic halo.

\acknowledgements

A.~F. acknowledges support through a Clay Fellowship administered by
the Smithsonian Astrophysical Observatory.  J.~E.~N. acknowledges
support from the Australian Research Council (grants DP03042613,
DP0663562, and DP0984924) for studies of the Galaxy's most metal-poor
stars and ultra-faint satellite systems.  It is a pleasure to thank
M.~Asplund, T.~C.~Beers, N.~Christlieb, G.~L.~Harris, A.~Karakas, and
H.~L.~Morrison for their perceptive comments, which led to improvement
to the manuscript.  The authors also thank N. Christlieb and E.~N.~Kirby for
supplying the high-resolution spectrum of {\hen} presented in
Figure~\ref{Fig:Hires} and Figure 10, respectively.

%\bibliography{paperI,fn_mps}
%\bibliographystyle{apj}

\end{document}